\documentclass[%
 reprint,
 amsmath,amssymb,
 aps,
superscriptaddress,]{revtex4-2}

\usepackage{graphicx}
\usepackage{dcolumn}
\usepackage{bm}

\usepackage{amsmath}
\usepackage{graphicx}
\usepackage{amssymb}
\usepackage[colorlinks=true, allcolors=black]{hyperref}
\usepackage[table]{xcolor}

\begin{document}

\preprint{APS/123-QED}

\title{High-Resolution Casimir Force Sensing Across a Superconducting Transition}%

\author{Minxing~Xu}
\altaffiliation{These authors contributed equally to this work.}
\affiliation{Department of Precision and Microsystems Engineering, Delft University of Technology, Delft, The Netherlands}
\affiliation{Kavli Institute of Nanoscience, Department of Quantum Nanoscience, Delft University of Technology, Delft, The Netherlands}

\author{Robbie~J.~G.~Elbertse}
\altaffiliation{These authors contributed equally to this work.}
\affiliation{Kavli Institute of Nanoscience, Department of Quantum Nanoscience, Delft University of Technology, Delft, The Netherlands}

\author{Ata~Keşkekler}
\affiliation{Department of Precision and Microsystems Engineering, Delft University of Technology, Delft, The Netherlands}

\author{Giuseppe~Bimonte}
\affiliation{Dipartimento di Fisica E. Pancini, Università di Napoli Federico II, Complesso Universitario di Monte S. Angelo, Via Cintia, Napoli, Italy}
\affiliation{INFN Sezione di Napoli, I-80126 Napoli, Italy}

\author{Jinwon~Lee}
\affiliation{Kavli Institute of Nanoscience, Department of Quantum Nanoscience, Delft University of Technology, Delft, The Netherlands}

\author{Sander~Otte}
\altaffiliation{a.f.otte@tudelft.nl}
\affiliation{Kavli Institute of Nanoscience, Department of Quantum Nanoscience, Delft University of Technology, Delft, The Netherlands}

\author{Richard~A.~Norte}
\altaffiliation{r.a.norte@tudelft.nl}
\affiliation{Department of Precision and Microsystems Engineering, Delft University of Technology, Delft, The Netherlands}
\affiliation{Kavli Institute of Nanoscience, Department of Quantum Nanoscience, Delft University of Technology, Delft, The Netherlands}



\begin{abstract}
The Casimir effect and superconductivity are foundational quantum phenomena whose interplay is an open question in physics, with significant implications for electron physics, quantum gravity, and high-temperature superconductivity. Determining how Casimir forces behave across a superconducting transition remains elusive due to the difficulty of realizing precise alignment, cryogenic operation, and isolating small force changes from competing effects. Recent theories predict milli-Pascal jumps in Casimir pressure across the transition, motivating experiments capable of reaching well below this regime. Here, we demonstrate an on-chip superconducting nanomechanical platform that overcomes these long-standing challenges, achieving the most parallel Casimir configurations to date. Our microchip-based parallel plates reach unprecedented area-to-separation ratios, exceeding past experiments across superconducting transitions by three orders of magnitude and yielding the strongest Casimir forces generated between compliant surfaces. Scanning tunneling microscopy (STM) directly detects the resonant motion of a suspended nanoscale plate with subatomic precision in lateral positioning and displacement, enabling suppression of van der Waals, electrostatic, and thermal effects. With verified micro-Pascal pressure resolution, our platform provides a credible entry point into a new field of quantum experiments, enabling exploration of Casimir–superconductivity interactions with the stability, parallelism, and sensitivity required to access this regime of physics. 
\end{abstract}

\maketitle

\section{Interplay of Casimir Forces and Superconductivity}

\begin{figure}[t]
\centering
\includegraphics[width=0.46\textwidth]{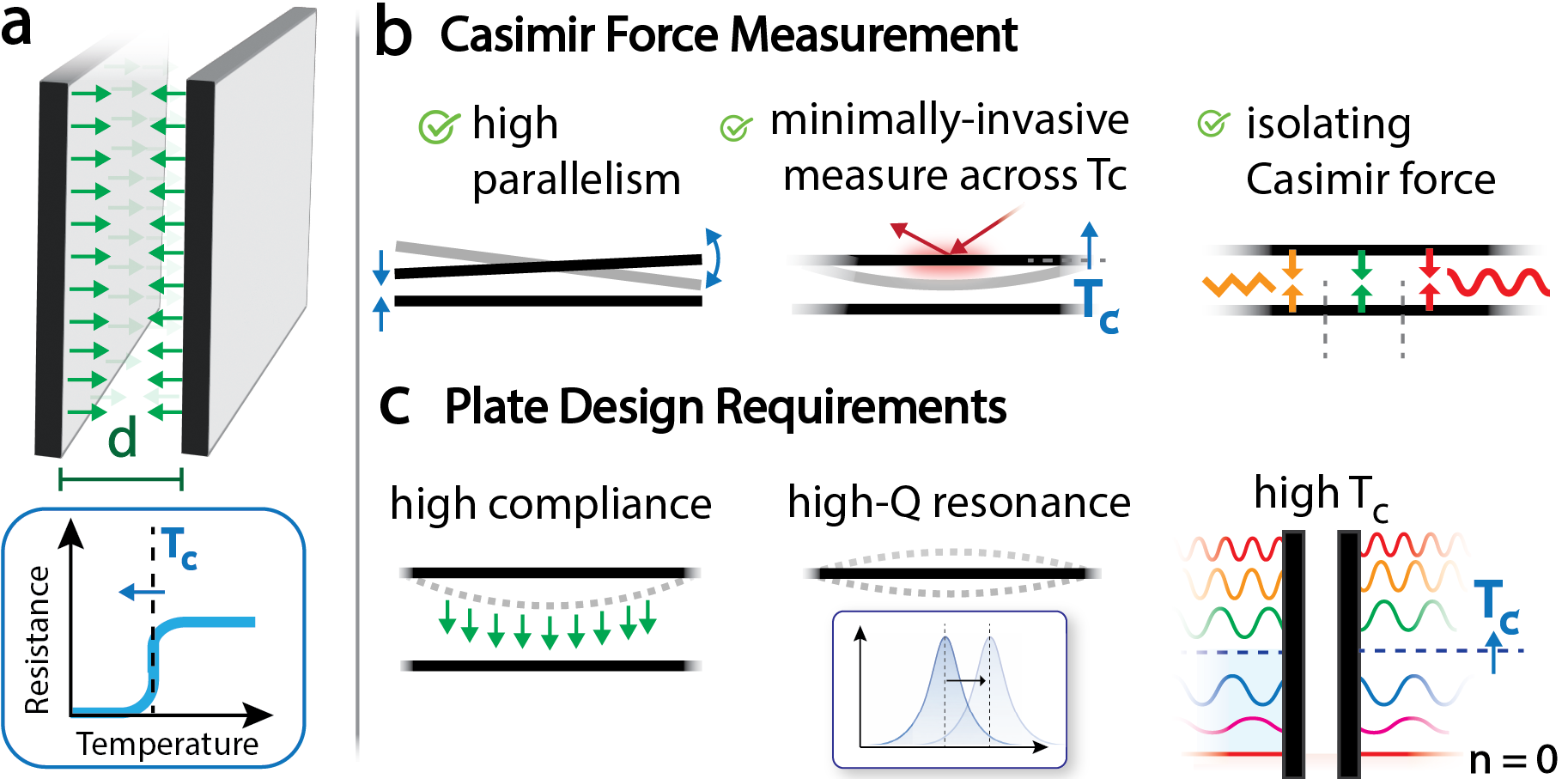} 
\caption{Superconducting Casimir Force (a) Schematic of a superconducting Casimir cavity with two plates of area $A$ separated by gap $d$. Inset: plates' superconducting transition. (b) Challenges in measuring Casimir shifts across $T_C$ include achieving high parallelism with compliant plates, minimally invasive readout, and isolating Casimir forces from electrostatic and measurement artifacts. (c) Plate design requirements are high compliance, high-$Q$ resonance, and high $T_C$ for sensitive measurements across $T_C$.}
\label{fig1}
\end{figure}

The Casimir effect and superconductivity are two landmark quantum phenomena that display striking macroscopic consequences of quantum mechanics. The Casimir effect~\cite{casimir1948attraction} arises from vacuum fluctuations of electromagnetic fields and produces an attractive force between closely spaced, uncharged surfaces. As shown in Figure~\ref{fig1}a, Casimir's original proposal consisted of two perfectly conducting parallel plates in vacuum experiencing a force that scales as $F \propto A/d^{4}$, where $A$ is plate area and $d$ the separation. Although difficult to isolate experimentally, Casimir forces were first confirmed half a century later~\cite{lamoreaux1997demonstration} and are now known to dominate nanoscale systems, particularly in micro-electromechanical systems. Superconductivity, discovered in 1911, represents an equally dramatic phenomenon: as shown in the resistance inset of Fig.~\ref{fig1}a, electrons above the critical temperature $T_C$ behave as individual particles, while below $T_C$ they condense into Cooper pairs that move without resistance, forming a macroscopic quantum state with coherence lengths spanning hundreds of nanometers~\cite{bardeen1957theory}. 

Together, these effects raise a central open question: how does the Casimir force change when the plates themselves become superconducting? Despite the fundamental importance of both of these well-known quantum phenomena, their interplay remains experimentally inaccessible due to their weak interaction and the stringent sensor requirements needed to measure small predicted Casimir pressure changes arising from superconductivity. The sharp onset of superconductivity at $T_C$ provides a natural switch to toggle this interaction, enabling precise differential measurements~\cite{gong2022electrically,torricelli2010switching}. Realizing this requires overcoming three key Casimir measurement challenges, outlined in Figure~\ref{fig1}b. \\

\noindent 1. \textbf{Extreme parallelism:} maintaining a nearly constant sub-micron gap over a large plate area at cryogenic temperatures; this demands precision alignment and nanometric feedback control~\cite{bressi2002measurement,fong2019phonon}, since slight tilt strongly distorts the effective Casimir interaction.

\noindent 2. \textbf{Minimally invasive readout across $T_C$:} measurements above and below $T_C$ must preserve both the superconducting state and the Casimir signal by minimizing probe-induced noise and back-action.

\noindent 3. \textbf{Force isolation:} Casimir contributions must be separated from competing effects, including thermal expansion, material changes at $T_C$~\cite{vsivskins2020magnetic,suchoi2017damping,binek2017elastic}, electrostatics~\cite{garcia2012casimir,liu2020elimination,garrett2020measuring}, and measurement artifacts. \\

Figure~\ref{fig1}c illustrates that the movable cavity plate used to read out Casimir force changes must satisfy several requirements: (1) high compliance to respond to weak forces; (2) high-$Q$ mechanical resonance to maximize force sensitivity; and (3) high superconducting transition temperature, which shifts superconductivity-induced changes in the electromagnetic response to higher frequencies and increases overlap with the Casimir-relevant spectrum~\cite{bimonte2019casimir}. These requirements involve coupled trade-offs, making them difficult to realize simultaneously in a single device. Together with cryogenic constraints, this has left the Casimir–superconductivity regime largely unexplored experimentally, with the most advanced prior experiments remaining orders of magnitude away from the pressure resolution needed to probe predicted Casimir shifts across $T_C$~\cite{norte2018platform,eerkens2017investigations,bimonte2019casimir}.

Theoretical expectations already converge on a demanding sensitivity scale. Existing calculations for broader temperature sweeps below $T_C$~\cite{bimonte2019casimir}, as well as our model-agnostic survey of hybrid Drude/plasma/BCS descriptions above and below $T_C$ (Methods), both place expected Casimir-pressure variations in the milli-Pascal range, with some scenarios smaller. This makes sub-milli-Pascal resolution a prerequisite: without it, genuine superconductivity-related Casimir changes cannot be cleanly separated from residual systematics or used to distinguish competing baseline models.

Here we present an experimental platform capable of resolving Casimir force changes in the \textit{micro}-Pascal regime across a superconducting transition, a sensitivity range that has remained experimentally inaccessible. Our approach uniquely employs scanning tunneling microscopy to read out on-chip, nanofabricated parallel-plate Casimir cavities. STM provides atomic-scale control between probe and plate, with minimal invasiveness that is difficult to achieve with other readout schemes. This allows precision cancellation of all other known sources of pressure change. Monolithic on-chip fabrication enables a degree of parallelism that is practically unachievable with precision-stage alignment, allowing the most parallel plate–plate Casimir geometry realized to date. 

The following sections describe the new fabrication techniques for our on-chip cavities, and introduce the STM-based detection method for nanomechanics. We then outline protocols for eliminating non-Casimir contributions and report initial calibrated frequency sweeps across $T_C$ to probe superconductivity-induced shifts in the Casimir force. We discuss potential theoretical implications and open questions; rather than a final answer, this work establishes a genuine entry point into the micro-Pascal regime needed to systematically explore Casimir-superconductivity physics.

\begin{figure*}[t]
\centering
\includegraphics[width=1\textwidth]{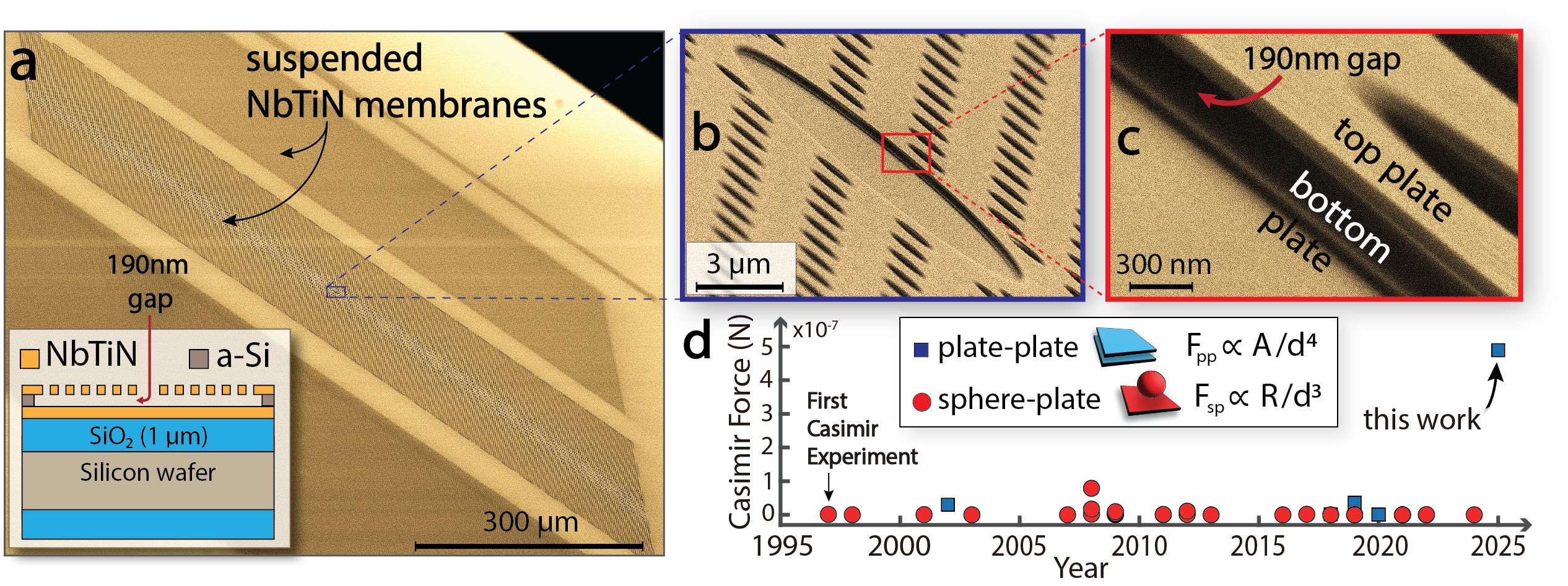}
\caption{Achieving extreme parallelism on-chip. (a) Colorized SEM of a microchip with suspended NbTiN (square) membranes above a NbTiN backgate; inset shows the layered structure. Membranes feature micron-scale holes for selective plasma etching of the a-Si layer. A $1~\mu \rm m$ SiO$_2$ layer electrically isolates the Casimir cavity. (b) A central hole in the top membrane offers visual access to the gap.
(c) Close-up of 190~nm gaps sustained over millimeter-scale areas, with $\sim$153~pm sag kept taut by high-stress NbTiN. (d) Comparison of area-over-gap ratios for published Casimir setups. Here, the ideal Casimir force between perfect conductors is used as a consistent metric to quantify effective parallelism. Blue squares: plate-plate ($F \propto A/d^4$); red circles: sphere-plate ($F \propto R/d^3$). Our chip platform's parallelism surpasses prior cryogenic and room-temperature experiments, achieving the highest Casimir force between compliant surfaces to date. See Supplementary Information~J for data sources.}
\label{fig2}
\end{figure*}

\section{Ultra-Parallel Superconducting Casimir Cavities}

Over the past decade, a variety of on-chip Casimir geometries have been explored~\cite{rodriguez2015classical, andrews2015quantum,tang2017measurement,norte2018platform,wang2021strong,perez2020system,pate2020casimir,de2025measurement}. Most conventional setups use a sphere–plate geometry for easier alignment, while plate–plate configurations, though closer to Casimir’s original theory, demand much higher parallelism and stability~\cite{bressi2002measurement,fong2019phonon,perez2020system}. To date, no superconducting Casimir experiments have reported measurable force shifts across $T_C$~\cite{norte2018platform,eerkens2017investigations,bimonte2008low,perez2020system}.

Maximizing sensitivity requires both large Casimir pressures ($F_{\text{Cas}} \propto A/d^4$) and high compliance. Thin membranes are more responsive but are prone to collapse, defining a trade-off between sensitivity and stability. We address this by nanofabricating plate–plate cavities directly on silicon chips using high-stress NbTiN films. As shown in Fig.~\ref{fig2}, a $709 \times 709~\mu$m$^2$ NbTiN membrane is suspended above a fixed NbTiN backplate with gaps as small as 190~nm. High tensile stress ($\approx$~680~MPa) keeps the membrane flat and stable while also enabling high-$Q$ resonances. The cavity is formed by depositing 155~nm-thick NbTiN layers above and below a sacrificial amorphous silicon (a-Si) layer, which is removed by a highly selective cryogenic SF$_6$ plasma etch that preserves 1–2~nm surface roughness (see Supplementary Information~D). A key discovery is that SF$_6$ at cryogenic temperatures ($-120\,^{\circ}\mathrm{C}$) selectively etches silicon over NbTiN, in contrast to its room-temperature behavior. After suspension, an adiabatic return from cryogenic vacuum prevents collapse, enabling high-yield fabrication of closely spaced compliant nanostructures. By adjusting a-Si thickness, we reliably co-fabricate both the small and big gaps, giving nearly identical devices that differ only in separation (as verified in the Supplementary Information~A).

At 190~nm, our gap is the smallest that can be sustained over millimeter-scale areas, pushing our fabrication methods to their limit (see Methods A). This distance is also large enough to avoid complex van der Waals interactions and roughness-induced noise, both of which become significant below 100~nm~\cite{van2008measurement}. The membranes combine high tensile stress, a superconducting transition temperature ($T_C \approx 14.2$~K), and atomically-smooth surfaces, all crucial for precise Casimir measurements. Achieving both high $T_C$ and high tensile stress is nontrivial, but material optimization preserves mechanical performance while ensuring a $T_C$ close to bulk NbTiN. This overlap with Casimir-relevant modes improves the chance of measurable effects.

The fixed-gap design eliminates the need for active alignment, keeping near-constant separation during temperature sweeps. Due to high-stress in the NbTiN membrane, the main deviation is an atomic-scale sag ($\sim$153~picometers) over a 709~$\mu$m span, caused primarily by Casimir attraction, corresponding to an angular misalignment of only $\sim2.5\times 10^{-5}$ degrees. A co-fabricated device with a 1.213~$\mu$m gap, where Casimir forces are negligible, serves as a control. Because fabrication is so reproducible, both large- and small-gap devices show nearly identical mechanical properties and $T_C$, enabling accurate comparison of different separations and isolation of Casimir-specific shifts from thermal or material effects (see Supplementary Information~H).

Figure~\ref{fig2} illustrates the achieved high aspect-ratio parallelism. A colorized SEM (Fig.~\ref{fig2}a) shows suspended NbTiN membranes with micron-scale release holes; the inset gives a side-view schematic. A central hole (Fig.~\ref{fig2}b) provides visual confirmation of suspension, though omitted in measurements where suspended modes are verified by vibrometry. Close-ups (Fig.~\ref{fig2}c) reveal 190~nm gaps maintained over millimeter spans. Fig.~\ref{fig2}d quantifies effective parallelism, showing our area-to-separation ratio surpasses prior Casimir experiments by orders of magnitude. Beyond extreme parallelism, the high-stress NbTiN membranes function as high-$Q$ mechanical resonators, enabling sensitive detection of Casimir-induced shifts with quality factors exceeding $10^6$ (Supplementary Information~I).

Materials engineering is central to this platform. High tensile stress keeps the membranes flat, and prevents collapse while enhancing dissipation-dilution~\cite{schmid2011damping} for high Q, making the membranes mechanically stable enough for STM interrogation. Years of cleanroom optimization produced films with both high $T_C$ and resistance to processing chemicals (see Supplementary Information~B). Crucially, NbTiN’s minimal oxidation allows effective in-situ annealing for tunneling, unlike aluminum which forms thick oxides. These combined properties enable both the generation of strong, well-defined Casimir forces and the atomically clean, mechanically stable surfaces required for scanning tunneling microscopy, as detailed in the next section.

\section{Experimental Setup and Measurement Protocol}
\begin{figure*}[t]
\centering
\includegraphics[width=1\textwidth]{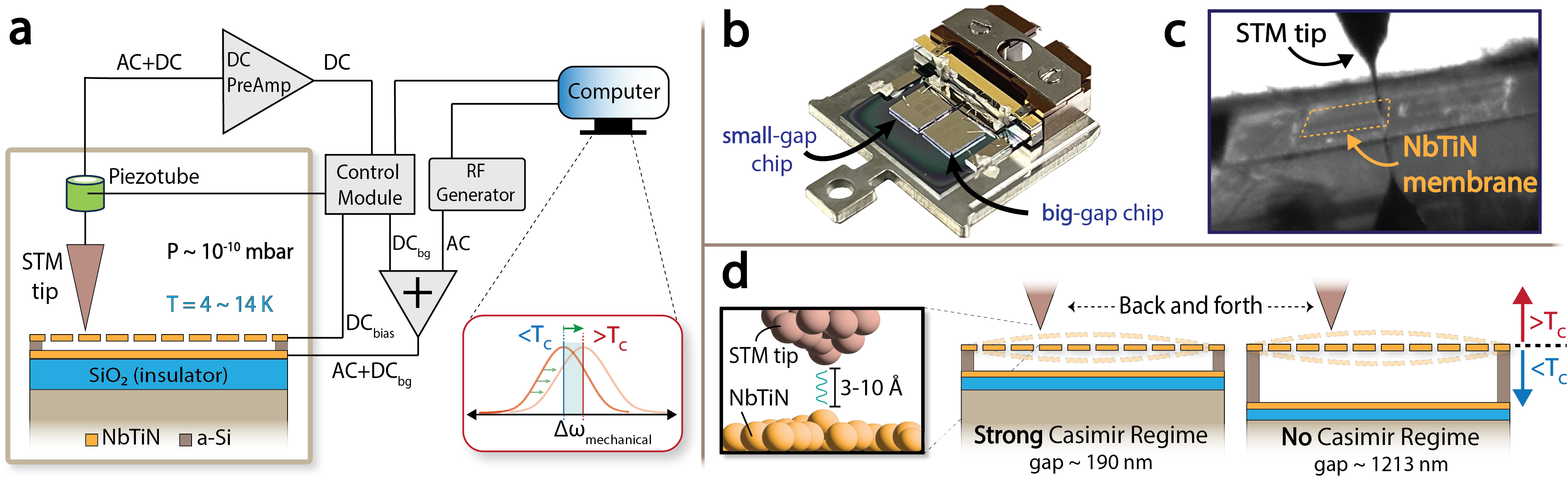}
\caption{STM readout of membrane motion (a) Schematic of the STM-based measurement setup, showing AC and DC signals applied to the backgate to drive the NbTiN membrane and detect resonance via tunneling with the STM tip. (b) Photo of the sample holder with wire-bonded chips featuring small (190~nm) and big (1213~nm) gaps for comparison. (c) In-situ optical image from the STM, highlighting the precise placement of the tip near the edge of the suspended membrane. (d) Protocol for detecting resonance frequency shifts induced by Casimir forces across $T_C$, comparing small-gap and big-gap configurations.}
\label{fig3}
\end{figure*}

Most Casimir experiments rely on optical or capacitive readout, which interrogates millions of atoms on a plate and can disturb superconductivity through heating or electromagnetic noise. While superconducting LC resonators offer extreme sensitivity~\cite{andrews2015quantum,de2025measurement}, they only operate below $T_C$, cannot probe changes across the transition, and require complex fabrication. Optical approaches~\cite{norte2018platform,eerkens2017investigations} also struggle near $T_C$ due to local absorptive heating and poor thermal anchoring in nanoscale suspended structures. To resolve superconductivity’s influence on Casimir forces, a minimally invasive method is desirable -- ideally avoiding optics, magnetic fields, and coupling only to a few atoms on the membrane at a time.

To this end, we use a scanning tunneling microscope (STM) to measure the mechanical resonance of suspended NbTiN membranes across their superconducting transition. The STM tip is positioned with sub-nanometer precision and, at separations of 0.3–1.0~nm, resolves plate oscillations in the regime of 30–100~picometers. Tunneling current changes by a factor of ten for every 0.1~nm variation in gap (about 1000$\times$ over a 0.3~nm atomic diameter), making STM exceptionally sensitive to atomic-scale motion. The tip interacts with only a few atoms of a $709 \times 709~\mu$m$^2$ membrane containing trillions in motion. This extreme localization and therefore minimal interaction enables precise cancellation of van der Waals and electrostatic forces, minimizes disturbance to superconductivity, and allows direct probing of superconducting gaps (See Supplementary Information~E).

Figure~\ref{fig3}a shows the experimental setup with detailed procedure in Methods and Supplementary Information. Measurements are conducted in ultra-high vacuum ($\sim$$10^{-10}$~mbar) and cryogenic temperatures (4.45–14.45~K). A suspended NbTiN membrane vibrates above a fixed NbTiN backgate, driven by an AC+DC voltage (AC for actuation, DC to cancel electrostatics). The STM tip is optically aligned with the membrane, gradually brought into tunneling near its stiffer edge to minimize perturbation while maintaining sensitivity (Fig.~\ref{fig3}c), and used to detect resonance by measuring tunneling current during a frequency sweep. The tip, mounted on a piezotube, maintains constant tunneling current and gap via feedback. At resonance, increased membrane motion modulates tip-sample distance, which influences the tunneling current. Since the STM’s low-noise amplifier has a cutoff frequency ($\sim$1~kHz) well below the membrane's resonance frequency, only the average current is recorded. Due to the exponential dependence of the tunneling current on the tip-sample distance, the oscillation results in an increase of the average current, triggering a retraction of the tip. The resulting piezotube displacement follows the oscillation amplitude. 

To isolate superconductivity-induced Casimir shifts from material effects and thermal expansion, we use two chips with distinct gaps: a small-gap cavity (190~nm) where Casimir forces are strong, and a large-gap cavity (1213~nm) where they are negligible (Fig.~\ref{fig3}d). Mounting both on the same holder (Fig.~\ref{fig3}b) ensures identical environmental conditions, enabling subtraction of material-dependent effects. Although the NbTiN films are co-deposited, we additionally verify that both chips exhibit matched mechanical frequencies and $T_C$ (see Supplementary Information~A). This differential approach allows us to subtract shared material responses across $T_C$ and isolate Casimir forces. All other extraneous forces are actively canceled or suppressed, leaving temperature sweeps around $T_C$ to reveal the Casimir contribution.

\section*{Measurements Across \texorpdfstring{$T_C$}{TC}}

\begin{figure*}
\centering
\includegraphics[width=1\textwidth]{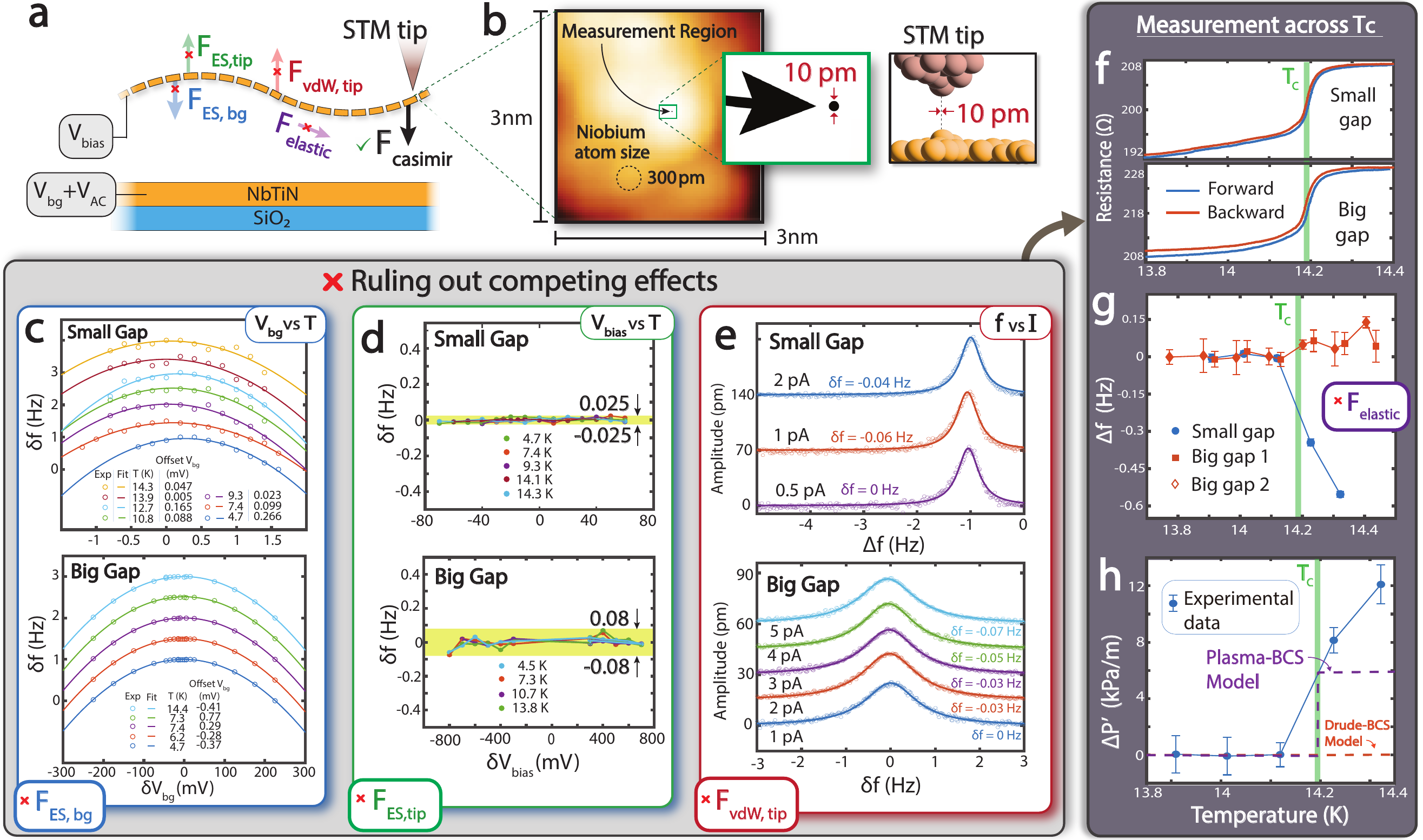}
\caption{STM-based Casimir measurement and cancellation of non-Casimir forces. (a) Schematic of thermo-elastic, electrostatic, and van der Waals interactions to be removed. (b) STM scan showing tip scanning within a $3\times 3$~nm$^2$ area at a topographic maximum with 10 pm localization. (c–e) Sequential cancellation of plate–plate electrostatics, tip–sample electrostatics, and van der Waals forces. (f) Large- and small-gap devices share $T_C \approx 14.2$ K; total resistance including non-superconducting leads (g) Small-gap device shows sharper frequency changes near $T_C$. (h) Background-subtracted Casimir pressure gradient agrees in magnitude and sign with plasma–BCS theory (See Methods).}
\label{fig4}
\end{figure*}

Our platform is designed to detect subtle changes in the Casimir force during slow, controlled sweeps across the superconducting transition temperature $T_C$. Resolving effects at the micro-pascal level requires exceptional frequency stability, high force sensitivity, and suppression of environmental noise. As shown in Fig.~\ref{fig4}a, isolating the Casimir contribution involves canceling electrostatic interactions between the membrane and backgate ($F_{\mathrm{ES},bg}$), electrostatic and van der Waals forces between the STM tip and membrane ($F_{\mathrm{ES},tip}$, $F_{\mathrm{vdW},tip}$), and elastic changes in the membrane ($F_{\mathrm{elastic}}$). The latter are removed using a co-fabricated large-gap cavity with negligible Casimir interaction.

To probe superconductivity-induced changes, we heat the membranes from 4.45~K to just above $T_C \sim 14.2$~K in stabilized steps. An open-flow helium system provides vibration-free cooling for long sweeps, while UHV preserves clean surfaces and minimizes mechanical damping. Each sweep lasts up to 16 hours, with the STM tip locked to within $\sim$10~pm to ensure reproducible readout.

As shown in Fig.~\ref{fig4}b, measurements begin with a $3 \times 3$~nm$^2$ atomic-scale scan to identify a topographic maximum. The STM tip is then locked to this site with $\sim$10~pm stability, minimizing position-dependent forces and ensuring remarkably reproducible conditions.This keeps the tunneling interaction confined to the same few atoms of the moving membrane throughout the full sweep across $T_C$ That atomic-scale control makes STM uniquely suited for precision Casimir measurements, because probe-induced contributions can be quantified and removed cleanly to isolate the Casimir signal.

Figures~\ref{fig4}c–e show sequential cancellation of competing forces. Backgate voltages are swept to cancel $F_{\mathrm{ES},bg}$, with no temperature dependence observed. Tip–membrane biases are adjusted to cancel $F_{\mathrm{ES},tip}$, with residual shifts far below the $\sim$0.5 Hz Casimir signal expected from theory. Van der Waals interactions are excluded by verifying that resonance frequency is independent of tunneling current. With parasitic forces minimized, Casimir changes can be isolated. See Methods for more detailed description of cancellation procedures.

To eliminate material-dependent effects, we differentially measure co-fabricated small-gap (190~nm) and large-gap (1213~nm) cavities mounted on the same holder. Both show the same superconducting transition temperature $T_C \approx 14.2$~K (Fig.~\ref{fig4}f), confirming identical superconducting on top of matching mechanical behavior (see Supplementary Information~A). The relative frequency shift $\Delta f$ between the two devices (Fig.~\ref{fig4}g) isolates the Casimir contribution and reveals an abrupt change in the Casimir pressure gradient near $T_C$.

Figure~\ref{fig4}h compares the extracted pressure gradient with theory, showing agreement in sign and magnitude with a hybrid plasma–BCS model, where the normal state follows the plasma prescription and the superconducting state adopts BCS permittivity. Only this combination predicts a discontinuous Casimir pressure across $T_C$, consistent with our measurements. The fits use $\Omega = 5.33$~eV$/\hbar$, consistent with prior estimates for NbTiN~\cite{bimonte2019casimir}, but reproducing the full jump requires $\Omega = 12.7$~eV$/\hbar$, suggesting additional physics or refinement of the theory may be involved.
 
Taken together, these measurements establish that the platform operates in the sensitivity regime required to probe superconductivity-dependent Casimir variations across \(T_C\), while suppressing known competing contributions \textit{in situ}. The system combines atomic-scale tip localization, high force responsivity, and systematic cancellation of electrostatic, van der Waals, and thermal backgrounds.

The noise-limited frequency resolution is $\langle \delta f\rangle \approx 4.7$~mHz, extracted from Lorentzian fits to the resonance peak using $\langle \delta f\rangle \approx \frac{f_0}{2Q \cdot \mathrm{SNR}}$, where $f_0$ is the membrane resonance frequency and SNR the signal-to-noise ratio (see Supplementary Information~F4 and Ref.~\cite{sansa2016frequency}). This demonstrated $\langle \delta f\rangle$ is nearly two orders of magnitude below the 0.29~Hz shift predicted by the plasma--BCS model. Simulations yield Casimir responsivities of 1.73~Hz/nN and 0.87~Hz/mPa, placing our device among the most responsive nanoresonator-based Casimir platforms reported to date~\cite{azgin2011resonant,salimi2024squeeze}. In pressure units, this corresponds to sensitivity at the level of tens of micro-Pascals, well below the milli-Pascal scale of leading predictions.

\section{Conclusions}

This work establishes a new experimental frontier for Casimir measurements, with unprecedented parallelism, sensitivity, and suppression of extraneous forces in cryogenic environments, where less than a handful of Casimir experiments have so far ventured. By combining micro-electromechanical systems (MEMS) cavity engineering with atomic-scale STM readout, we realize ultra-parallel, compliant, high-$Q$ superconducting membranes and differential cancellation protocols that access the micro-Pascal regime across $T_C$. This architecture brings Casimir measurements closer to the original parallel-plate limit under superconducting conditions, reducing geometric assumptions and enabling more direct comparisons between theory and experiment in regimes where sphere-plate approximations may be insufficient.

Because superconductivity is a second-order (continuous) phase transition that primarily modifies low-frequency electrodynamics, it has long been unclear whether crossing $T_C$ should produce any abrupt or measurable change in the Casimir force~\cite{bimonte2025proposal,eerkens2017investigations}. If confirmed as Casimir in origin, such shifts could open a rich field of new physics where superconductivity modifies vacuum fluctuations. The apparent discontinuity, in both direction and magnitude, qualitatively matches with the plasma–BCS model presented in Methods, which uniquely predicts a jump across $T_C$. Both the plasma model and BCS theory describe dissipationless electronic responses, but in different senses: the plasma model is dissipationless only by neglecting relaxation in a free-electron gas, while BCS dissipationlessness arises from quantum coherence of the superconducting condensate, which also enforces the Meissner effect. Recent theory and experimental work suggest the plasma model better captures low-frequency transverse electric responses in the normal state~\cite{klimchitskaya2023casimir}, while coherence and non-local effects may further modify this behavior~\cite{klimchitskaya2022theory}, highlighting importance of experiments in this uncharted territory of physics.

Looking ahead, the next crucial step is to resolve potential shifts at $T_C$ and determine their evolution by systematically varying plate separation. This will require series of nominally identical chips with progressively tuned gaps, extending beyond the two extremes probed here: smallest gaps where Casimir forces dominate and largest gaps where they are negligible. A further step is to separate competing theories using their distinct scaling with parameters such as plate thickness and superconducting $T_C$~\cite{villarreal2019casimir}. Establishing this baseline would move the field beyond long-standing debates and enable credible tests of more speculative directions, from mechanisms in high-$T_C$ superconductors~\cite{kempf2008casimir,orlando2018correlation,strongin1968enhanced,strongin1968superconductive} to possible gravitational signatures~\cite{avino2018archimedes,calloni2002vacuum,fulling2007does,bimonte2006energy,quach2015gravitational}. Any confirmed signal along these directions would be a major advance in our understanding of quantum materials and vacuum energy. Our platform sets a new benchmark for entering this regime and provides a foundation for future discoveries.

\section{Methods}
\subsection{On-Chip Casimir Cavity Fabrication}

\begin{figure*}[t]
\centering
\includegraphics[width=1\textwidth]{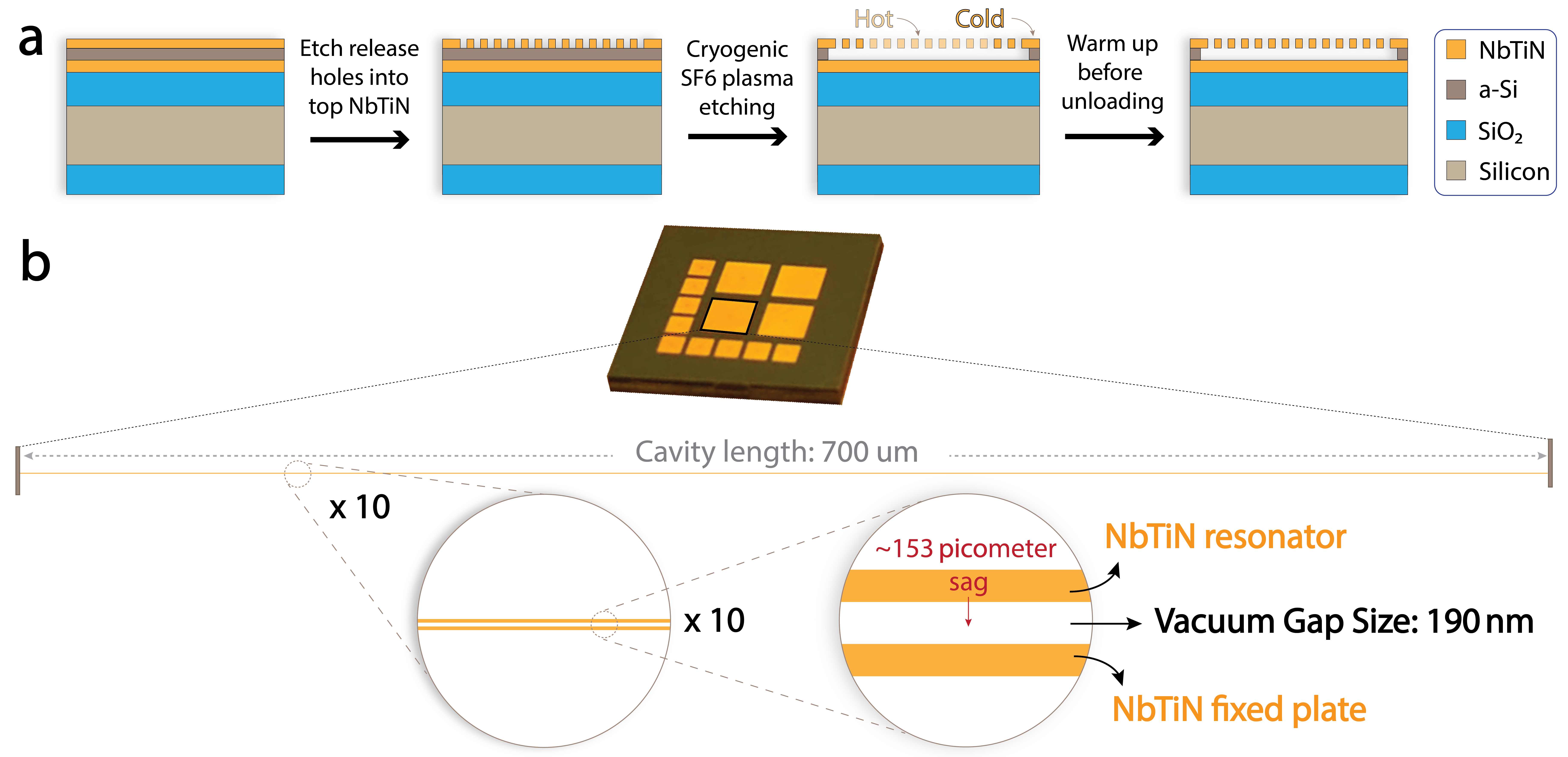}
\caption{Fabrication of the on-chip superconducting Casimir cavity. (a) Key steps for patterning and releasing a freestanding NbTiN membrane. (b) Photograph of a 10$\times$10~mm$^2$ chip containing multiple cavities, with schematic zooms to scale showing the achieved parallelism: a 700~$\mu$m span separated by a 190~nm gap, with an analytically estimated central deflection of only $\sim$153~pm due to Casimir forces.}
\label{fig5}
\end{figure*}

Figure~\ref{fig5}a outlines the fabrication process. A completed 10 × 10 mm$^2$ chip (Fig.~\ref{fig5}b) hosts multiple Casimir cavities, each formed from a suspended NbTiN membrane spanning 700~µm over a 190~nm vacuum gap. The scale illustration shows the extreme aspect ratio; the membrane’s central deflection is not visible but is calculated to be only $\sim$153~pm (See Supplementary Information~D), set almost entirely by the Casimir force, and would require roughly a $10^3$ zoom beyond the gap view to visualize.

Our platform achieves area-to-gap ratios exceeding prior Casimir configurations by orders of magnitude, enabled by monolithic chip-based parallel plates rather than active alignment. This yields nanometer-scale gaps over hundreds of micrometers with robustness, scalability, and reproducibility. Stability, compliance, and smoothness are engineered directly into the structure using high-stress superconductors, sacrificial amorphous silicon, and cryogenic plasma etching.

Layered chips are produced by sequential deposition: thermal oxide for insulation, a bottom NbTiN electrode, an ICP-CVD amorphous silicon sacrificial layer, and a top NbTiN membrane, followed by annealing and dicing. The flat, tensile trilayer geometry avoids thermally induced out-of-plane distortions seen in drum-type membranes, simplifying temperature sweeps and analysis.

Casimir cavities are defined by etching release holes in the top NbTiN via electron beam lithography and fluorine plasma, then removing the a-Si with a cryogenic SF$_6$ plasma undercut at –120~$^{\circ}$C. The etch uses ICP power only (no RF bias) to protect the superconducting membrane; low temperature slows NbTiN etching, preserves extreme aspect ratios, and improves surface smoothness. Devices are then warmed slowly and adiabatically from cryogenic temperatures to room temperature under vacuum, an essential step to prevent collapse, before being transferred to the STM lab. Even slight temperature gradients across a strained membrane can cause deformations that can lead to runaway collapse. This was a key ingredient for reliably fabricating smaller gap devices.

Two key innovations enable this platform: (1) ICP-CVD deposition of low-hydrogen a-Si with high uniformity, ensuring predictable gaps and avoiding blistering; and (2) cryogenic SF$_6$ etching with $\sim$1000:1 selectivity to NbTiN, releasing ultra-compliant membranes without damage. Together, these methods allow large-span metallic resonators to be suspended at Casimir-dominated gaps without active alignment, eliminating drift and enabling quantum-limited parallelism for Casimir studies at any temperature.

\subsection{STM measurement protocol}

The STM offers several key advantages for this application. Unlike optical or capacitive readouts, which perturb the membrane over large areas and billions of atoms, STM tunneling is highly localized and equally effective above and below $T_C$, making it suitable for studying the superconducting transition. The ultra-high vacuum (UHV) environment required for cryogenic STM operation provides strong control over surface conditions. Our STM setup includes annealing capabilities that reduce surface oxidation, an often-overlooked factor in cryogenic nanomechanical setups. NbTiN resists oxidation more effectively than aluminum~\cite{andrews2015quantum, norte2018platform, de2025measurement} due to its stable nitride structure, which limits oxygen diffusion. Unlike aluminum oxides, which are chemically stable and difficult to remove, NbTiN oxidation is minimal and can be annealed in UHV at 200~$^{\circ}$C, reducing electrostatic patch potentials~\cite{liu2020elimination,garrett2020measuring} and improving tunneling current signal. The STM tip can also probe superconducting properties by identifying the superconducting gap and $T_C$, providing a diagnostic tool alongside mechanical measurements (See Supplementary Information~E). In addition to its high $T_C$, NbTiN’s surface compatibility with STM enables coupling between the membrane and tunneling readout. Casimir forces are surface-dominated, as are many noise sources, making STM well-suited for both mechanical detection and surface characterization.

Our protocol detects changes in the Casimir force across the superconducting transition by tracking shifts in the mechanical resonance frequency of suspended membranes. Instead of measuring absolute Casimir forces, which carry large theoretical uncertainties from unknown superconducting corrections and thermal effects, we measure relative changes across $T_C$. This differential approach is more sensitive to superconductivity-induced modifications of the Casimir interaction. Small-gap (190~nm) and large-gap (1213~nm) devices are compared; the large-gap device serves as a control to subtract non-Casimir contributions such as thermal expansion, material changes, and tunneling variations across $T_C$. STM’s high spatial precision enables stable, reproducible readout on both devices, ensuring confident removal of non-Casimir effects.

As described in the main text, the STM provides high-sensitivity, tunneling current readout of membrane resonance across the superconducting transition. Here, we detail the protocol used to achieve the stability, localization, and noise suppression required for the results in Fig.~\ref{fig4}. We use a scanning tunneling microscope (STM) to measure membrane dynamics with atomic-scale precision. The tunneling current depends exponentially on tip–sample separation, with 0.1~nm changes producing an order-of-magnitude variation. This extreme sensitivity allows us to track the mechanical resonance of suspended superconducting NbTiN membranes, extending STM from atomic-scale imaging to mesoscopic devices.

Two aspects of the STM’s precision are essential. First, its high vertical ($z$) resolution detects membrane resonance by tracking oscillation-induced changes in tunneling current within tunneling distance. Second, its lateral ($x$/$y$) precision, maintained by dual feedback loops, keeps the tip fixed at the same membrane location, avoiding spatial variation in local forces.

The STM’s signal control is handled by a Nanonis\texttrademark{} system, which applies both a bias voltage ($V_{\rm bias}$) and a backgate voltage ($V_{\rm bg}$), while positioning the tip based on the measured tunneling current. The backgate voltage is combined with an oscillating voltage $V_{\rm AC}$ from an RF generator (Rohde \& Schwarz SMB100A) via a bias tee (Mini-Circuits ZFBT-4R2GW+) and applied to the bottom NbTiN film. The bias voltage is applied to the suspended NbTiN membrane, producing a tunneling current when the STM tip is nearby. Because the tip–membrane conductance depends exponentially on their separation, membrane oscillations modulate the tunneling current. This oscillating current is then amplified by a low-noise current amplifier (NF Corp. SA-607F2). Because the amplifier’s cutoff frequency is well below the membrane’s resonance frequency, the amplifier effectively converts the oscillating tunneling current into an averaged DC current output that is recorded by the control system.

By setting fixed values for the tunneling current and bias voltage, the system controls the distance between the tip apex and the membrane. When $V_{\rm AC}$ is applied to the backgate, the membrane is driven, and its vibrational amplitude peaks as the AC signal approaches the membrane's mechanical resonance frequency. To maintain a constant averaged tunneling current, the STM tip retracts, adjusting its $z$-position via the piezoceramic tube. This $z$-position is recorded by the control system as a direct measure of the membrane's vibrational amplitude. When the AC signal frequency nears the resonant frequency of the nanomembrane, the increased vibrational amplitude is clearly resolved in the tip’s $z$-position, indicating both the resonance position and the mode amplitude as seen in Fig.~\ref{fig4}e.

The experimental setup consists of a loadlock, a preparation chamber, and a measurement chamber. Samples are first loaded into the loadlock and annealed at 200$^\circ$C under a pressure of \(1.7 \times 10^{-8}\)~mbar for two days to remove contaminants. They are then transferred to the preparation and measurement chambers, where experiments are performed at 4.45~K and \(1.5 \times 10^{-10}\)~mbar. As shown in Fig.~\ref{fig3}b, the sample holder accommodates chips with both small- and big-gap cavities, allowing side-by-side comparison. AC and DC signals are applied to the backgate beneath the membranes (via wire bonding) to drive mechanical oscillations and cancel electrostatic forces. A bias voltage on the suspended membrane enables tunneling current detection by the STM tip. Tip placement is guided by an in-situ optical camera (Fig.~\ref{fig3}c), enabling coarse positioning near the membrane edge (60 microns away) where vibrational amplitude is minimized. Resonance frequency measurements are then performed at different temperatures across $T_C$. 

By combining STM readout with microchip-based Casimir cavities, we create a platform for minimally invasive, high-sensitivity measurements across the superconducting transition. To obtain meaningful results, it is essential to remove all extraneous forces that could mask the Casimir effect, including those from the STM, electrostatic interactions between the plates, and changes from thermal expansion or material properties at $T_C$ during temperature sweeps. While the main text describes these cancellation strategies conceptually, the following subsections detail their exact implementation in our experiments.

\subsection{Experimentally isolating Casimir forces}

To isolate the Casimir force in our solid-state parallel-plate cavities, we systematically eliminate other forces acting on the superconducting membrane, including membrane–backgate electrostatics, tip–membrane electrostatics, van der Waals (VdW) interactions, and thermal–elastic effects from temperature sweeps across $T_C$. Our protocol cancels membrane–backgate electrostatics first, then tip–membrane electrostatics, and finally minimizes VdW interactions. With the STM tip fixed at a single membrane location and temperature carefully controlled, this sequence enables reproducible, high-sensitivity measurements of Casimir forces across the superconducting transition.

\subsubsection{Positioning and Minimizing Tip–Membrane Forces}

As shown in Fig.~\ref{fig4}b, the STM can scan a $3 \times 3$~nm$^2$ area of the membrane, imaging the resonator with atomic resolution. This capability is unique in that the same instrument both resolves the membrane’s resonant motion and provides in-situ atomic-scale images of the exact spot being measured. Due to sputtering growth, the surface has a 1–2~nm roughness, producing local high and low points. Using the STM’s current-to-height feedback and a local gradient search algorithm, the system autonomously locks onto the highest point with 10~pm precision, nearly 30 times smaller than the 300~pm radius of a niobium atom on the NbTiN membrane (inset Fig.~\ref{fig4}b). This sub-atomic spatial precision ensures the tip remains fixed at one location, minimizing positional drift and keeping the local force environment stable. Holding the same position is essential since identical current at a different location would not guarantee the same forces~\cite{elbertse2026detection}. Fixing the tip with sub-atomic precision minimizes variations in local interactions and allows precise cancellation of perturbative effects introduced by the measurement itself.

\subsubsection{Mitigating Electrostatic Forces}

Electrostatic interactions — arising from residual surface charges and contact potential differences — pose a significant challenge in Casimir experiments. Because electrostatic forces scale as \( F_{\mathrm{ES}} \propto A/d^2 \) in plate–plate geometries, they directly compete with the Casimir force and must be experimentally eliminated. These forces stem from sources such as surface contamination, grain boundaries, or static charges from fabrication or air exposure, making them difficult to predict or model~\cite{garcia2012casimir,liu2020elimination,garrett2020measuring}. Therefore, in Casimir experiments they must be removed experimentally, as they contribute in unknown ways at small separations.

To cancel electrostatic forces between the membrane and the backgate, we measure the membrane’s resonance frequency as a function of backgate voltage for both small- and big-gap cavities (Fig.~\ref{fig4}c). The apex of the resulting parabola identifies the voltage that minimizes the electrostatic interaction. This compensation voltage is used throughout temperature sweeps to keep the membrane–backgate force balanced.

Temperature sweeps and corresponding measurements on both small- and big-gap membranes show that the apex voltage shifts by less than 0.2~mV (small-gap) and 1.1~mV (big-gap). Finite element simulations translate these into pressure differences of \( 4.9 \times 10^{-6} \)~Pa and \( 3.6 \times 10^{-6} \)~Pa, and frequency shifts of \( 2.6 \times 10^{-3} \)~Hz and \( 2.6 \times 10^{-4} \)~Hz, respectively — close to the sensitivity limit of our experiment. These small shifts indicate that the electrostatic potential remains stable during temperature sweeps. For subsequent measurements, we fixed the backgate voltage at 257.2~mV (small-gap) and 223.6~mV (big-gap), ensuring electrostatic cancellation without compromising sensitivity.

To ensure minimal electrostatic forces between the STM tip and the membrane, the tip–membrane bias voltage is swept while maintaining constant tip–sample separation. Here we clearly show that irrespective of the compensation voltage between tip and membrane, the frequency shift will be small compared to the measured signal changes in Fig.~\ref{fig4}g. This procedure ensures that tip-induced electrostatic forces remain negligible throughout measurements.

\subsubsection{Minimizing Tip–Membrane van der Waals Forces}

As the STM tip approaches the membrane, van der Waals (VdW) forces can become significant. To reduce these effects, the tip is sharpened and cleaned by dipping into a copper sample and applying voltage pulses. Imaging atomic steps on the copper surface verifies a well-defined apex, although sub-apex features such as dangling atoms may still remain. 

The optimal landing site on the membrane is chosen as the highest and most stable location, minimizing position-dependent VdW interactions during temperature sweeps. The membrane’s resonance frequency is then measured at different tip–membrane currents (i.e., separations) to confirm Lorentzian mechanical peaks. As shown in Fig.~\ref{fig4}e, the absence of frequency shifts across these distances indicates that distance-dependent VdW contributions are negligible within our measurement uncertainty.

\subsubsection{Managing Thermal Drift and Determining Casimir Pressure Gradient Variations}

Temperature sweeps introduce both thermal drift and elastic effects that can distort Casimir measurements if not carefully managed. We begin each run at the base temperature (4.45~K) with a precise $3 \times 3$~nm$^2$ scan of the membrane, using the STM’s atomic-scale scanning and 50~Hz feedback to maintain the tip position within a 10~pm region during resonance sweeps. Incremental heating with stabilization periods (1~hour for 0.1~K, 2.5~hours for 1~K, and 3~hours for 5~K) minimizes drift from thermal expansion.

Long-range sweeps from 4.45~K to $T_C \sim 14.2$~K are performed with open-flow helium cooling (0.45~L/hour) to eliminate vibrations that would otherwise degrade STM performance. The STM holds 9~L of liquid helium — 1.8~L for overnight tip and temperature stabilization and 7.2~L for the $T_C$ sweep. Ultra-high vacuum (\(\sim 10^{-10}\)~mbar) preserves clean surfaces after annealing and suppresses squeezed-film damping at nanoscopic separations.

After systematically eliminating electrostatic forces (membrane–backgate and tip–membrane) and minimizing van der Waals forces, we track resonance frequency shifts across $T_C$. Small-gap (190~nm) and big-gap (1213~nm) membranes are measured and placed side-by-side in the STM holder; the large-gap device serves as a reference to remove non-Casimir effects from material property changes, thermal expansion, or tunneling variations. For consistency, we denote temperature-dependent variations by $\Delta \psi$ and fixed-temperature variations by $\delta \psi$ for any quantity $\psi$ (e.g., frequency, pressure).

Independent resistivity sweeps confirm $T_C \sim 14.2$~K for both devices (Fig.~\ref{fig4}f). This resistance includes non-superconducting leads and thus does not go to zero, but allows one to observe where the abrupt transition happens. Following calibration in Supplementary Information~G, we linearly fit the elastic force below $T_C$ to extract the residual $\Delta \omega^2$ above $T_C$, where $\omega = 2\pi f$. Because gap size is the only difference between the membranes, differences in $\Delta \omega^2$ are attributed to a separation-dependent force. With other distance-dependent forces removed, this shift is assigned to the Casimir force. Using $\Delta \omega^2$ from the big-gap device, we estimate mechanical changes in NbTiN and compute the Casimir-induced $\omega^2$ shift in the small-gap membrane.

\subsection{Converting Mechanical Frequency Change to Pressure Gradient}
The relation $\Delta \omega^2 = -\Delta P'(d)/\rho t$ links frequency shift to the pressure gradient, where $d$ is separation, $\rho$ density, and $t$ thickness (Supplementary Information~F). From the data, we obtain a Casimir pressure gradient variation of $12.10 \pm 1.38$~kPa/m for a $0.20 \pm 0.02$~K temperature change (Fig.~\ref{fig4}c). For our parallel-plate geometry, $\Delta f/f \approx -\Delta P'_{\rm Cas} L^2 / (4\pi^2 \sigma t) = -\Delta F'_{\rm Cas} / (4\pi^2 \sigma t)$, where $\sigma$ is stress (Supplementary Information~F).

Based on Lifshitz theory~\cite{bimonte2019casimir}, the Casimir pressure at $d = 190$~nm is
\[
P_{\rm Cas}|_{d=190\ \mathrm{nm}} = -1.081 \times 10^{-24} / d^{3.507} \ \mathrm{[Pa]} \approx -0.403\ \mathrm{Pa}
\]
($d$ in meters; negative sign indicates attraction). Finite element simulations then yield a Casimir force change $\Delta F_{\rm Cas} = -0.33 \pm 0.04$~nN, or $\Delta P_{\rm Cas} = -0.65 \pm 0.07$~mPa across $T_C$. The negative signs indicate increased attraction above $T_C$.

The abrupt pressure gradient change in Fig.~\ref{fig4}h is not predicted by existing theory~\cite{bimonte2019casimir} which focuses on long-term drifts below $T_C$. The Casimir force, as the derivative of the Helmholtz free energy $F_{\rm Cas} = -\partial {\cal F}/\partial d$~\cite{bimonte2017nonequilibrium}, should remain continuous through a second-order phase transition. However, our data agree only with a hybrid plasma-BCS model, which combines plasma permittivity above $T_C$ with BCS below $T_C$, altering the zero-frequency ($n = 0$) Matsubara term. Only this combination produces a discontinuous pressure gradient quantitatively consistent with our measured signal in both direction and order-of-magnitude.

In Fig.~\ref{fig4}h, we use $\Omega = 5.33$~eV/$\hbar$ (NbTiN estimate~\cite{bimonte2019casimir}), though matching the full jump requires $\Omega = 12.7$~eV/$\hbar$. Because the zero-frequency term is poorly constrained, $\Omega$ may differ from the IR plasma frequency. Additional contributions from superconducting Casimir effects beyond the plasma–Drude transition could vary strongly with parameters such as gap, plate thickness, and $T_C$. Addressing these questions will require next-generation experiments with carefully fabricated chip series, liquid-helium stability, and extended STM measurement times. As shown in Supplementary Information~F4, we can confirm a sensitivity to micro-pascal pressure changes about 10 times smaller than the observed pressure shift meaning we are well within experimental range to measure mPa shifts predicted by plasma-BCS theory, with high resolution. The micro-pascal sensitivity enabled by combining STM with on-chip nanomechanics opens a new frontier for quantum experiments.

\section{Theoretical Expectations and Sensitivity Targets}

\begin{table}[h]
\centering
\small 
\rowcolors{2}{white}{white} 
\begin{tabular}{|l|c|c|c|}
\hline
Model & Above \(T_C\) & Below \(T_C\) & Trend, \(\Delta P'\) (kPa/m) \\
\hline
\rowcolor{yellow!25}
Plasma--BCS     & Plasma & BCS    & jump,\ 6.0 \\
Plasma--Plasma  & Plasma & Plasma & smooth,\ 0.065 \\
Drude--BCS      & Drude  & BCS    & kink,\ 0.019 \\
\hline
\end{tabular}
\vspace{0.3em}

\small Table 1:~Variation of Casimir pressure gradient \(\Delta P'\) values correspond to changes across $T_C$ ($\pm 0.1~\rm K$).
\end{table}

Understanding how superconductivity influences the Casimir effect could open new directions in condensed matter physics~\cite{klimchitskaya2023casimir,klimchitskaya2022theory} and superconductivity~\cite{kempf2008casimir,orlando2018correlation}, and may eventually inform more speculative proposals~\cite{avino2018archimedes,calloni2002vacuum,fulling2007does,bimonte2006energy,quach2015gravitational}. Before such questions can be meaningfully addressed, experiments must first reach the sensitivity required to resolve changes predicted by conventional electrodynamic models.

Two baseline targets already impose stringent requirements. Over a broad temperature sweep from $T_C \approx 14~\mathrm{K}$ down to a few kelvin, the predicted difference between plasma and Drude prescriptions for NbTiN is on the order of $0.2~\mathrm{mPa}$~\cite{bimonte2019casimir}. Across a narrow sweep just above and below $T_C$, conventional models span outcomes ranging from no detectable change to shifts of a few milli-Pascals, depending on how superconducting electrodynamics enters the zero-frequency contribution.  

Although superconductivity primarily modifies the optical response at low frequencies, finite-temperature Casimir forces are uniquely sensitive to this regime through the Matsubara zero-frequency (n=0) term. We therefore introduce a model-agnostic approach, analyzing all plausible combinations of Drude, plasma, and BCS permittivity models above and below the superconducting transition. Nearly all combinations predict only smooth or negligible changes across $T_C$; the sole exception is the plasma--BCS case, which predicts an abrupt milli-Pascal-scale jump in the pressure gradient (Table~1). This prediction defines the experimental sensitivity threshold that superconducting Casimir platforms must reach.

\begin{acknowledgments}
We thank Peter Steeneken, Shahrukh Ashruf, and Noud Derks for help with analysis; Martin Lee for early cleanroom support, along with Charles de Boer, Marinus Fischer, Bas van Asten, Pauline Stevic, and Roald van der Kolk. We are grateful to Allard Katan, Alexandre Artaud, Evert Stolte, Ronald Bode, and Tino Kool for measurement support. We also thank Yufan Li, Zichao Li, Hanqing Liu, Ruben Guis, Martijn Veen, Andrea Cupertino, Vidharshana Sivakumar, and Junho Suh for their contributions, and Martin Tajmar, James Quach, Nathan Inan, Farbod Alijani, Achim Kempf, Gerard Verbiest, and Peter Steeneken for valuable feedback on the manuscript. This publication is part of the project, Probing the physics of exotic superconductors with microchip Casimir experiments (740.018.020) of the research programme NWO Start-up which is partly financed by the Dutch Research Council (NWO). Funded/Co-funded by the European Union (ERC, EARS, 101042855). Views and opinions expressed are however those of the author(s) only and do not necessarily reflect those of the European Union or the European Research Council. 
\end{acknowledgments}

\appendix









\newpage
~ 
\newpage


\begin{widetext}

\begin{center}
\large{\textbf{SUPPLEMENTARY INFORMATION}}
\end{center}

\renewcommand{\theequation}{S.\arabic{equation}}
\renewcommand{\thesection}{}
\renewcommand{\thesubsection}{}
\renewcommand{\thetable}{S\arabic{table}}  
\renewcommand{\thefigure}{S\arabic{figure}}

\section*{Supplementary Information A: Parameter table}

To highlight the strength of our fabrication approach, Table~\ref{tab:research outline} compares the mechanical properties of membranes suspended over the two extremes of gap size used in this work. Despite a more than fivefold increase in gap, introduced by a thicker silicon spacer, the devices remain nearly identical in resonance frequency, stress, and density, showing that the release process introduces minimal stiction or distortion. This reproducibility is crucial, since it ensures that any variations in response can be attributed to Casimir forces rather than uncontrolled changes in membrane mechanics. The ability to fabricate such matched devices provides a solid foundation for systematically probing Casimir effects across a wide range of separations.

\begin{table*}[ht]
\centering
\caption{\label{tab:research outline}
Mechanical properties of the NbTiN membranes over small- and big-gaps, and STM setting during the temperature sweep measurements. The length of the device is by design. The gap sizes are measured with a scanning electron microscope (SEM) and with a atomic force microscope (AFM). The tensile stress and the film density of the NbTiN membranes are measured in Supplementary Information~\textbf{C3}. The thermal expansion coefficients (CTE) of the membranes are fitted from $AT+BT^3$ in the Equation~\ref{eq:CTE}. For membranes suspended over varying gap sizes, distinct AC driving powers yield vibrational amplitudes consistently on the order of tens of picometers.}
\vspace{2mm}
{\renewcommand{\arraystretch}{1.6}%
\begin{tabular}{|c|c|c|}
  \hline\hline
 \textbf{Parameters}   & \textbf{Small gap} & \textbf{Big gap}   \\
   \hline 
   Length $L$ (um)  & 709 & 709 \\ \hline
   Gap size $d$ (nm) & 190 & 1213   \\ \hline
  Membrane thickness $h$ (nm) & 155 & 155   \\ \hline
  Backgate thickness (nm) & 155 & 155   \\ \hline
  Resonance frequency $f$ (Hz)  & 352800 & 343008   \\ \hline
  Quality factor $Q$ ($\sim 4.45$~K) ($\times 10^3$) & 720 & 200 - 300 \\ \hline
  Quality factor $Q$ ($\sim 14.2$~K) ($\times 10^3$) & 1000 & 130 - 170 \\ \hline
  Tensile stress $\sigma$ (MPa)  &  677 &  683  \\ \hline
  Density $\rho$ (kg/m$^3$)  & 4992  & 5332  \\ \hline
  Spring constant $k_{0}$~(N/m) & 1804  & 1821 \\ \hline
  CTE-A (1/K$^2$)  & $2.001\times 10^{-10}$  &  $4.289\times 10^{-10}$   \\ \hline
  CTE-B (1/K$^4$)  & $9.159\times 10^{-13}$  & $3.181\times 10^{-13}$  \\ \hline
  $T_C$ (K)  & 14.195  & 14.195  \\ 
  \hline\hline
  AC driving power (dBm)  & -85  &  -40 \\ \hline
  Tunneling current (pA)  &  1 &  1    \\ \hline
  Bias voltage $V_{\rm bias}$ (mV)  &  300 &  300   \\ \hline
  Backgate voltage $V_{\rm bg}$ (mV)   &  300+257.2 & 300+223.6    \\ \hline
  Averaging time (ms)   &  200 &  200     \\ \hline
  \hline
\end{tabular}}
\end{table*}

\newpage

\section*{Supplementary Information B: On-chip Casimir Cavity Fabrication}

\subsection*{Supplementary Information B1: Layered Chip Fabrication Procedure}

\begin{figure}[ht]
\centering
\includegraphics[width=0.8\textwidth]{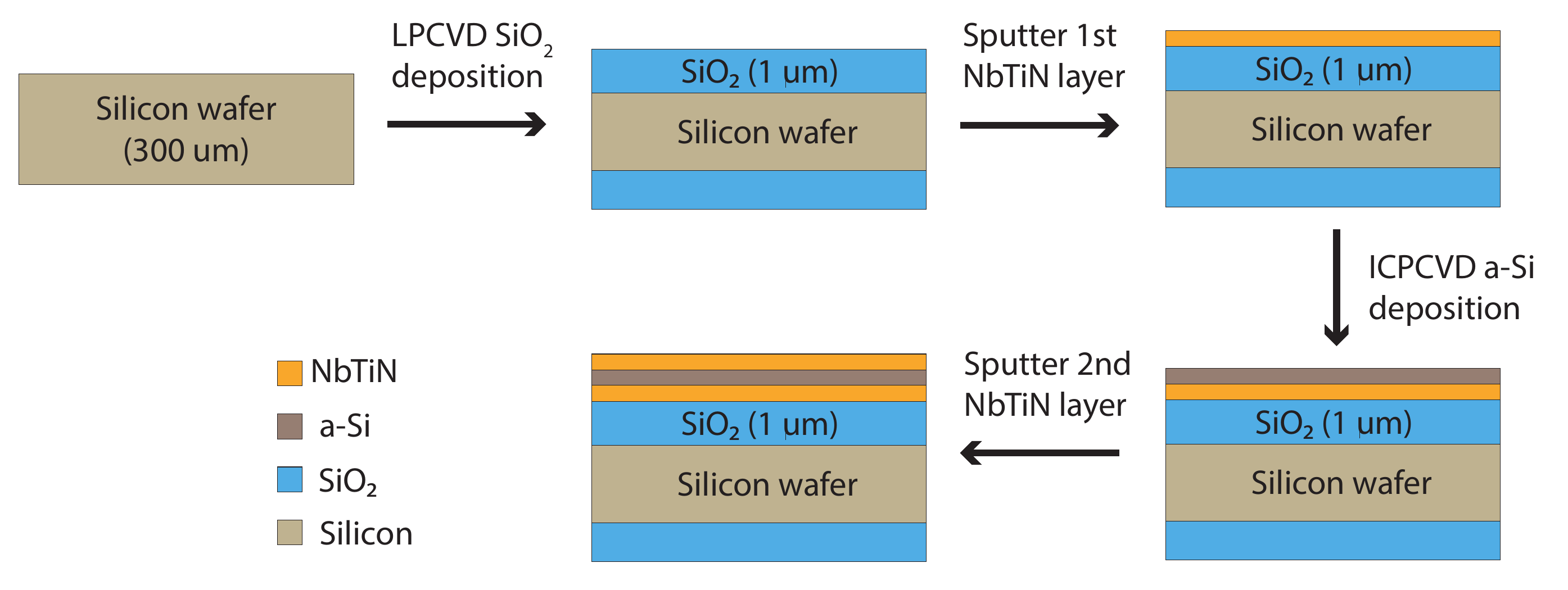}
\caption{Fabrication procedure of the layered structure that is later used for fabricating the Casimir cavity.}
\label{fig_supp_Fab_layer}
\end{figure}

In this study, the layered chips were fabricated using the following procedure:

\begin{enumerate}

\item \textbf{Wafer Preparation:} A double-side polished 300~$\mu$m bare Si wafer was selected. This thickness was chosen to avoid fragility associated with wafers thinner than 200~$\mu$m, while minimizing the measurement uncertainty of thin-film stresses, which increases in wafers thicker than 500~$\mu$m. 

\item \textbf{Insulating Layer Deposition:} A 1~$\mu$m thick layer of low pressure chemical vapor deposition (LPCVD) wet thermal oxide (SiO$_2$) was deposited on the Si wafer to form a robust insulating barrier.

\item \textbf{First NbTiN Film Deposition:} A 155~nm layer of niobium titanium nitride (NbTiN) was sputtered onto the SiO$_2$-coated front side of the wafer using a Sputter system from AJA Int. (SuperAja) at the Kavli Nanolab. Deposition parameters were: Chamber temperature 20~$^\circ$C; Initial chamber pressure lower than $5\times 10^{-7}$~mbar; Target composition: NbTi (Nb:Ti=7:3); Deposition time 300~s; Power 250~W; N$_2$ flow 3.5~sccm; Argon flow 50~sccm; Voltage 300~V; Vacuum level during deposition 2.3~mTorr; Pre-sputtering time 2~min.

\item \textbf{Amorphous Silicon Deposition:} An amorphous silicon (a-Si) layer was deposited onto the NbTiN film by Inductively Coupled Plasma Chemical Vapor Deposition (ICP-CVD) using a PlasmaPro 100 system (Oxford Instruments). The deposition conditions were: Temperature 300~$^\circ$C; Plasma power 1000~W; SiH$_4$ flow 20~sccm; Argon flow 20~sccm; Pressure 10~mTorr.

\item \textbf{Second NbTiN Film Deposition:} A second 155~nm NbTiN film was sputtered onto the a-Si layer using the same parameters as described in step 3.

\item \textbf{Annealing:} The wafer was annealed on the ICP-CVD table at 300~$^\circ$C for 20~min in a high-vacuum environment.

\item \textbf{Dicing:} Finally, the wafer was coated with dicing resist, spin-coated at 1000~rpm, and baked at 110~$^{\circ}$C for 5~min. The processed wafer was subsequently diced into $8\times 10$~mm$^2$ chips for further processing.

\end{enumerate}

\subsection*{Supplementary Information B2: On-Chip Superconducting Casimir Cavity Fabrication}

\begin{figure}[ht]
\centering
\includegraphics[width=1\textwidth]{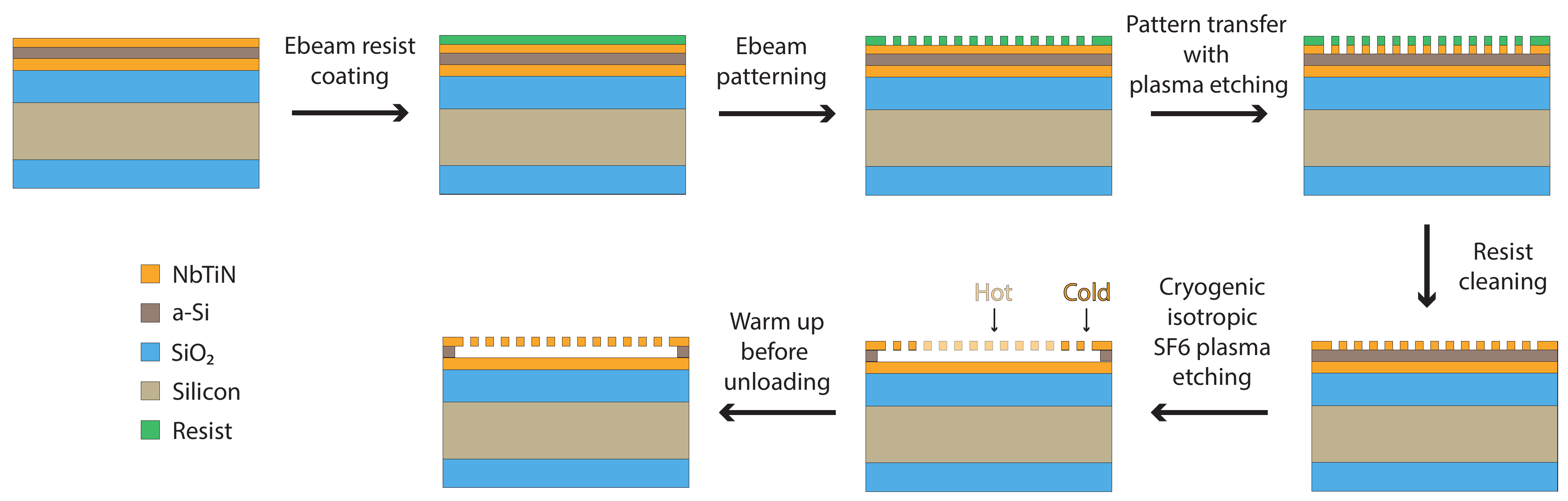}
\caption{Fabrication of the superconducting Casimir cavity.}
\label{fig_supp_Fab_cavity}
\end{figure}

Following the dicing of the wafers into microchips, the on-chip superconducting Casimir cavity is fabricated using the process outlined below:

\begin{enumerate}

\item \textbf{Resist Removal:} The dicing resist on each microchip is removed by sequentially rinsing with acetone and isopropanol (IPA).

\item \textbf{E-beam Resist Coating:} The microchips are spin-coated with the e-beam resist ARP-6200-13 (CSAR13) at 1300–1500~rpm, followed by a bake at 150~$^{\circ}$C for 3~min.

\item \textbf{Pattern Writing:} The desired pattern is written on the resist using electron beam lithography.

\item \textbf{Resist Development:} The e-beam pattern is developed sequentially in the following solutions (each for 1~min): pentyl acetate, a 1:1 mixture of MIBK:IPA, and IPA.

\item \textbf{Plasma Etching:} The developed pattern is transferred onto the top NbTiN layer by plasma fluorine etching using CHF$_3$ (25~sccm) and Ar (25~sccm) at 60~W power, 0.01~mbar pressure, for 45~min. This step was performed using the Sentech F2 system at the Kavli Nanolab Delft.

\item \textbf{Resist Removal:} The e-beam resist is removed by first immersing the microchips in a hot Dimethylformamide (DMF) bath (75~$^{\circ}$C) for 10~s, followed by a second hot DMF bath (75~$^{\circ}$C) in an ultrasonic bath at maximal power (level 9) for 15~min. Residual DMF is subsequently removed by rinsing with acetone and IPA.

\item \textbf{Cryogenic SF$_6$ Undercut:} The final undercut is performed at –120~$^{\circ}$C using SF$_6$ plasma. Prior to etching, the sample is cooled down in a high-vacuum environment with a 10~sccm helium flow for 15~min. The etch is then conducted for 3~min and 10~s with SF$_6$ plasma, where the RF power is set to 0~W and the ICP power to 1000~W. After etching, the sample is warmed to 40~$^{\circ}$C, pumped out, and then unloaded from the process chamber.

\end{enumerate}

\subsection*{Supplementary Information B3: Choice of materials for Casimir cavity}

The reasons for choosing niobium titanium (NbTiN) and inductively-coupled-plasma chemical vapor deposition (ICP-CVD) amorphous silicon (a‑Si) as the cavity layers are summarized below.

For the bottom layer, one could, in principle, use either a gold film or a superconducting film. However, gold films tend to be relatively inert and adhere strongly to subsequently deposited layers (e.g., plasma-enhanced chemical vapor deposition (PECVD) SiO$_2$/a-Si and ICP-CVD a-Si layers). To avoid these issues, we opted for a superconducting film instead. Although aluminum (Al) and NbTiN are both commonly used superconductors, we selected NbTiN for several reasons \cite{bimonte2019casimir}. First, its high critical temperature ($T_C$) enhances the sensitivity of the Casimir effect to the superconducting transition. Second, most cryogenic setups with low mechanical vibrations, such as the STM, cannot be cooled below 1~K, making NbTiN an ideal choice for both the bottom and top layers of the superconducting cavity. Lastly, the cryogenic SF$_6$ dry etching process (performed at -120~$^\circ$C) exhibits high etch selectivity between NbTiN and a‑Si (exceeding 1000), and the small gap configuration ensures that the top NbTiN layer remains well‑thermalized (except at the final undercut), allowing the superconducting layers to remain parallel and unetched during the undercutting process.

Compared to other superconducting thin films like Al, the NbTiN film forms only a relatively thin native oxide when exposed to air, unlike the dense 1–3~nm oxide that typically forms on Al surfaces \cite{revie2008corrosion}. This native oxide on NbTiN can be removed by annealing at elevated temperatures \cite{zhang2018characterization}, thereby cleaning the surface prior to STM measurements. Removing the surface oxide is crucial for Casimir force experiments, as residual electrostatic charges on the oxide can limit the measurement resolution.

ICP‑CVD a‑Si was chosen as the sacrificial layer because it can be undercut using cryogenic SF$_6$ dry etching. Dry etching is preferred for undercutting small-gap structures since wet etching introduces surface tension effects that limit the aspect ratio of the suspended structure. Although SiO$_2$ can also serve as a sacrificial layer and be undercut with vapor hydrofluoric acid (HF), we found that vapor HF aggressively attacks NbTiN. Moreover, we opted for ICP‑CVD instead of PECVD because PECVD‑deposited SiO$_2$/a‑Si films tend to develop bubbles after the deposition of the top NbTiN layer. These bubbles likely arise from residual hydrogen in the amorphous film, a consequence of using NH$_3$ during the PECVD process. In contrast, the high plasma power available in ICP‑CVD allows the use of N$_2$ instead of NH$_3$ during deposition, even at lower temperatures, thereby avoiding bubble formation.

\subsection*{Supplementary Information B4: Microchip Design and Sample Preparation for STM Measurement}

The microchip used to measure the variation of the Casimir force between superconductors during the phase transition is designed for optimal compatibility with the scanning tunneling microscopy (STM) system. The microchip layout consists of four key components:

\begin{enumerate}

\item \textbf{Suspended Membranes:} NbTiN membranes are suspended over small gaps by introducing arrays of holes on top, forming the core measurement region.

\item \textbf{Recognition Pattern Box:}  A reference pattern is incorporated to assist the in-situ STM optical camera in identifying the exact landing position of the STM tip on the microchip. This recognition is facilitated by the bending of relatively large pattern edges after SF$_6$ undercut etching, which alters the reflection intensity detected by the camera. The objective is to ensure precise STM tip alignment at the very edge of the suspended membrane, minimizing tip-sample interactions during measurement.

\item \textbf{Electrical Isolation Rings:} Electrical isolation structures encircle both the suspended membranes and the recognition pattern boxes. These rings prevent unintended electrical shorting between the top and bottom metallic layers. The isolation rings consist of four concentric rings with varying widths, from 50~nm to 200~nm, ensuring that the NbTiN top layer ($t= 155$~nm) is completely etched through while simultaneously restricting the SF$_6$ undercut size. This design minimizes potential electrical shorting in narrow-gap structures.

\item \textbf{Connection Boxes:} Dedicated connection regions facilitate wire bonding between the bottom NbTiN layer and the STM holder, establishing reliable electrical contact.

\end{enumerate}

Following fabrication of the microchip with the superconducting Casimir cavity, the chips are mounted onto the STM holder for integration into the STM system. The preparation steps are as follows:

\begin{enumerate}

\item \textbf{Electrical Short Testing:} The electrical connectivity between the back gate and the suspended membrane is verified using a probe station at the Kavli Nanolab.

\item \textbf{Chip Mounting with Conductive Adhesive:} The chip is affixed to the holder using EPOTEK ELECTRICAL ADHESIVE P.N. H20E (prepared by mixing equal masses of part~A and part~B). The adhesive is cured by baking at 150~$^{\circ}$C for 5~min.

\item \textbf{STM Holder Assembly:} The holder chip consist of two $1\times 12$~mm$^2$, 1-mm-thick bars (one coated with Au layer on top, the other coated with thick PECVD SiO$_2$ to achieve electrical isolation), a $12\times 10$~mm$^2$ 300~um bottom chip (coated with thick PECVD SiO$_2$), and two $1\times5~$mm$^2$ 300~um small pieces. The assembly process involves the following steps: (1) The isolation bar is first glued onto the base chip, followed by the Au-coated bar on top of the isolation bar. This structure is then baked. (2) The assembled dummy chip is glued onto the STM holder, with the four small pieces positioned on the sides and baked again. (3) The small piece chips (two of them) are electrically connected to both the STM holder and the Au-coated bar, as well as to the remaining two small chips using epoxy, followed by another baking step. (4) The electrical probe components are screwed onto the STM holder. Finally, the sample chips are glued onto the dummy chip and baked.

\item \textbf{Wire Bonding:} The membranes and back gates of the sample chip are wire bonded to the dummy chip holder, ensuring stable electrical connections.

\item \textbf{High-Vacuum Baking:} The STM holder is placed inside the STM load lock and baked at 200~$^{\circ}$C under high vacuum ($<10^{-8}$~mbar) for two days. This step removes contaminants such as native oxides and residual organic materials.

\item \textbf{STM System Integration and Measurement:} The sample is transferred into the STM head. The STM tip is sharpened using a flat copper sample before commencing measurements.

\end{enumerate}

\begin{figure*}[ht]
\centering
\includegraphics[width=0.9\textwidth]{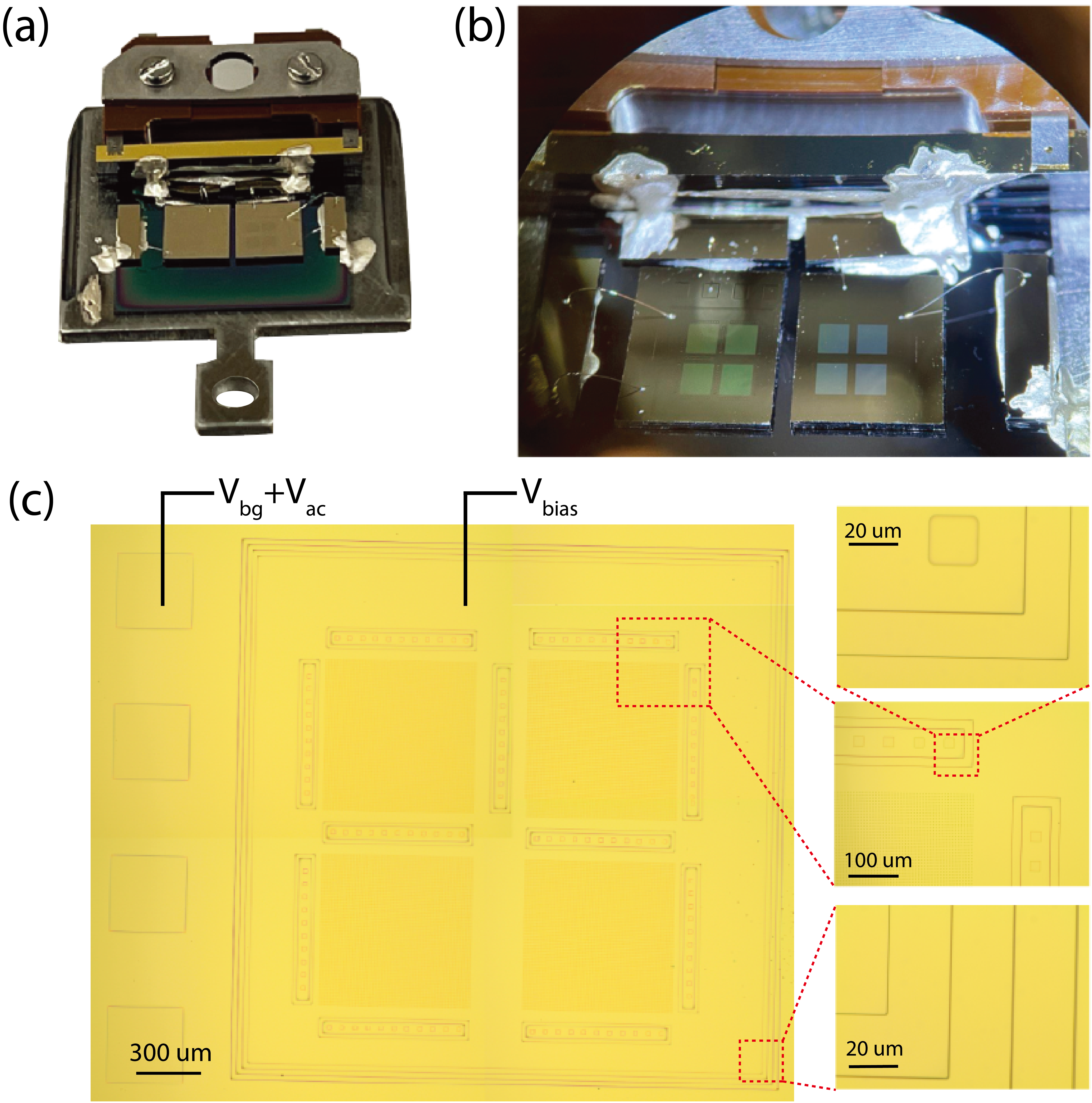}
\caption{\textbf{(a)} Photo of the STM holder with small- and big-gaps chips on top. \textbf{(b)} Optical microscope image of the STM holder. \textbf{(c)} Optical images of the small-gap chip, with suspended membranes, electrical isolation patterns, and boxes for accessing electrically to the backgate electrodes.}
\label{fig_supp_STMholder2}
\end{figure*}

\newpage
~
\newpage

\section*{Supplementary Information C: Additional information on mitigating unwanted forces}

\subsection*{Supplementary Information C1: Signature of minimized tip-membrane interaction}

\begin{figure*}[ht]
\centering
\includegraphics[width=1\textwidth]{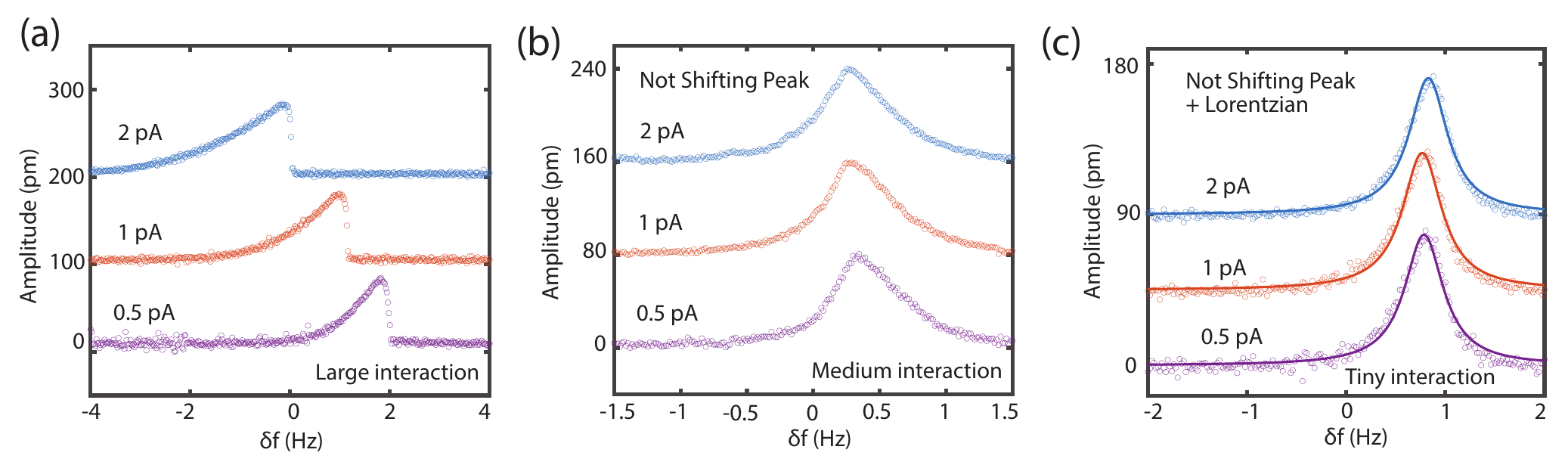}
\caption{Confirmation of minimizing the tip-membrane interaction by checking the 1-1 mode peak of the membrane. \textbf{(a)} With relatively strong interaction between the tip and the membrane, the peak does not have a Lorentzian shape and shifts as the tunneling current increase. \textbf{(b)} With relatively weak interaction between the tip and the membrane, the peak does not have a Lorentzian shape but does not shift as much when the tunneling current increases. \textbf{(c)} With minimized interaction between the tip and the membrane, the peak has a Lorentzian shape and does not shift when the tunneling current increases.}
\label{fig_supp_MinimizeTipSample}
\end{figure*}

\subsection*{Supplementary Information C2: STM tip preparation before measurement}

\begin{figure}[ht]
\centering
\includegraphics[width=1\textwidth]{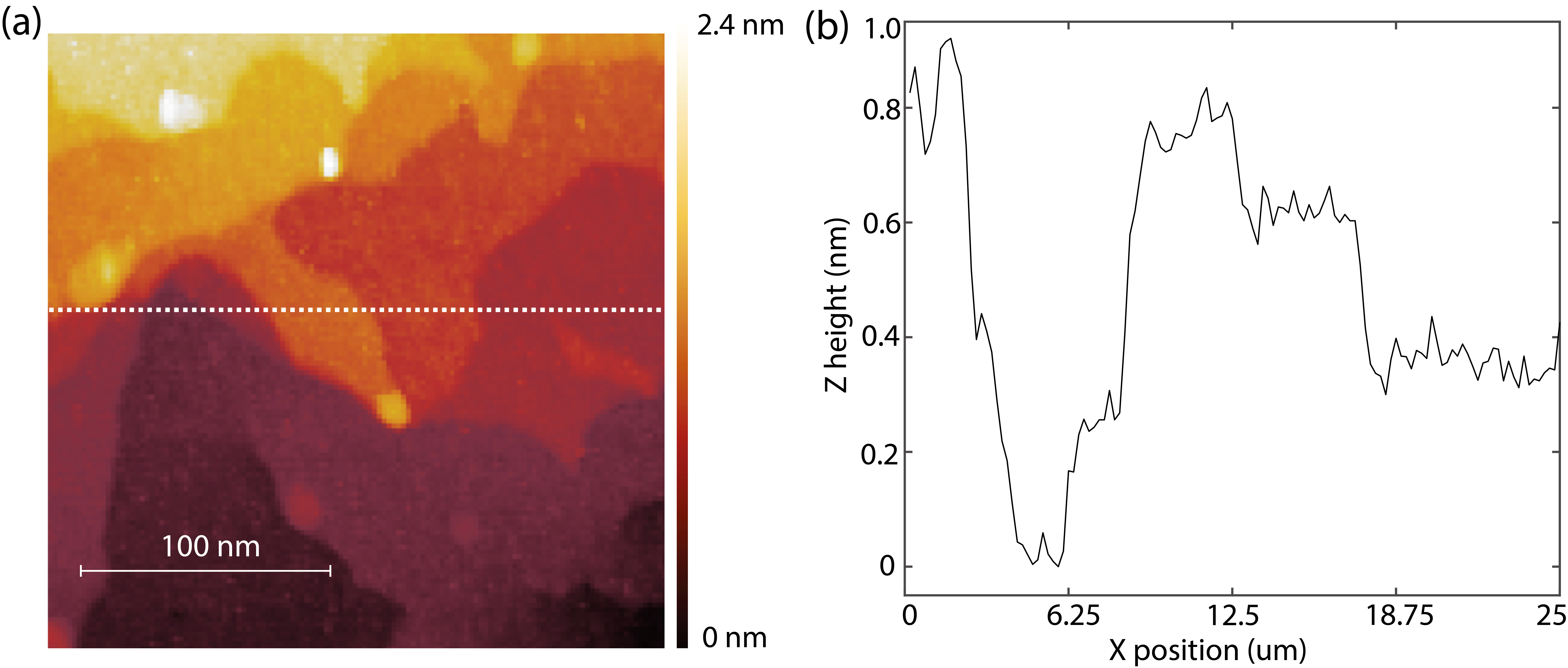}
\caption{\textbf{(a)} STM scan on an atomically-flat Copper sample. The tip is prepared to be sharp enough, such that \textbf{(b)} atomic-scale steps can be observed.}
\label{fig_supp_CuSample}
\end{figure}

A well-shaped STM tip is essential for atomic-scale imaging and achieving high-resolution data. The fabrication and conditioning process for the STM tip is as follows:

\begin{enumerate}

\item \textbf{Tip Fabrication and Cleaning:} The STM tip, made of tungsten (W), is fabricated by electrochemically etching tungsten wires in a KOH solution. Once the tip is made and dropped into an in-situ catcher, it is rinsed in distilled water and further cleaned by ultrasonic agitation in isopropanol (IPA) before installation into the tip holder.

\item \textbf{Tip Installation and Conditioning:} Following installation, the tip holder undergoes annealing and sputtering in the STM’s preparation chamber. The holder is then transferred to the STM measurement stage and is ready for use.

\item \textbf{Tip Sharpening on an Atomically Flat Copper Sample:} Prior to measuring the Casimir samples, an atomically flat copper sample is used to pre-condition the STM tip. This involves: (1) Pulsing the tip with appropriate voltages, and (2) Dipping the tip into the copper sample. These steps refine the tip’s profile, yielding atomic resolution as confirmed by imaging a step height portfolio on the copper sample, as shown in Figure~\ref{fig_supp_CuSample}\textbf{(b)}.

\item \textbf{Positioning the STM Tip on Casimir Cavities:} The in-situ optical camera is employed for recognition of the membrane's location, then the STM tip can be positioned on the spot, where is approximately 50~$\mu$m from the edge of the Casimir cavity. This placement minimizes unwanted interactions between the tip and the suspended membrane. Once positioned, the tip is gently dipped into the membrane by approximately 2~nm for a few seconds. This ensures that the measurement spot conforms well to the apex of the tip, facilitating high-resolution scans over a $100\times 100$~nm$^2$ area. When the STM tip can perform a high-resolution scan over $100 \times 100~\rm nm^2$ around a chosen spot, we consider the chosen spot to be stable enough for measurements.

\item \textbf{Resonance Measurement and Spot Selection:} Identifying a suitable measurement spot is nontrivial, particularly since the NbTiN surface lacks the flatness of the copper sample. Instabilities and surface imperfections may necessitate the exploration of multiple areas until an ideal spot is found. With a stable measurement spot, resonance peaks of the suspended membrane are determined by sweeping the AC drive signal applied to the bottom NbTiN film. The peaks are verified to exhibit a normal Lorentzian shape, free from nonlinear effects such as hardening or softening, which is critical for confirming the minimization of tip–membrane interaction, thereby preserving the membrane’s mechanical quality (Q) factor and enhancing the Casimir force measurement resolution.

\item \textbf{Temperature Sweep Considerations:} During measurements of the membrane’s resonance frequency as a function of temperature, it is important that the STM tip cannot be retracted out of the tunneling range in the $z$-direction, since the piezo-actuator’s expansion or contraction in the $x$/$y$-directions with temperature variations can cause the loss of the original measurement spot. As a result, the atomic tracker, a feedback system in the STM for tracking the same spot on the topography, is activated not only during high-resolution measurement (with a radius of 10~pm) but also during temperature changes (with a tracking radius of approximately 100~pm).

\end{enumerate}

\newpage

\subsection*{Supplementary Information C3: Film stress measurement based on electrostatic force pulling}\label{sec:StressMeasurement}

\begin{figure*}[ht]
\centering
\includegraphics[width=1\textwidth]{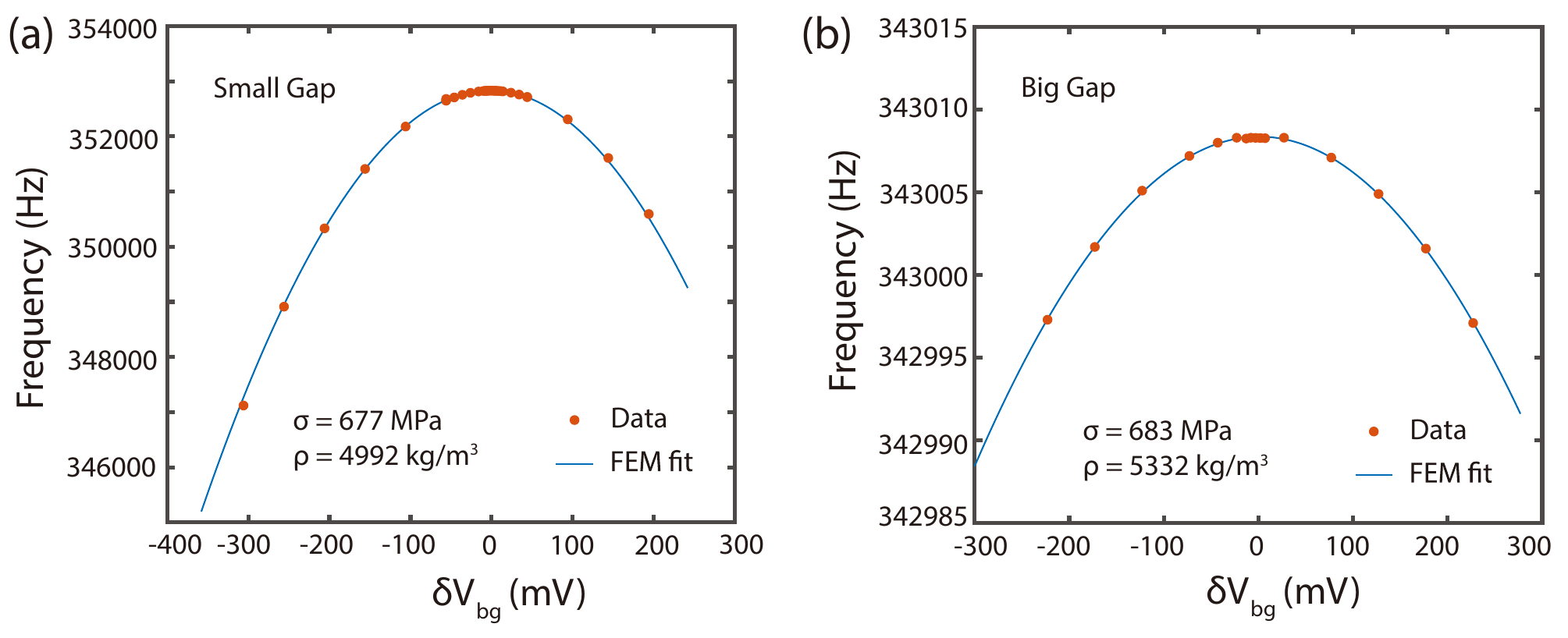}
\caption{Local contact potential difference (LCPD) measurements for checking the equilibrium backgate voltages of the membrane over \textbf{(a)} small- and \textbf{(b)} big-gaps. From the experimental results (orange dots), one can fit with the finite element model (FEM) result (blue line), and find the tensile stresses and densities of the membranes. The results obtained from FEM simulation is matching very well with the analytical formula Equation~\ref{eq:FreqshiftKeff}. From the LCPD measurement, film densities and film stresses of NbTiN membranes suspended over small- and big-gap cavities can be calculated.}
\label{fig_supp_VbgFreq}
\end{figure*}

Local contact potential difference (LCPD) measurements on both small- and large-gap membranes allow us to record resonance frequencies as a function of backgate voltage ($V_{\rm bg}$). Applying Equation~\ref{eq:FreqshiftKeff}, the angular frequency square shift becomes
\begin{equation}
    \Delta \omega^2 = -\frac{\epsilon_0}{\rho h}\frac{(V_{\rm bg}-V_0)^2}{d^3}-\frac{\epsilon_0}{\rho h}\frac{\delta V_{\rm bg}^2}{d^3}~,
\end{equation}
where $\epsilon_0$ is the vacuum permittivity, $\rho$ is the film density, $h$ is the membrane thickness, $V_{\rm bg}$ is the backgate voltage, and $V_0$ is the residual electrostatic voltage. Fitting the experimental data with the above formula, yields the densities ($\rho$) of the suspended NbTiN membranes, while the peak frequencies of the parabolic response curves provide the film stresses.

\newpage

\subsection*{Supplementary Information C4: Spectroscopy (\texorpdfstring{$I$-$V$}{I-V}) measurement for checking the bias voltage's influence on the membrane frequency}

\begin{figure}[ht]
\centering
\includegraphics[width=1\textwidth]{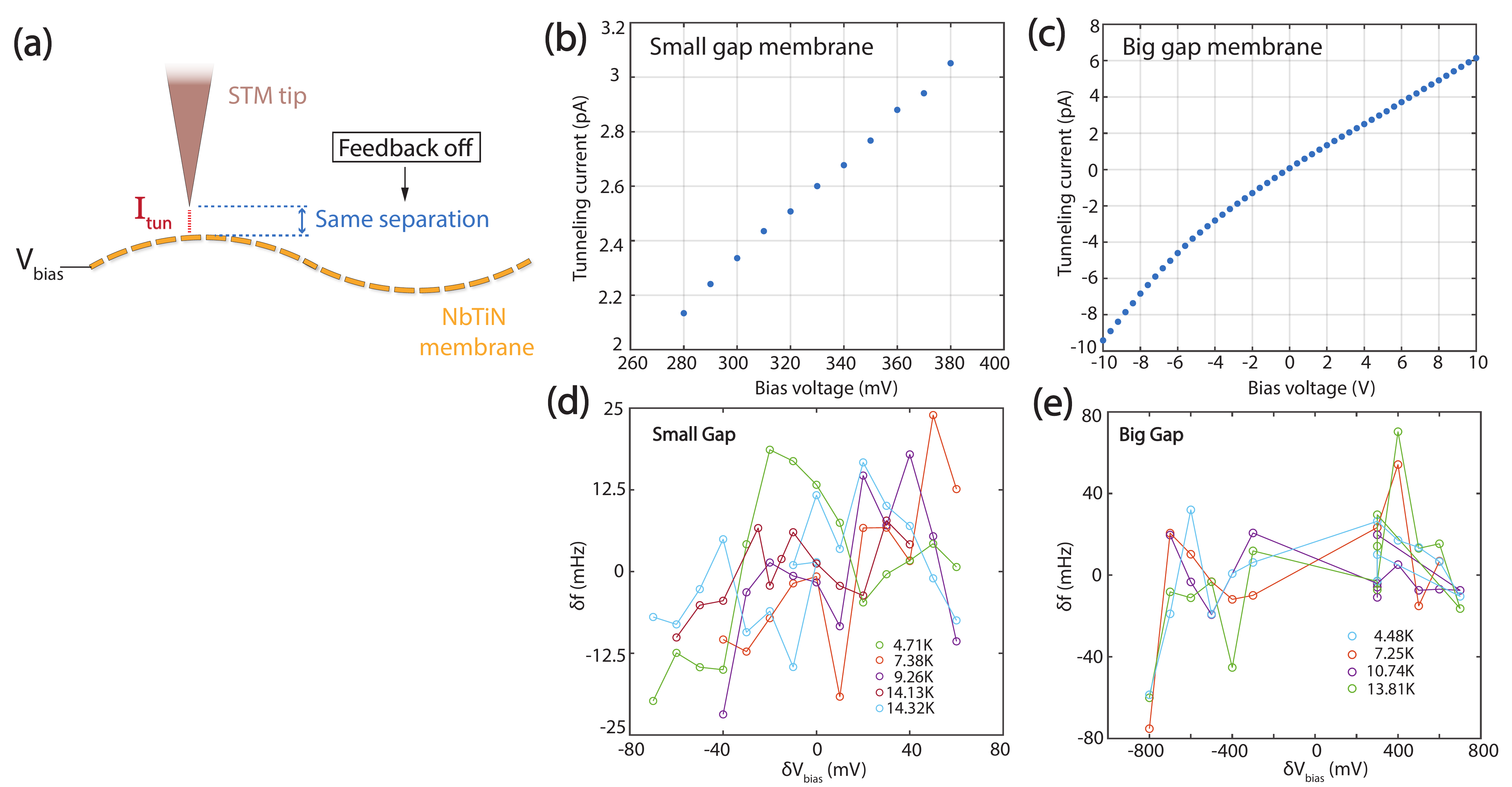}
\caption{\textbf{(a)} Schematic of the spectroscopy ($I$-$V$) measurement, where the separation between the STM tip and the suspended membrane is kept the same, by varying the bias voltage between the STM and the membrane, and turning off the feedback control of the tunneling current. Tunneling current $I$ as a function of the bias voltage ($V$), between the STM tip and the \textbf{(b)} small- and \textbf{(c)} big-gap membranes, for each pair of $I$-$V$ values, the tip-sample separation are kept. These values are used for checking whether the tip-sample interaction depends on the bias voltage, as well as the dependence on the temperature, as shown in \textbf{(d)} for small-gap membrane, and in \textbf{(e)} for big-gap membrane.}
\label{fig_supp_didvTc}
\end{figure}

Experimental results on the nanomembranes' resonant frequencies dependence on the bias voltage is shown in Figure~\textbf{3(d)} in the main text. The experiment is performed by setting the separation between the STM tip and the membrane unchanged, and varies the bias voltage and thus the tunneling current between the tip and the membrane. The corresponding tunneling current values at different voltages are shown in Figure~\ref{fig_supp_didvTc}\textbf{(b)} for small-gap and \textbf{(c)} for big-gap membranes. The insets in Figure~\textbf{3(d)} are shown with more details in Figure~\ref{fig_supp_didvTc}\textbf{(d)} for small-gap and \textbf{(e)} for big-gap membranes

\newpage

\section*{Supplementary Information D: Measurements with atomic force microscopy and white light interferometry}

\subsection*{Supplementary Information D1: Roughness and thickness measurement of the NbTiN films}

We performed atomic force microscopy (AFM) measurement to identify the roughness of the NbTiN suspended membranes and backgates (bottom films) after the cryogenic SF$_6$ plasma undercut. In Figure~\ref{fig_supp_AFMbottom}, we show the AFM scan of the backgate films on the small-gap (Panel~\textbf{(a)}) and big-gap (Panel~\textbf{(c)}) chips, whose RMS roughness are $(0.90\pm 0.05)$~nm and $(0.96\pm 0.07)$~nm, respectively. The area exposed to the plasma etchant is roughened up to $(1.73\pm 0.55)$~nm and $(1.78\pm 0.80)$~nm for the small- and big-gap bottom films, respectively. The zoom-in images of the bottom films under the small- and big-gap cavities, are shown in Panel~\textbf{(b)} and Panel~\textbf{(d)}, respectively, showing the patch sizes of the NbTiN films to be around $\ell=20\sim 30$~nm in radius.

\begin{figure*}[h]
\centering
\includegraphics[width=0.7\textwidth]{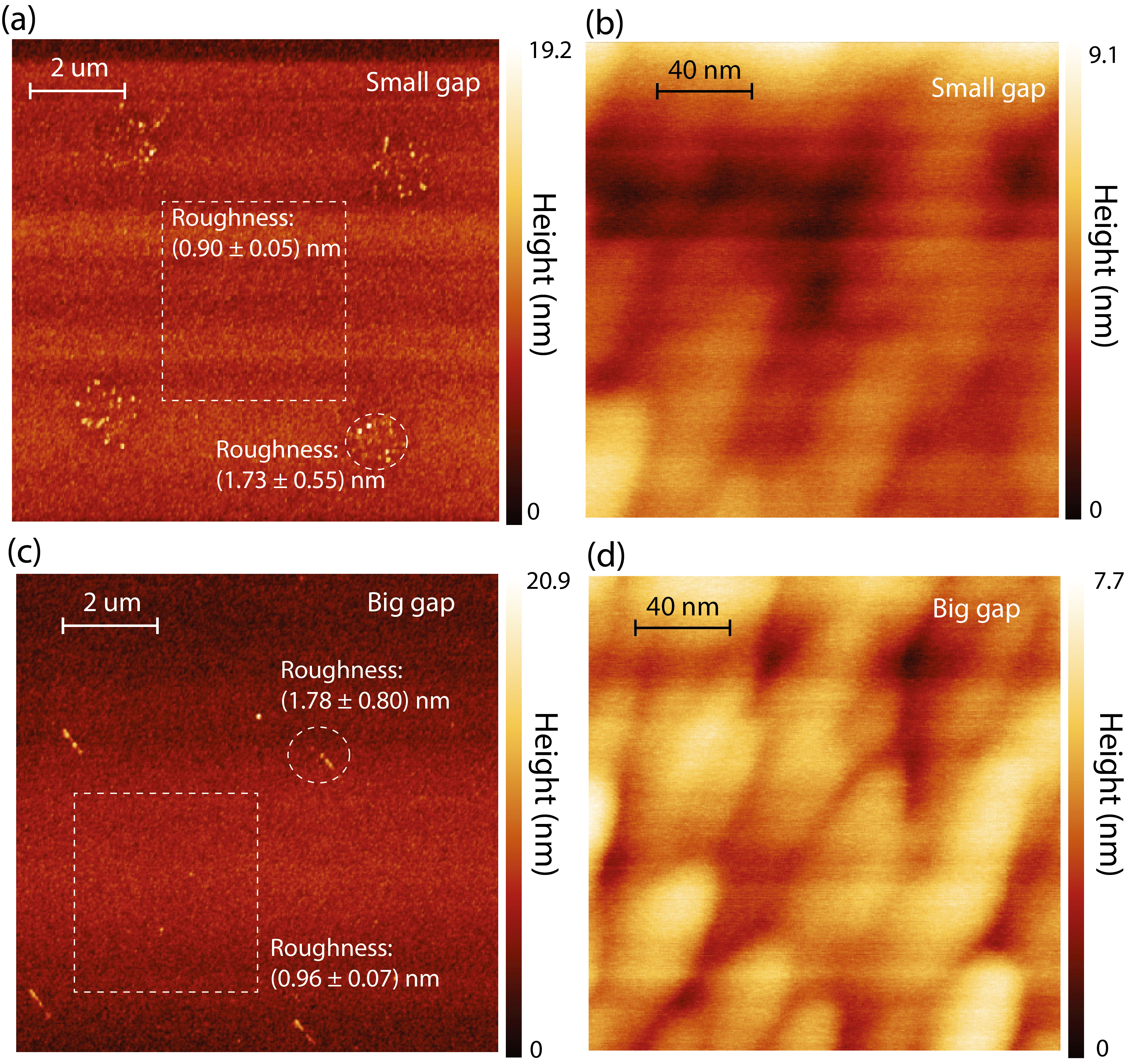}
\caption{Atomic force microscope scan on the bottom superconducting films. The Panels~\textbf{(a)} and \textbf{(c)} represent the large span topography of the bottom backgate films of the small- and big-gap cavities, respectively. Panels~\textbf{(b)} and \textbf{(d)} are topography over a small span, respectively, i.e. the right figures are the zoom-in scans of the figures on the right. From these AFM scans, the patch sizes of the NbTiN films are found to be around $\ell=20\sim 30$~nm in radius.}
\label{fig_supp_AFMbottom}
\end{figure*}

In Figure~\ref{fig_supp_AFMtop}, we scan the top of the membranes on small- (Panel~\textbf{(a)}) and big-gaps (Panel~\textbf{(b)}), after they collapsed to the bottom. The roughness of the top membrane over the small-gap cavity is $(1.50\pm 0.27)$~nm, while the one over the big-gap cavity is $(2.73\pm 0.26)$~nm, from which we know that depositing a thicker amorphous silicon layer (1213~nm instead of 190~nm), will increase the roughness of the top NbTiN film a bit.

\begin{figure*}[h]
\centering
\includegraphics[width=0.56\textwidth]{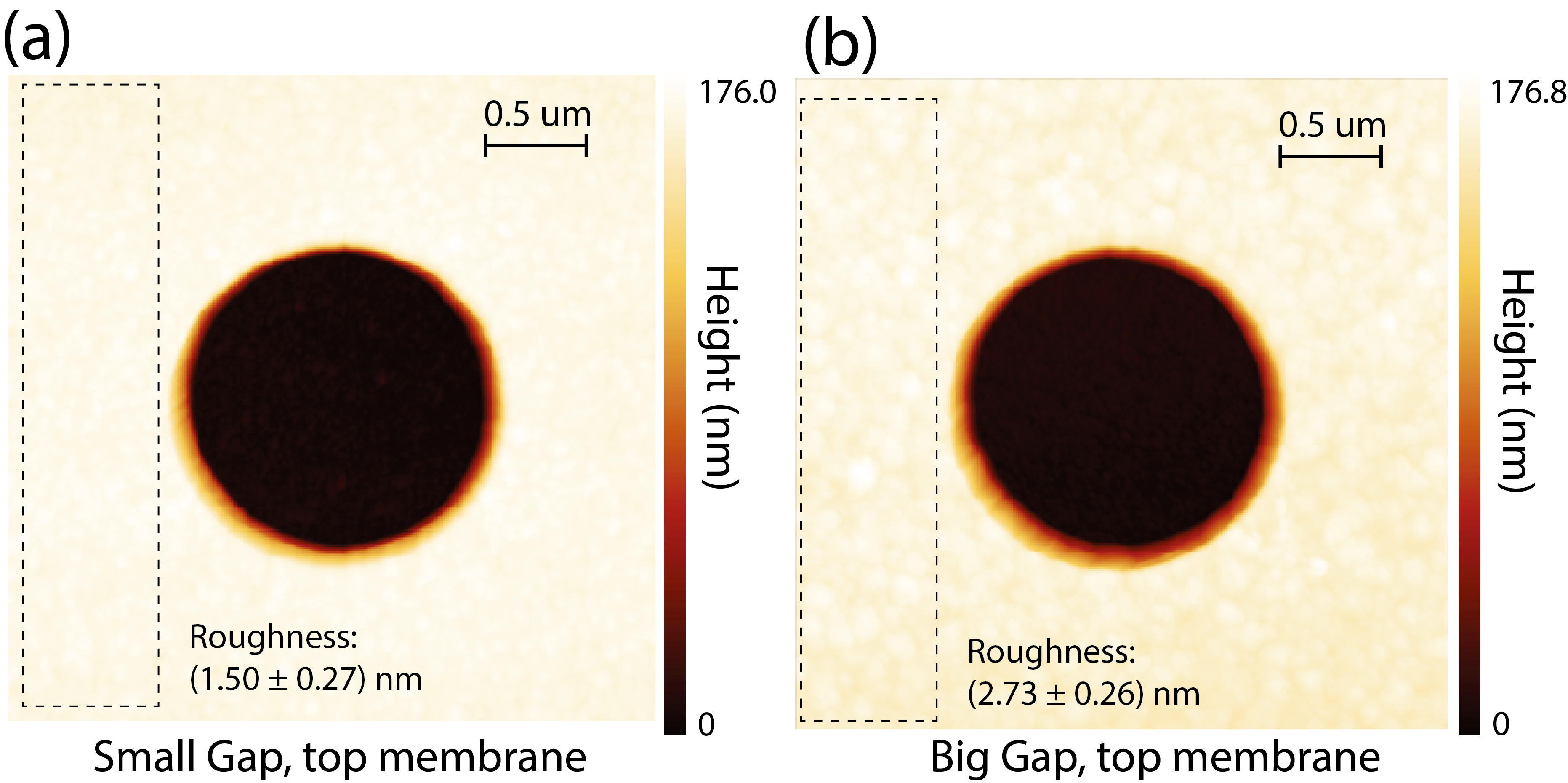}
\caption{Atomic force microscope scan on top of the NbTiN membranes after they are collapsed to the bottom film. Panels~\textbf{(a)} and \textbf{(b)} represent the topography of the suspended membrane over the small- and big-gap cavities, respectively. The thickness of the membranes are both around 155~nm.}
\label{fig_supp_AFMtop}
\end{figure*}

\newpage

\subsection*{Supplementary Information D2: Gap size measurement of the Casimir cavities}

By comparing the heights of the collapsed membranes over the small-gap (Panel~\textbf{(b)}) and big-gap (Panel~\textbf{(d)}) cavities in Figure~\ref{fig_supp_AFMgap}, we found both the thicknesses of the membranes over the small- and big-gap to be around 155~nm. From Panels~\textbf{(a)} and \textbf{(c)}, the small-gap size is found to be (345-155)=190~nm, and the big gap size is found to be (1368-155)=1213~nm, both matching the values we measured with the SEM. Here the suspended films are fabricated with large opening windows on the top NbTiN films, thus large overhangs are realized and the film stresses are released. for the convenience of AFM scan,

\begin{figure*}[ht]
\centering
\includegraphics[width=0.7\textwidth]{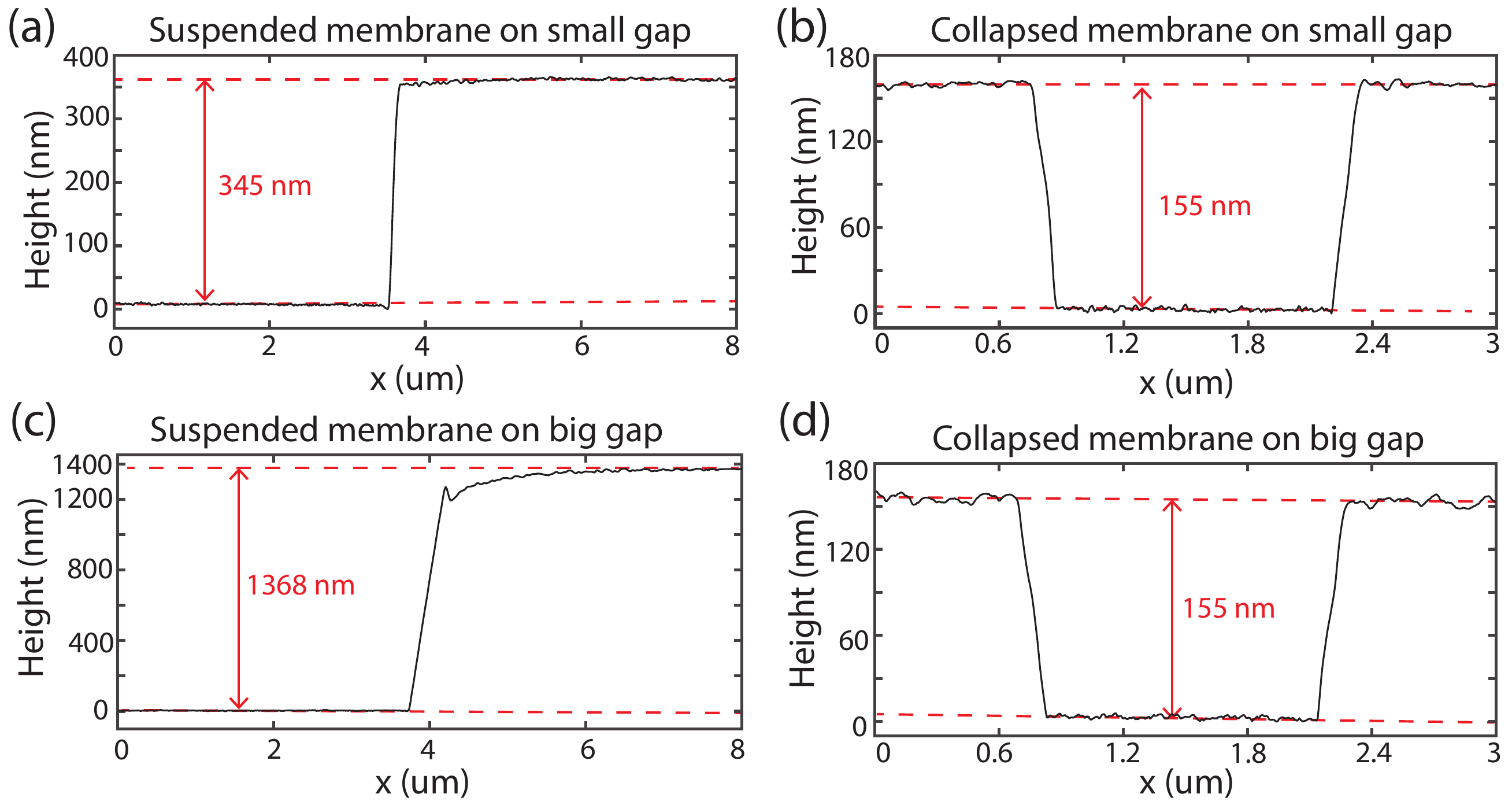}
\caption{Atomic force microscope scan on the top NbTiN suspended membranes before, and after (right column) they are collapsed to the bottom film. The top and bottom images represent the topography of the suspended membrane over the small- and big-gap cavities. The thickness of the NbTiN membranes are both found to be around 155~nm for small- and big-gap membranes.}
\label{fig_supp_AFMgap}
\end{figure*}

\newpage

\subsection*{Supplementary Information D3: Flatness estimation of the small-gap cavity}

In order to estimate the bending amplitude of the NbTiN membrane suspended over the small-gap, we use both theoretical estimation and FEM simulation for the approximation. The static deflection of the membrane due to external forces, such as the electrostatic force and the Casimir force, can be approximated by the formula describing the membrane's deflection under an uniform pressure $P(z_0)$ \cite{maier1995new}:

\begin{equation}
    P(z_0) = C_1 \frac{4t\sigma}{L^2}z_0 + C_2 \frac{16E}{L^4}z_0^3,
\end{equation}
where $C_1$ is a constant and $C_1=3.45$ for square membrane, $C_2$ is a Poisson-ratio-dependent constant, $E$ is the Young's modulus, $z_0$ is the deflection amplitude, $t$ is the thickness of the membrane, $\sigma$ is the film stress, $L$ is the lateral size of the membrane. For cases of the membranes under high tensile stress, the first term dominates and $P(z_0)\approx 4 C_1 t \sigma z_0 / L^2$. By comparing it with the static pressure $P_{Cas}(d)$ induced by the Casimir force is
\begin{equation}
    z_{0,Cas}(d) \approx P_{Cas}(d)\cdot \frac{L^2}{4C_1 t\sigma},
\end{equation}
where $d$ is the gap size. When the square membrane is patterned with holes, an correction factor $C_{hole}$ is required for correcting the deflection:
\begin{equation}
    z_{0,Cas}(d) \approx C_{hole} P_{Cas}(d)\cdot \frac{L^2}{4C_1 t\sigma}.
\end{equation}
In our case, $C_{hole}\approx 1.086$ (FEM simulation), $P_{Cas}(190~nm)\approx -1.081\times 10^{-24}\times d^{-3.507}$, thus the deflection can be estimated as
\begin{equation}
    z_{0,Cas}(190~\rm nm) = -1.085\times \frac{1.081\times 10^{-24}}{(190\times 10^{-9})^{3.507}}\times\frac{(709\times 10^{-6})^2}{3.45\times 4 \times 155*10^{-9}\times 677\times 10^6}[m]= -152~pm,
\end{equation}
which is close to the FEM simulation result as shown in Figure~\ref{fig:fig_supp_CasimirBending}, whose estimated deflection is 153~pm. For comparison, the big-gap membrane has a maximum deflection under the Casimir pressure $P_{Cas}(1213~nm)\approx -1.013\times 10^{-26}\times d^{-3.829}$, to be $z_{0,Cas}(1213~\rm nm)=0.17~pm$.

\begin{figure*}[h]
\centering
\includegraphics[width=1\textwidth]{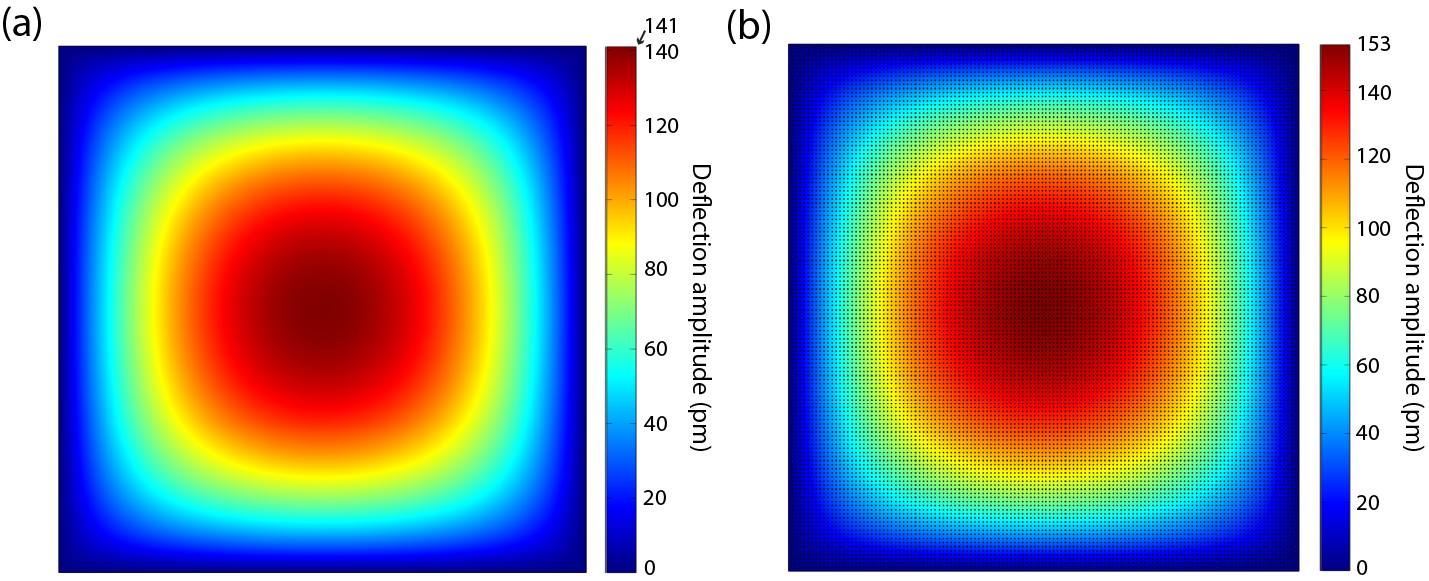}
\caption{Membrane deflection profiles simulated with finite element method. (a) Deflection of the simple $709\times 709\rm ~um^2$ square membrane, with a maximum deflection of 141~pm. (b) (a) Deflection of the $709\times 709\rm ~um^2$ square membrane patterned with arrays of holes, with a maximum deflection of 153~pm. A ratio of $C_{hole}=1.085$ is found between the two cases. The FEM simulations are performed by executing a separation-dependent Casimir pressure on the membrane, such that the balance between }
\label{fig:fig_supp_CasimirBending}
\end{figure*}

\newpage

\subsection*{Supplementary Information D4: White-light interferometry upper bound on membrane deflection}

To place an upper bound on the maximum deflection of the small- and large-gap membranes, we measured their surface profiles using a white-light interferometer (Bruker K1).The measured height maps are shown in Figure~\ref{fig:fig_supp_WhiteLightFlatness}. Cross-sectional profiles were taken along the blue dashed lines in Figure~\ref{fig:fig_supp_WhiteLightFlatness}(a,b) and fitted with a cosine function. The fitted amplitudes are $(0.327 \pm 0.535)$~nm for the small-gap membrane and $(0.032 \pm 0.451)$~nm for the large-gap membrane, both consistent with near-flat profiles within uncertainty. Unlike indirect approaches such as the balanced-capacitance method~\cite{fong2019phonon}, this direct interferometric measurement provides a conservative upper bound on membrane deflection and supports ultrahigh parallelism between the compliant nanomechanical membrane and the fixed plate. The measurement precision is limited by the interferometric wavelength, diffraction/interference from the membrane hole array, surface roughness, and instrument noise.

\begin{figure*}[h]
\centering
\includegraphics[width=1\textwidth]{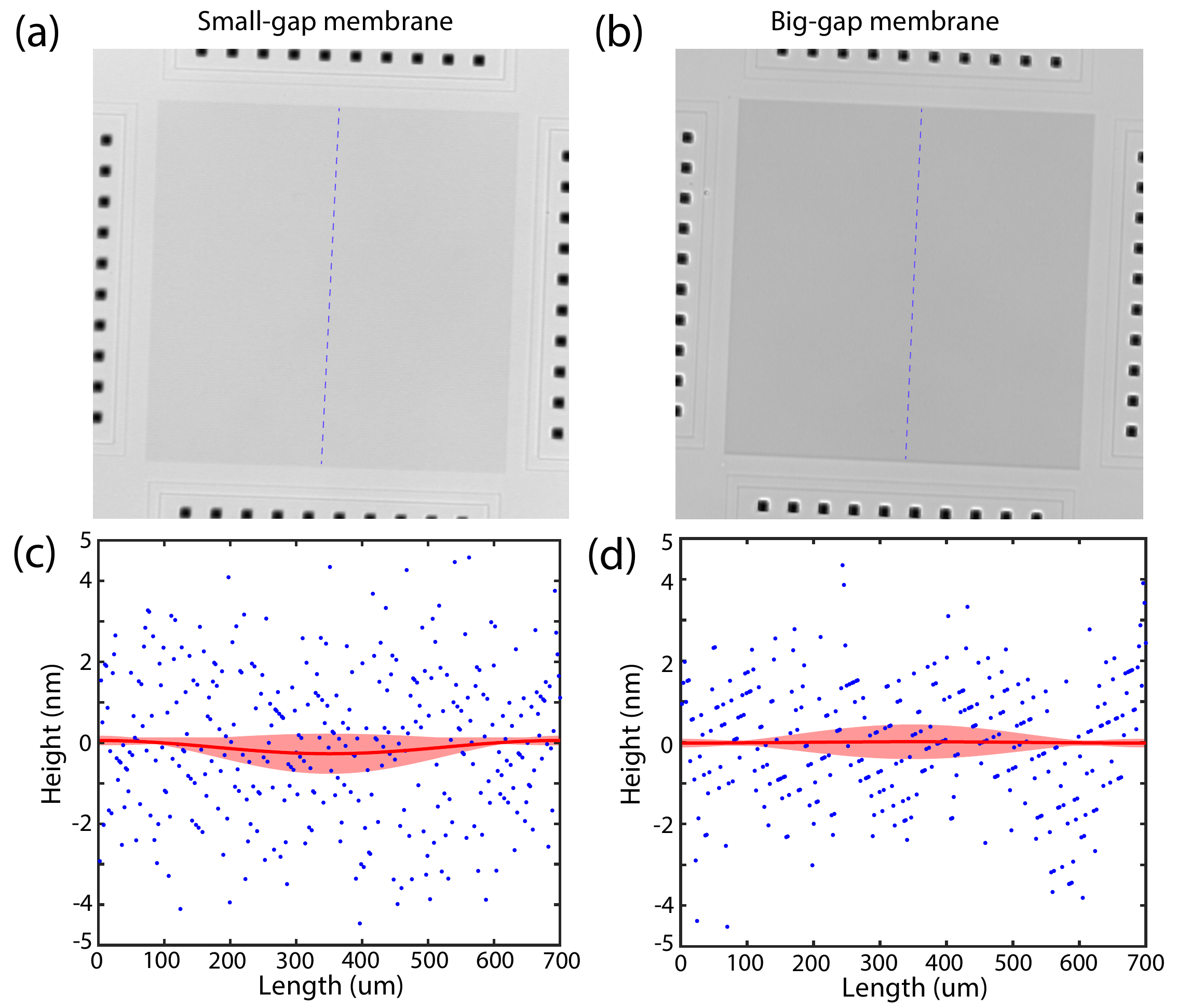}
\caption{White-light interferometry of membrane flatness (Bruker K1). (a,b) Surface-height maps of the small-gap and large-gap membranes. (c,d) Cross-sectional height profiles extracted along the blue dashed lines in (a,b), respectively. Red curves are cosine fits; red shaded bands denote fit uncertainty. Extracted deflection amplitudes are $(0.327 \pm 0.535)$~nm (small-gap) and $(0.032 \pm 0.451)$~nm (large-gap), both consistent with near-flat profiles within uncertainty. Negative height corresponds to deflection toward the backgate electrode.}
\label{fig:fig_supp_WhiteLightFlatness}
\end{figure*}

\newpage

\section*{Supplementary Information E: Scanning tunneling spectroscopy the superconducting gap}

\begin{figure*}[h]\label{fig_supp_IVspec}
\centering
\includegraphics[width=0.6\textwidth]{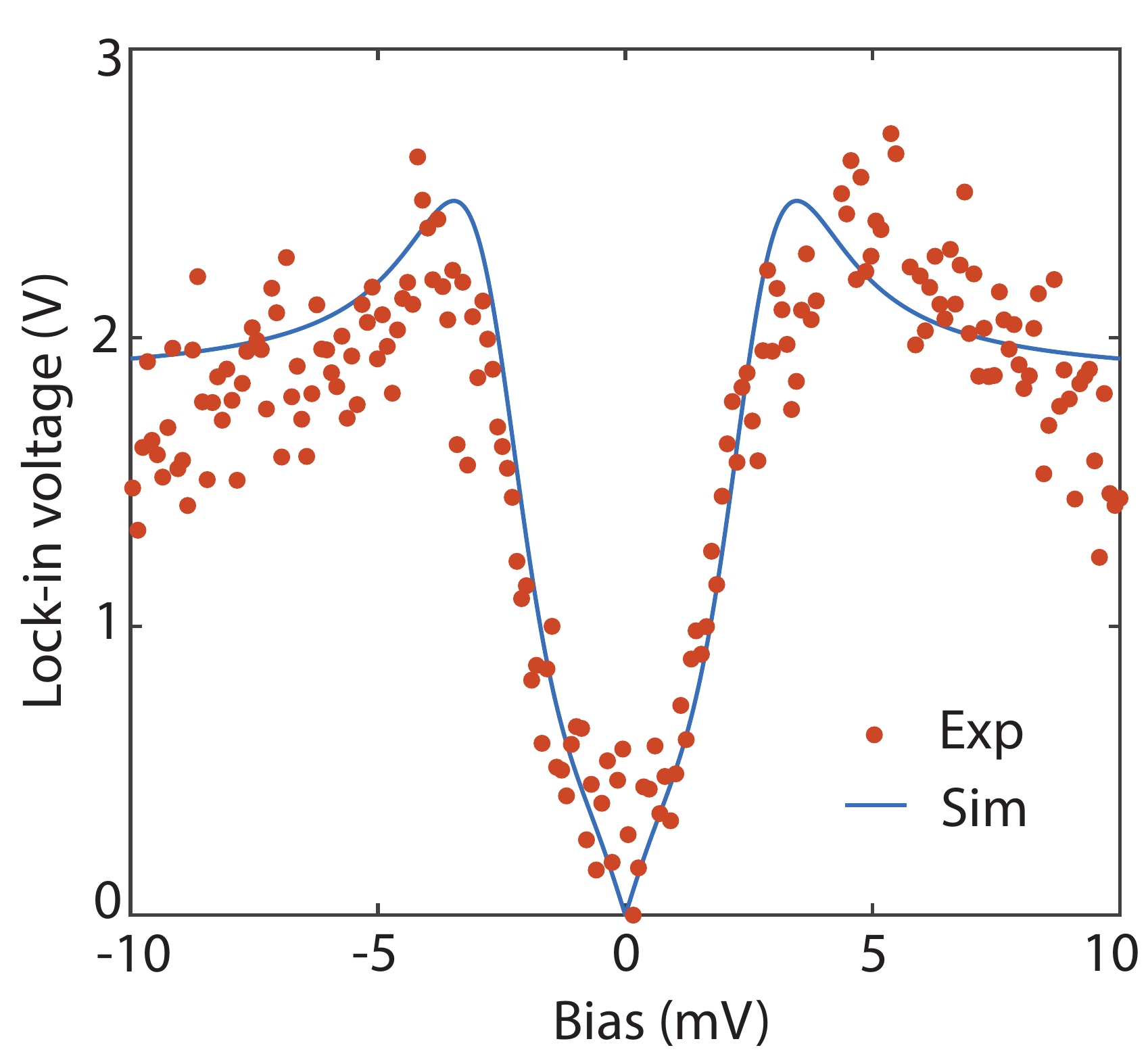}
\caption{Scanning Tunneling spectroscopy of the superconducting gap on the suspended membranes over the big gap cavity. The orange dots are experimental data and the blue line is the analytical fit, by using the superconducting gap $\Delta = 2.6$~meV, $\gamma=0.465$~meV, $T=4.6$~K and $T_c=14.2$~K.}
\end{figure*}

Following the superconducting gap calculation in \cite{khestanova2018unusual}, we fitted the measured tunneling conductance $G(V)$ using the standard expression:
\begin{equation}
    G(V_{\rm bias}) = \frac{dI}{dV_{\rm bias}}=A\int^{+\infty}_{-\infty} N(E,\gamma, \Delta)\frac{\partial f(E+e\cdot V_{\rm bias}, T)}{\partial (e\cdot V_{\rm bias})}dE,
\end{equation}
where $V_{\rm bias}$ is the bias voltage, $I$ is the tunneling current, $\Delta$ is the superconducting gap, $\gamma$ is the quasi-particle lifetime broadening (relaxation), $N$ is the density of state (DoS) at the Fermi level for the superconducting electrode, and $f(E+e\cdot V, T)$ is the Fermi-Dirac distribution at temperature $T$. The superconducting DoS is given by the Dynes formula that applies for thin films
\begin{equation}
    N(E,\gamma,\Delta) = \text{Re}\Bigl[ \frac{E-i\gamma}{(\sqrt{(E-i\gamma))^2-\Delta^2} } \Bigl].
\end{equation}

For the superconducting films used in this paper, the experimental data can be fitted with $\Delta=2.6$~meV and $\gamma=0.465$~meV, similar to the values used in Bimonte's work \cite{bimonte2019casimir}.

\newpage

\section*{Supplementary Information F: Theoretical analysis}

\subsection*{Supplementary Information F1. Frequency responses to electrostatic and Casimir forces}

\begin{figure*}[h]
\centering
\includegraphics[width=0.3\textwidth]{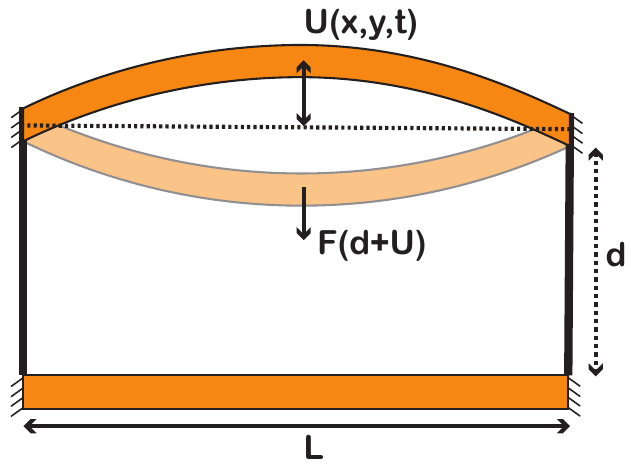}
\caption{Schematic of a parallel plate capacitor under an external force.}
\label{fig:ParallelPlate}
\end{figure*}

The resonance frequency of the parallel plate capacitor, consists of a suspended NbTiN membrane under the electrostatic and the Casimir interaction with the fixed NbTiN backgate, is calculation in this Supplementary Information. The schematic of the device is illustrated in Figure~\ref{fig:ParallelPlate}. We focus on analyzing the fundamental mode frequency $\omega_0$ (or $\omega_{\rm (1,1)}$). The thin membrane is under high tension, and the height of the membrane is deviated from the equilibrium (not forced) position only by a negligible amount compared to the separation. Based on these considerations, the membrane’s dynamics are described by the following equation of motion (EoM) \cite{schmid2016fundamentals}:
\begin{equation}
    \rho h \frac{\partial^2 H(x,y;t)}{\partial t^2}-\sigma h \triangledown^2 H(x,y;t)=P_{\rm ES}(H(x,y;t))+P_{\rm Cas}(H(x,y;t))~,
\end{equation}
where the height function $H(x,y;t)$ describes the separation between the membrane and the backgate, $\rho$ the density, $\sigma$ the tensile stress, and $h$ the thickness of the membrane. $P_{\rm ES}(H(x,y;t))$ and $P_{\rm Cas}(H(x,y;t))$ are the pressures coming from the electrostatic force and the Casimir force between the membrane and the backgate, respectively. The electrostatic and Casimir forces can be estimated using the additive Derjaguin Proximity Force Approximation (PFA) \cite{derjaguin1934untersuchungen}. This method decomposes the NbTiN surfaces into parallel patch pairs and sums the corresponding pairwise force contributions, thus
\begin{equation}
    \rho h \frac{\partial^2 H(x,y;t)}{\partial t^2}-\sigma h \triangledown^2 H(x,y;t)= -\frac{\epsilon_0}{2}\Bigl[ \frac{(V_{\rm bg}-V_0)^2}{H^2(x,y;t)} + \frac{V^2_{\rm rms}(H^2(x,y;t))}{H^2(x,y;t)} \Bigl]  +P_{\rm Cas}(H(x,y;t))~.
\end{equation}
where $\epsilon_0$ is the  permittivity of vacuum, $V_{\rm bg}$ is the backgate voltage, $V_0$ is the residual electrostatic potential, $V_{\rm rms}$ accounts for the force originated by the patches that are smaller than the separation between the plates \cite{garcia2012casimir}. By decomposing $H(x,y;t)$ as $d+U(x,y;t)$, where $d$ is the initial separation between the plates, $U(x,y;t)$ is the displacement under the external force, and noting that $\vert U(x,y;t)\vert \ll d$. We perform a first-order Taylor expansion of the (unit-area) forces on the right-hand side of the preceding equation, yielding:
\begin{equation} \label{eq:mainEoM}
    \rho h \frac{\partial^2 U(x,y;t)}{\partial t^2}-\sigma h \triangledown^2 U(x,y;t) - k_{\rm eff}^2 U(x,y;t)  = -\frac{\epsilon_0}{2}\Bigl[ \frac{(V_{\rm bg}-V_0)^2}{d^2} +\frac{V_{\rm rms}^2(d)}{d^2} \Bigl] + P_{\rm Cas} (d)~,
\end{equation}
where the effective spring constant
\begin{equation}
    k_{\rm eff}^2 = \epsilon_0 \Bigl[ \frac{(V_{\rm bg}-V_0)^2}{d^3} +\frac{V^2_{\rm rms}}{d^3} - \frac{1}{2 d^2} \frac{\partial V_{\rm rms}^2}{\partial d} \Bigl] + P'_{\rm Cas} (d)~.
\end{equation}
The displacement $U(x,y;t)$ can be further decomposed into $U_0(x,y)+U_f(x,y;t)$, where $U_0(x,y)$ is the static displacement, and $U_f(x,y;t)$ is the time-dependent membrane displacement under the external force. In the case, the membrane is only statically pulled down ($U_f(x,y;t)=0$), the Equation~\ref{eq:mainEoM} can be simplified into
\begin{equation} \label{eq:EoMstatic}
    -\sigma h \triangledown^2 U_0(x,y) - k_{\rm eff}^2 U_0(x,y)  +\frac{\epsilon_0}{2}\Bigl[ \frac{(V_{\rm bg}-V_0)^2}{d^2} +\frac{V_{\rm rms}^2(d)}{d^2} \Bigl] - P_{\rm Cas} (d) = 0~,
\end{equation}
implying that the static terms in the force only cause static deflection. Meanwhile, if $U_f(x,y;t)\neq 0$, the Equation~\ref{eq:mainEoM} can be written as
\begin{equation} \label{eq:EoMdynamic}
    \rho h \frac{\partial^2 U_f(x,y;t)}{\partial t^2}-\sigma h \triangledown^2 \Bigl( U_0(x,y) + U_f(x,y;t) \Bigl) - k_{\rm eff}^2 \Bigl( U_0(x,y) + U_f(x,y;t) \Bigl)  +\frac{\epsilon_0}{2}\Bigl[ \frac{(V_{\rm bg}-V_0)^2}{d^2} +\frac{V_{\rm rms}^2(d)}{d^2} \Bigl] - P_{\rm Cas} (d) = 0~.
\end{equation}
By putting together Equations~\ref{eq:EoMstatic} and \ref{eq:EoMdynamic}, the EoM is reduced into
\begin{equation} \label{eq:EoMpre}
    \rho h \frac{\partial^2 U_f(x,y;t)}{\partial t^2}-\sigma h \triangledown^2 U_f(x,y;t)  - k_{\rm eff}^2 U_f(x,y;t) = 0~.
\end{equation}
As shown in \cite{schmid2016fundamentals}, $U_f(x,y;t)$ can be written in the form:
\begin{equation}
    U(x,y,t)= \sum^{\infty}_{n=1}\sum^{\infty}_{j=1}u_{0,n,j}\phi_{n,j}(x,y)e^{i\omega_{n,j}t},
\end{equation}
where $\phi_{n,j}(x,y)=\text{sin}\frac{n\pi x}{L_x}\text{sin}\frac{j\pi y}{L_y}$ are the normalized modeshape function, $u_{0,j}$ are the coefficients that weight how much mode $(n,j)$ contributes to the overall vibration, and $\omega_{n,j}$ is the eigenfrequency of the spatial mode $\phi_{n,j}$.
The fundamental mode EoM of Equation~\ref{eq:EoMpre} can then be obtained by multiplying $\phi_{1,1}$ and integrating over the entire membrane area $A=L\times L$ with $L_x=L_y=L$, following the Galerkin's method, and gives
\begin{equation}
    \iint_A \Bigl( \rho h \frac{\partial^2 U_f(x,y;t)}{\partial t^2}-\sigma h \triangledown^2 U_f(x,y;t)  - k_{\rm eff}^2 U_f(x,y;t)
 \Bigl) \phi_{1,1} dxdy=0,
\end{equation}
then the equation can be written as
\begin{equation}
    -\rho h \omega_{0}^2\iint_A \phi_{1,1}^2 dxdy + 2\frac{\pi^2}{L^2}\sigma h \iint_A \phi_{1,1}^2 dxdy - k^2_{\rm eff} \iint_A \phi_{1,1}^2 dxdy=0.
\end{equation}
The EoM results in
\begin{equation} \label{eq:freqvsF}
    \omega_{0}^2= 2\frac{\pi^2}{L^2}\frac{\sigma}{\rho} - \frac{k^2_{\rm eff}}{\rho h}~.
\end{equation}
When $k_{\rm eff}=0$, the natural eigenfrequency is $\omega_n=\frac{\sqrt{2}\pi}{L}\sqrt{\frac{\sigma}{\rho}}$, and the membrane frequency under a voltage is written as
\begin{equation}\label{eq:freqES}
    \omega_{0}^2 = \omega_n^2\Bigl(1-\frac{L^2 k^2_{\rm eff}}{2\pi^2 \sigma h}\Bigl)~.
\end{equation}
From Equation~\ref{eq:freqvsF}, we can calculate the membrane's frequency response under a given external force to be 
(For clarity, we denote temperature-dependent variations in frequencies, pressures, and similar parameters by $\Delta$, while $\delta$ indicates changes at a fixed temperature $T$.)
 
\begin{equation}\label{eq:FreqshiftKeff}
   \delta \omega^2= \omega_{0}^2 - \omega_{_n}^2 = - \frac{k^2_{\rm eff}}{\rho h}~,
\end{equation}
or more intuitively, since the effective spring constant equals to the gradient of the external pressure, i.e. $k^2_{\rm eff}=P_{\rm ext}'$, when the membrane is under a separation-dependent external pressure $P_{\rm ext}$ or force $F_{\rm ext}$, the frequency shift is
\begin{equation}\label{eq:forceGradient}
   \delta \omega^2=  - \frac{1}{\rho h}  P'_{\rm ext} (d) =  - \frac{1}{\rho h L^2}  F'_{\rm ext} (d) ~,
\end{equation}
implying that the change in $\omega^2$, is proportional to the pressure (or force) gradient.

While sweeping the temperature of the membrane from $T_1$ to $T_2$, the variation of the external force gradient on the membrane, after calibrating out the frequency shift due to thermal expansion (contraction), is
\begin{equation}
    \Delta \omega^2 (T_1,T_2) = \delta \omega^2(T_2) -  \delta \omega^2(T_1) = - \frac{1}{\rho h L^2}\bigl( F'_{\rm ext}(T_2) -F'_{\rm ext}(T_1) \bigl)=- \frac{1}{\rho h L^2} \Delta F'_{\rm ext}~.
\end{equation}
In our experiment, the electrostatic force is canceled out in the experiment, i.e. $V_{bg}-V_0=0$. Meanwhile, the effect of $V_{\rm res}$, originated from the patches with the average size $\ell$ ($\ell \ll d$), can be estimated by the pressure of the form $P_{\rm patch}\approx 0.9\cdot \epsilon_0 V_{\rm rms}^2 \ell^2/d^4$, as shown in \cite{behunin2012modeling}. In our case, $\ell \approx 30$~nm is measured via the AFM scan, and reasonably assuming that $V_{\rm rms}$ is less than 10~mV, we have $P_{\rm patch}< 5.5\times 10^{-4}$~Pa, which is much smaller than the estimated Casimir pressure $P_{\rm Cas}\approx -0.4021$~Pa, as estimated in Supplementary Information~\textbf{H}, thus the pressure contribution from $V_{\rm rms}$ can be safely neglected. Given the above analysis, the effective spring constant $k_{\rm eff}^2=P'_{\rm Cas}(d)$, indicating that the small-gap membrane frequency shift in our experiment, after calibrating out the elastic force dependence on temperature, is a result of the variation in Casimir pressure. In this case, Equation~\ref{eq:freqES} can be rewritten for $\omega_n-\omega_0\ll \omega_n$, as
\begin{equation}
    \frac{\delta \omega}{\omega_0} = \frac{\delta f}{f} \approx - \frac{L^2}{4\pi^2 \sigma h}P'_{\rm Cas}= - \frac{1}{4\pi^2 \sigma h}F'_{\rm Cas}~.
\end{equation}
While checking the frequency shift induced by the temperature dependence on the Casimir force, the above equation can be rewritten as
\begin{equation}
    \frac{\Delta f}{f} = \frac{\delta f(T_2)-\delta f(T_1)}{f}=\frac{1}{4\pi^2 \sigma h} \Bigl( F'_{\rm Cas}(T_2)- F'_{\rm Cas}(T_1)\Bigl)~,
\end{equation}
indicating that the larger the temperature variation of the Casimir force gradient is, the higher percentage of the membrane frequency will shift. When the absolute Casimir gradient has a larger value, the temperature-induced influence on it will be greater, thus the frequency response will be more announced. \\

Moreover, the frequency response of a parallel plate capacitor, when it is under a generalized force of the form
\begin{equation}
     F(d)=\frac{B\cdot A}{\bigl(d+U(x,y;t)\bigl)^n},
\end{equation}
where $B$ is a separation independent constant, $A=L\times L$ is the surface area of the square membrane, $n>0$ is the power law factor, and $d$ is the separation. Assuming the membrane is deflecting only slightly, following the calculation precedure presented above, the frequency shift of the membrane under the external force will become
\begin{equation}
    \omega_{0}^2=2\frac{\pi^2}{L^2}\frac{\sigma}{\rho}+\frac{B\cdot n}{\rho h d^{n+1}},
\end{equation}
and the variation of $\omega^2$ is proportional to the gradients of the external force $F'(d)$ and pressure $P'(d)$:
\begin{equation}
    \delta \omega_0^2 = -\frac{P'(d)}{\rho h} = -\frac{F'(d)}{\rho h L^2}~.
\end{equation}

\subsection*{Supplementary Information F2: Modifications from the holes on the membranes}

For the purpose of releasing the nanomembranes, patterned holes on the membranes are required. By performing finite element method (FEM) simulation, we show the relationships of responses between the nanomembranes with holes and without holes here.

For membranes simulated with identical parameters including lateral size, thickness, Young's modulus, Poisson ratio, tensile stress, and gap size, their responses towards external forces are shown with the subscript '$h$' (with holes) and '$nh$' (without holes):
\begin{equation}
    \Bigl( \frac{f_{h}}{f_{nh}} \Big)^2= \frac{z_{nh}}{z_{h}} = \Bigl( \frac{\delta f_{nh}}{\delta f_{h}} \Big)^2 = Y_{\rm ratio}~,
\end{equation}

where $f_{\theta} = \omega/(2\pi)$ is the linear resonant frequency, $z_{\theta}$ is the center displacement due to the external force, $\delta f_{\theta}$ is the variation in the linear frequency ($\theta=h, nh$), and $Y_{\rm ratio}$ is the hole-dependent ratio. In our case, the radius of the holes is 0.75~$\mu$m and the center-to-center separation is 5.5~$\mu$m, $Y_{\rm ratio}=0.923$ is obtained from finite element method (FEM) simulation.

As for the change in the angular resonant frequency $\omega$, it is found that
\begin{equation}
    \delta \omega_{h}^2 = \delta \omega_{nh}^2~.
\end{equation}

As for the electrostatic and Casimir pressures acting on the membrane, the relation between the membranes with holes and without holes, in our case, is found to be
\begin{equation}
    \frac{\rm Membrane~area - Hole~area}{\rm Membrane~area}=\frac{\delta F_{h}}{\delta F_{nh}}\approx 0.945~.
\end{equation}

With FEM simulation, we also confirm that the effective mass of our squared membranes with holes, is still $m_{\rm eff}\approx 0.25m_0$, where $m_0$ is the total mass of the membranes after subtracting the masses of holes. 

The above relationships are taken into account while analyzing the experimental data.

\subsection*{Supplementary Information F3: Thermal expansion of the nanomembranes}

In order to find out the thermal expansion coefficients of the nanomembranes suspended over the small- and big-gap cavity, as shown in Figure~\textbf{5(f)} in the main text, we derive the membrane frequency dependence on the temperature here. For squared tensile nanomembranes, the linear resonant frequency $f=\omega/(2\pi)$ of the fundamental mode during the temperature ($T$) sweep is
\begin{equation}
    f(T)=\frac{1}{\sqrt{2}L}\sqrt{\frac{\sigma(T)}{\rho}},
\end{equation}
where $\sigma(T)$ is the temperature-dependent film stress, $\rho$ is the film density, and $L$ is the length of the membrane. We assume that the temperature variation is small $T=T_0+\Delta T$, where $T_0$ is the base temperature and $\Delta T$ is the temperature variation, this temperature change will lead to a change in stress $\sigma = \sigma_0 +\Delta \sigma$, where $\Delta \sigma$ can be written as \cite{zhang2020radiative}
\begin{equation}
    \Delta \sigma = -\frac{E}{1-\nu}\cdot \alpha(T)\cdot \Delta T,
\end{equation}
where $E$ is the Young's modulus, $\nu$ is the Poisson ratio of the membrane, and $\alpha (T)$ is the temperature-dependent thermal expansion coefficient of the membrane. With this and by using the Taylor expansion, the above equation can be rewritten as
\begin{equation}
    f(T)=\frac{1}{\sqrt{2}L}\sqrt{\frac{\sigma_0}{\rho}}\cdot\sqrt{1+\frac{\Delta \sigma}{\sigma}}=f_0 + \frac{f_0}{2}\frac{\Delta \sigma}{\sigma_0},
\end{equation}
where $f_0\approx\frac{1}{\sqrt{2}L}\sqrt{\sigma_0/\rho}$, and the change in frequency becomes
\begin{equation}
    \Delta f = f-f_0 = \frac{-f_0}{2\sigma_0}\cdot\frac{E\cdot\alpha(T)}{1-\nu}\Delta T.
\end{equation}
Since $\sigma_0 = 2L^2 f_0^2 \rho$, the frequency variation becomes
\begin{equation}
    \Delta f = \frac{-\alpha(T)\cdot E}{4L^2 f_0 \cdot \rho \cdot(1-\nu)}\Delta T,
\end{equation}
when the change of the temperature is very small, then $f_0\rightarrow f(T)$, and the above equation can be written as
\begin{equation}
    \Delta f^2 =\frac{-\alpha(T)\cdot E}{2L^2\cdot\rho\cdot (1-\nu)}\Delta T,
\end{equation}
From previous studies, the thermal expansion coefficient of superconductors at cryogenic temperature are usually formulated as \cite{smith1976thermal,simpson1978thermal}
\begin{equation}\label{eq:CTE}
    \alpha (T) = A\cdot T + B\cdot T^3,
\end{equation} 
where $A$ and $B$ are the temperature-independent coefficients. The the stress coming from the thermal expansion of the material becomes \cite{nelson2013atomic}

\begin{equation}
    \sigma_{\text{th}} = \frac{E_{\text{film}}}{1-\nu_{\text{film}}}\int^{T_2}_{T_1}(\alpha_{\text{film}}(T)-\alpha_{\text{sub}}(T))dT,
\end{equation}

where $E_{\text{film}}$ and $\nu_{\text{film}}$ are the Young's modulus and Poisson ratio of the thin film. From \cite{zhang2020mechanical}, we have $E_{\text{film}}=375$~GPa and $\nu_{\text{film}}=0.2949$, $\alpha_{\text{film}}$ and $\alpha_{\text{sub}}$ are the temperature-dependent thermal expansion coefficient of the thin film and the deposited substrate, respectively.

With the temperature data points shown in Figure~\textbf{5(f)} in the main text, one can calculate the thermal expansion coefficients of membranes suspending over the small- and big-gap cavities using Equation~\ref{eq:CTE}. For the small-gap membrane
\begin{equation}
    \alpha_{sg}(T) = (2.001\times 10^{-10})\cdot T+(9.159\times 10^{-13})\cdot T^3~[\rm K^{-1}]~,
\end{equation}
and for the big-gap membrane
\begin{equation}
    \alpha_{bg}(T) = (4.289\times 10^{-10})\cdot T+(3.181\times 10^{-13})\cdot T^3~[\rm K^{-1}]~.
\end{equation}

At $T=14.2$~K, $\alpha_{sg}=5.46\times 10^{-9}~\rm K^{-1}$ and $\alpha_{bg}=7.00\times 10^{-9}~\rm K^{-1}$. The difference in the thermal expansion coefficients between the two membranes over different gap sizes, can come from the different substrates the two membranes are deposited on, i.e. the different amorphous silicon sacrificial layer thicknesses, which can contribute to the overall membrane frequency dependence on the temperature.

\begin{figure*}[h]
\centering
\includegraphics[width=1\textwidth]{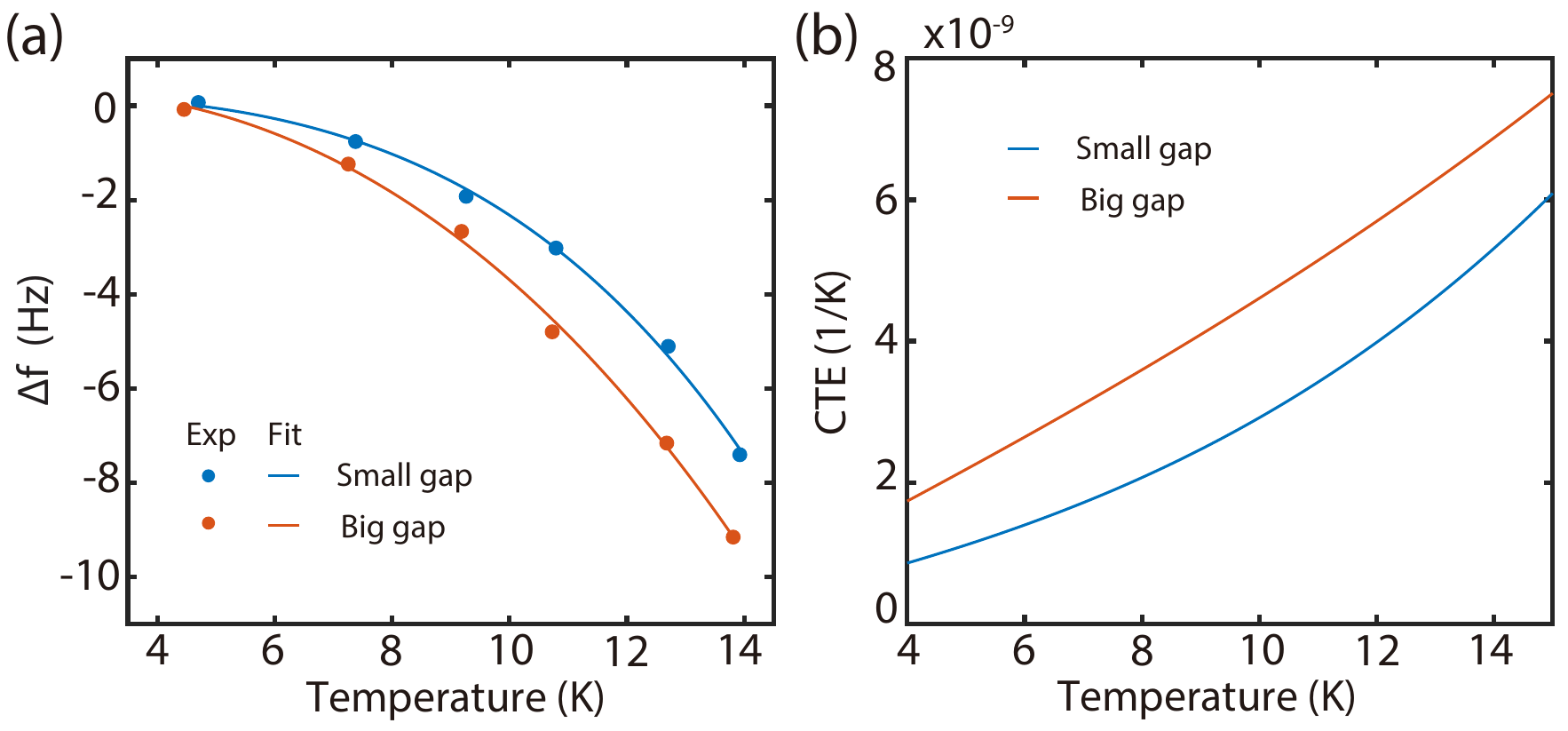}
\caption{Thermal expansion coefficient (CTE) of NbTiN membranes. \textbf{(a)} Frequency dependence on temperature for NbTiN membranes suspended over the small- (blue) and big-gap (orange). The dots are experimental data, and the solid lines are the fitting curve. \textbf{(b)} Fitted CTE of NbTiN membranes suspended over the small- (blue) and big-gap (orange). The membranes suspended over cavities of different gap sizes have different CTE, due to the different thicknesses of the amorphous silicon (a-Si) sacrificial layer.}
\label{fig_supp_Thermalexpansion}
\end{figure*}

\subsection*{Supplementary Information F4: Frequency noise in resonance frequency measurement}

For a mechanical resonator driven at an angular frequency $\omega=2\pi f$, the steady-state response $H(\omega)$ is proportional to
\begin{equation}
    H(\omega) = \frac{1}{\omega_0^2 - \omega^2 + i \frac{\omega \omega_0}{Q}},
\end{equation}
where $\omega_0 = 2\pi f_0$. Let $y=\omega/\omega_0=f/f_0$. The phase lag of the response is
\begin{equation}
    \phi(f) = \arg (H(\omega))= -\arctan (\frac{\omega \omega_0 /Q}{\omega_0^2 - \omega^2}) = -\arctan (\frac{y/Q}{1-y^2}).
\end{equation}
Use $\arctan (u) +\arctan (1/u) = \pi/2$ for $u>0$,
\begin{equation}
    \phi(f) = \frac{\pi}{2} - \arctan[Q(\frac{1}{y}-y)]=\frac{\pi}{2} - \arctan[Q(\frac{f_0}{f}-\frac{f}{f_0})].
\end{equation}
When $\vert f-f_0\vert \ll f_0$, one has
\begin{equation}
    \frac{d\phi}{df} = -\frac{Q}{1+[Q(\frac{f_0}{f}-\frac{f}{f_0})]^2} \cdot (-\frac{f_0}{f^2}-\frac{1}{f_0}),
\end{equation}
and
\begin{equation}
    \frac{d\phi}{df}\vert_{f=f_0} = \frac{2Q}{f_0}.
\end{equation}
Thus
\begin{equation}
    \delta f \approx \frac{f_0}{2Q}\delta \phi.
\end{equation}
A lock-in measuring the signal $S_T$ with the white noise $N_T$ over a bandwidth $B_T$, yields a rms phase uncertainty \cite{sansa2016frequency}
\begin{equation}
    \sigma_{\phi} \approx \frac{N_T}{S_T}\sqrt{B_T} = \frac{N_T}{S_T}\sqrt{\frac{1}{2\pi \tau}},
\end{equation}
here $\tau$ is the integrating time. Converting the phase noise into the frequency noise gives
\begin{equation}
    \langle \delta f\rangle \approx \frac{f_0}{2Q}\sigma_{\phi} = \frac{f_0}{2Q}\frac{N_T}{S_T}\sqrt{\frac{1}{2\pi \tau}}.
\end{equation}

\newpage

\section*{Supplementary Information G: Procedure on calculating the pressure and force variations from experimental frequency shifts}\label{sec:Procedure}

\begin{figure*}[h]
\centering
\includegraphics[width=1\textwidth]{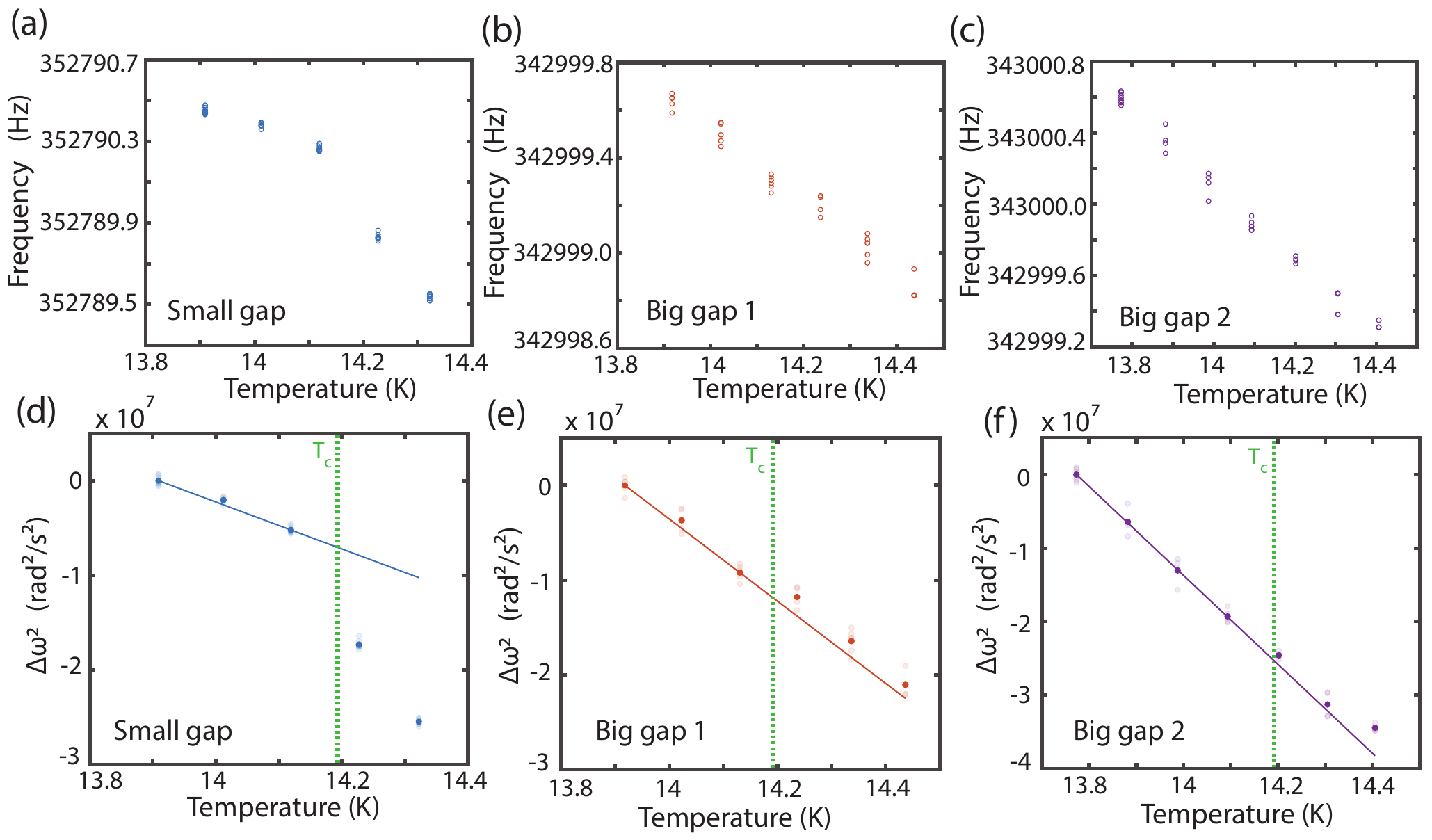}
\caption{Calibration of the temperature dependence of the elastic force near $T_C$. \textbf{(a)} Small-gap membrane, \textbf{(b)} and \textbf{(c)} big-gap membranes resonant frequencies ($f$) as a function of temperature. After converting the linear frequency to angular frequency $\omega$, we show the variation in $\omega^2$ as a function as temperature in \textbf{(d)} for small-gap membrane, and in \textbf{(e)} and \textbf{(f)} for big-gap membranes. Linear fits are performed on the variation of $\omega^2$ for the small-gap and big-gap membranes, with the aim of calibrating out the thermal expansions of the membranes.}
\label{fig_supp_PressureCalLinearFit}
\end{figure*}

In this paper, the nanomembrane suspended over a 190~nm gap size experiences a variation of resonant frequency that is different from the variation of the big-gap membrane, in order to find the corresponding abrupt pressure/force change on the small-gap membrane near $T_C$, we first perform linear fits on temperature-dependent frequency data points ($\omega=2\pi f$) below $T_C$ for both small-gap membrane (Figure~\ref{fig_supp_PressureCalLinearFit}\textbf{d}) and big-gap membranes (Figure~\ref{fig_supp_PressureCalLinearFit}\textbf{(e)} and Figure~\ref{fig_supp_PressureCalLinearFit}\textbf{(f)}), in order to accurately cancel out the thermal expansion of the membranes. Linear fits are used, because for such small temperature span the thermal expansion coefficients can be approximated as constant values. Then we subtract the different between the frequency square values $\omega^2$ at a temperature above $T_C$ from the linear fit, and obtain the variation of angular frequency square $\Delta\omega^2$ and linear frequency $\Delta f$ for membranes on both gap sizes, which is shown in Figure~\ref{fig_supp_PressureCal}\textbf{(a)} and Figure~\ref{fig_supp_PressureCal}\textbf{(b)}, respectively. Later we compare the $\Delta\omega^2$ difference between the two membranes, and use Equation~\ref{eq:forceGradient} to check the corresponding force gradient and pressure gradient variations, as shown in Figure~\ref{fig_supp_PressureCal}\textbf{(c)} and \ref{fig_supp_PressureCal}\textbf{(d)}, respectively, without knowing the exactly behaviors of the external pressure/force. The error bars in Figure~\ref{fig_supp_PressureCal}\textbf{(a)} and Figure~\ref{fig_supp_PressureCal}\textbf{(b)} are obtained with the different frequency sweep data points for every temperature setting, while the error bars in Figure~\ref{fig_supp_PressureCal}\textbf{(c)} and \ref{fig_supp_PressureCal}\textbf{(d)} are calculated by adding up the error bars of $\Delta \omega^2$ for small- and big-gap membranes shown in Figure~\ref{fig_supp_PressureCal}\textbf{(a)}. Here we combined the datasets from the big-gap membranes into one dataset. The error bars of the $\Delta \omega^2$ data on membranes over both gap sizes are obtained from the frequency sweep datasets at each temperature data point, and the error bars of the force gradient and pressure gradient variations are calculated by adding up the error bar size of the small-gap membrane at a certain temperature and the two error bar sizes of the big-gap membrane around the corresponding temperature.

\begin{figure*}[h]
\centering
\includegraphics[width=0.7\textwidth]{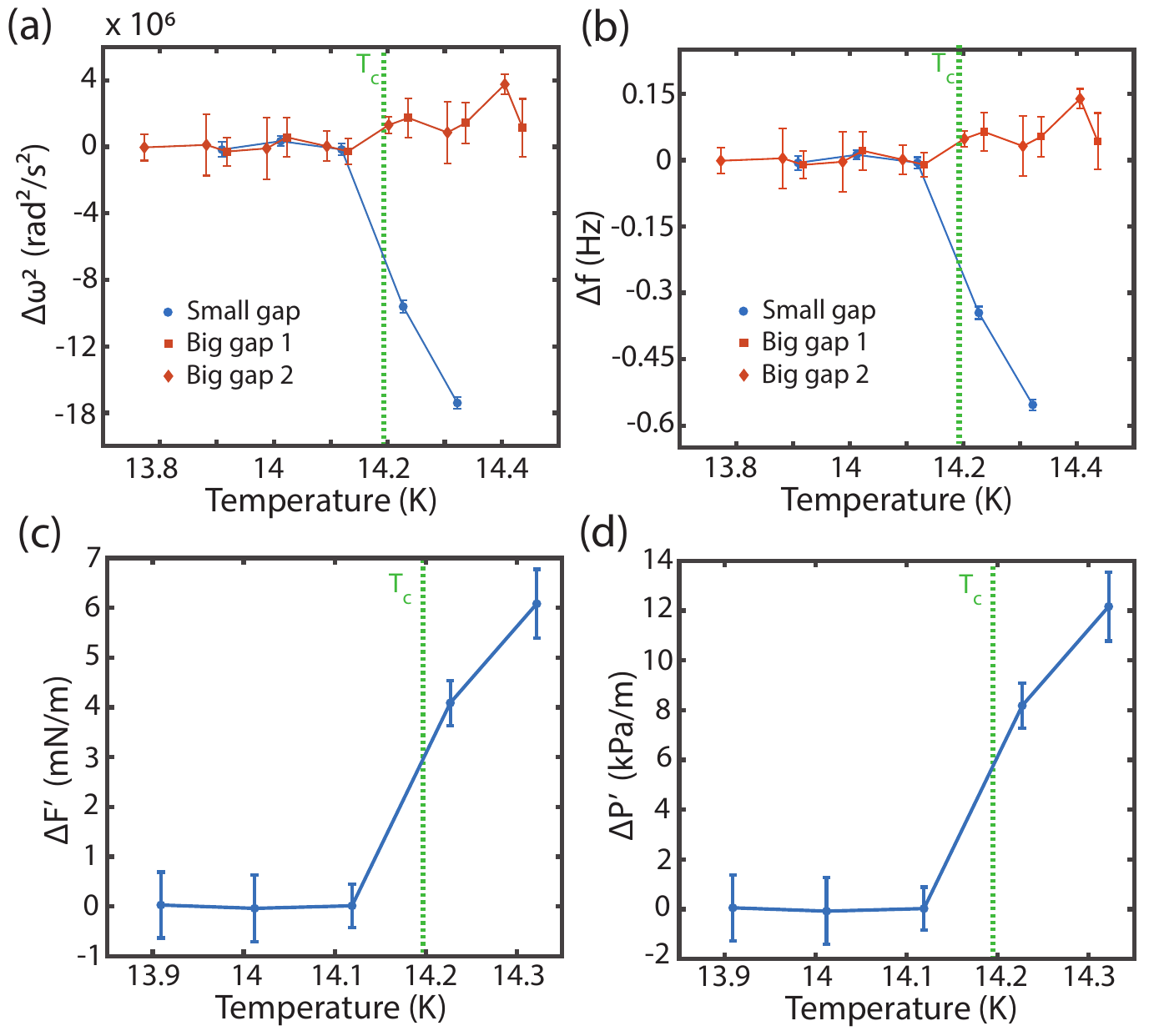}
\caption{Procedure on calculating the pressure and force variations from experimental frequency shifts.After performing linear fits on the variation of $\omega^2$ for the small-gap and big-gap membranes, the derivations of $\omega^2$ from the linear fits are shown in Panel~\textbf{(a)} for small- and big-gap membranes, where the two big-gap membrane datasets are combine. The derivations of $f$ from the linear fits are shown in Panel~\textbf{(b)}. After subtracting the difference in $\omega^2$ between small- and big-gap membranes, we used Equation~\ref{eq:forceGradient} to obtain the corresponding force gradient and pressure gradient variations, as shown in Panels~\textbf{(c)} and \textbf{(d)}. Here the exact form of the external force/pressure is not required to be known beforehand for the calculation.}
\label{fig_supp_PressureCal}
\end{figure*}

\newpage
~
\newpage


\section*{Supplementary Information H: Computing the Casimir pressure between superconductors and explaining its abrupt Change across \texorpdfstring{$T_C$}{TC}} \label{sec:bimonte}

The most striking finding of our experiment is the abrupt change in the membrane resonance frequency observed across the superconducting transition $T_C$, as displayed in Fig. \ref{fig_supp_BimontePressure_n0change}. This shift corresponds to a sudden increase of approximately 12~kPa/m in the pressure gradient acting on the membrane, as its temperature is raised through $T_C$. The central question is to elucidate the physical cause of this unexpected result. Strong evidence that the thermal Casimir effect is responsible comes from the observation that the measured increase in the pressure gradient is of the same order of magnitude as the classical Casimir pressure gradient between two conducting plates at distance $d$:
\begin{equation}
    P'_{\rm cl} = \frac{ 3 k_B T}{8\pi d^4}\zeta (3)~,\label{class}
\end{equation}
where $\zeta(3)=1.202$ is Riemann's zeta function. Indeed, at $T=T_C$ and for a separation of $d=$ 190 nm, the classical Casimir pressure gradient $P'_{\rm cl}$ is calculated to be 21.6 kPa/m, approximately double the measured increase.

In this Section, we present the theory of the Casimir effect for superconductors, based on the general Lifshitz theory and the BCS theory, which describes their optical properties. It is important to note that precision room-temperature Casimir experiments with metallic test bodies over the last 20 years have shown that computing the classical contribution to the Casimir force is a complex matter. It has been realized that the straightforward approach based on the over 100-year-old Drude model for a metal's conductivity, leading to Equation (\ref{class}), is inconsistent with experimental data. Excellent agreement, however, is achieved when the classical term is computed by modeling a metal as a dissipationless electron plasma. The situation with superconductors remains largely unknown, as no prior experiments have successfully provided information on the influence of the superconducting transition on the Casimir force. We will use our experimental data to guide the development of a theory capable of reproducing our observations. Our analysis indicates that the most promising theoretical framework employs the plasma approach for $T>T_C$ and the BCS theory for $T<T_C$. 
This approach provides a qualitative explanation for the observed discontinuity in the pressure gradient and, with an appropriate choice of the plasma frequency $\Omega$, can reproduce the observed magnitude of the discontinuity. 


\subsection*{Supplementary Information H1: Lifshitz formula for superconductors}

\begin{figure}[h]
\centering
\includegraphics[width=1\textwidth]{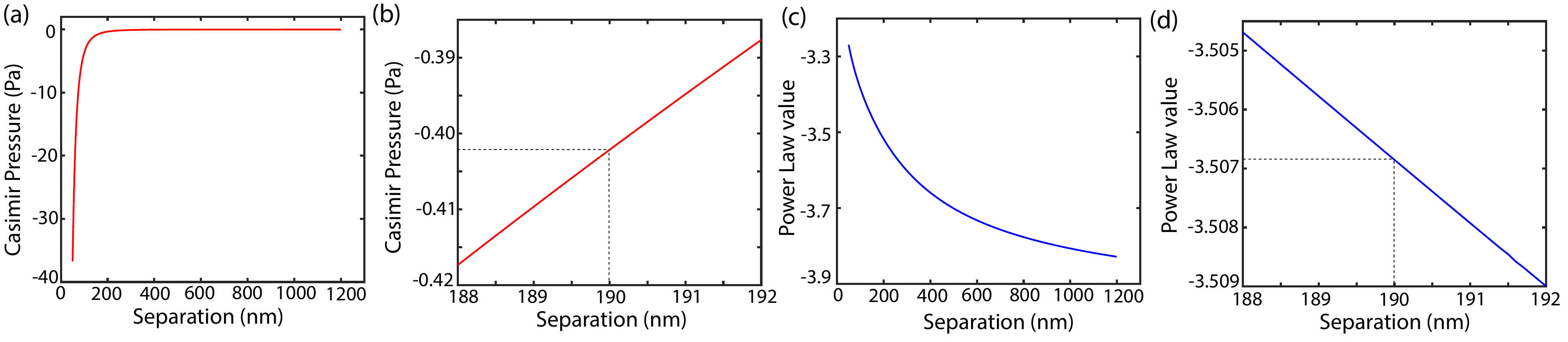}
\caption{Estimation of the Casimir pressure between superconductors, at the temperature of $T=0.99\cdot T_C$. \textbf{(a)} Casimir pressure between superconducting NbTiN surfaces. \textbf{(b)} Power law value $x$ in the Casimir pressure $P(d)\sim 1/d^{n}$ between superconducting NbTiN surfaces. \textbf{(c)} Casimir pressure between superconducting NbTiN surfaces, with a separation around 190~nm. \textbf{(d)} Power law value $x$ in the Casimir pressure $P(d)~d^{n}$ between superconducting NbTiN around $T_C$, around 190~nm separation. At 190~nm, the absolute Casimir pressure between NbTiN near $T_C$, is $P=-0.4021$~Pa, and the power law value is $n=3.5069$, leading to the explicit expression of the Casimir pressure to be $P_{\rm Cas}\bigl|_{d=190\rm nm}= (-1.081\times 10^{-24})/d^{3.507}$~[Pa].}
\label{fig_supp_BimontePressure}
\end{figure}

We now calculate the Casimir pressure between superconducting surfaces. It is observed that the considerable thickness ($t=150$~nm) of the NbTiN membranes permits their modeling as infinitely thick slabs. The Casimir pressure, $P(d, T)$, between two identical parallel slabs is given by the Lifshitz formula as (negative pressures indicate attraction):
\begin{equation}\label{eq:bimontePressure}
    P(d,T) = -\frac{k_B T}{\pi}\sum^{\infty}_{l=0}{}^\prime\int^{\infty}_0 dk_{\bot} k_{\bot} q_l \sum_{\alpha} \Bigl[  \frac{e^{2dq_l}}{r_{\alpha}^{2}(i\xi_l,k_{\bot})} -1 \Bigl]^{-1},
\end{equation}
where $k_B$ represents the Boltzmann constant, $k_{\bot}$ the in-plane momentum, the prime on the summation signifies that the $l = 0$ term is assigned a weight of one-half,  $\xi_l=2\pi l k_B T/\hbar$ are the imaginary Matsubara frequencies, $q_l=\sqrt{\xi^2_l/c^2+k_{\bot}^2}$, and the summation over $\alpha = \rm TE, TM$ encompasses the independent polarization states of the electromagnetic field, namely transverse magnetic and transverse electric. The symbols $r_{\alpha}(i \xi_l, k_{\bot})$ denote the Fresnel reflection coefficients of the slab:
\begin{equation}\label{eq:TEterm}
    r_{\rm TE}(i \xi_l, k_{\bot})=\frac{q_l - s_l}{q_l+s_l},
\end{equation}
\begin{equation} \label{eq:TMterm}
    r_{\rm TM}(i \xi_l, k_{\bot})=\frac{\epsilon_{l} q_l - s_l}{\epsilon_{l} q_l+s_l},
\end{equation} 
where $s_l = \sqrt{\epsilon_l \xi^2_l/c^2 +k_{\bot}^2}$ and $\epsilon_l \equiv \epsilon (i\xi_l)$. 
To compute the frequency shift, we require the derivative $P'(d)$, which is given by Equation~\ref{eq:bimontePressure} as:
\begin{equation}\label{eq:bimontePressureGradient}\
    P'(d,T) = 2\frac{k_B T}{\pi}\sum^{\infty}_{l=0}{}^\prime\int^{\infty}_0 dk_{\bot} k_{\bot} q_l^2 \sum_{\alpha} \frac{e^{2dq_l} r_{\alpha}^{2}(i\xi_l,k_{\bot})}{[e^{2 d q_l}- r_{\alpha}^{2}(i\xi_l,k_{\bot})]^2} ,
\end{equation}
In our computational model, core electron contributions to the NbTiN permittivity are disregarded, and only intraband transitions are taken into account. For temperatures exceeding $T_C$, these transitions are accurately modeled by a Drude dielectric function:
\begin{equation}\label{eq:Drude}
    \epsilon (i\xi)=1+\frac{\Omega^2}{\xi(\xi+\gamma)}
\end{equation}
where $\Omega$ represents the plasma frequency for intraband transitions, and $\gamma$ is the relaxation frequency. We have assumed the plasma frequency, $\Omega$, to be temperature-independent, using its room-temperature value. The relaxation frequency, $\gamma$, on the other hand, varies with temperature, typically decreasing as temperature decreases. At cryogenic temperatures, $\gamma$ approaches a constant residual value that depends on the specific sample. Following standard practice, we relate the residual relaxation frequency to the room-temperature value, $\gamma_0$, using the formula $\gamma = \gamma_0/\rm RRR$, where RRR is the residual resistance ratio. The following Drude parameters were used: $\Omega = 5.33$~eV/$\hbar$ and $\gamma_0 = 0.465$~meV/$\hbar$. Moreover, we took $\text{RRR}=1$.
The permittivity of NbTiN in the superconducting state was calculated using the Mattis-Bardeen formula for the conductivity $\sigma_{\rm BCS} (\omega)$ \cite{mattis1958theory}, known to accurately represent the optical response of BCS superconductors \cite{tinkham2004introduction}. While the general Mattis-Bardeen formula depends on both frequency ($\omega$) and wavevector ($q$), the $q$-dependence is negligible in the dirty limit ($\ell/\xi_0 \ll 1$), where $\ell=v_F/\gamma$ the dirty limit is well satisfied ($\ell/\xi_0 = 1.4\times 10^{-2}$). This is confirmed by optical measurements of NbTiN films in the THz region, that are in excellent agreement with the local dirty-limit of the Mattis-Bardeen formula \cite{hong2013terahertz}. Analytic continuation to imaginary frequencies \cite{bimonte2010optical}, gives:
\begin{equation}
    \sigma_{\rm BCS}(i \xi) = \frac{\Omega^2}{4\pi} \Bigl[ \frac{1}{(\xi+\gamma)}+\frac{g(\xi ; T)}{\xi}\Bigl],
\end{equation}
The first term is the Drude contribution, and the second is the BCS correction. The explicit expression of $g(\xi)$ and its properties are discussed in \cite{bimonte2019casimir}. The permittivity of superconducting NbTiN is then given by:
\begin{equation}\label{eq:DrudeBCS}
    \epsilon_{\rm BCS}(i\xi) = 1+\frac{4\pi}{\xi}\sigma_{\rm BCS}(i\xi) = 1 + \frac{\Omega^2}{\xi}\Bigl[\frac{1}{\xi+\gamma} + \frac{g(\xi;T)}{\xi} \Bigl],
\end{equation}
where $\Omega$ is the plasma frequency for intraband transition, and $\gamma$ is the relaxation frequency. For NbTiN we adapt the values used from \cite{bimonte2019casimir,eerkens2017investigations}, i.e. $\Omega=5.33~\text{eV}/\hbar$ and $\gamma=0.465~\text{eV}/\hbar$. The function $g(\xi ;T)$ that describes the modification as a result of BCS theory, can be expressed as
\begin{equation}
    g(\xi;T)=\Theta(T_C - T)\int^{\infty}_{-\infty}\frac{d\epsilon}{E}\tanh \Bigl(\frac{E}{2k_B T} \Bigl) \text{Re}[G_{+}(i\xi , \epsilon)]
\end{equation}
where $T_C$ is the superconducting phase transition temperature, $E=\sqrt{\epsilon^2 +\Delta^2}$, $\Theta(x)$ is the Heaviside step function, i.e. $\Theta(x)=1$ for $x>0$, and $\Theta(x)=0$ for $x\leq 0$, and
\begin{equation}
    G_{+}(z,\epsilon)=\frac{\epsilon^2Q_{+}(z,E)+\bigl[Q_{+}(z,E)+i\hbar \gamma\bigl]\cdot A_{+}(z,E)}{Q_{+}(z,E)\cdot\{ \epsilon^2 - \bigl[Q_{+}(z,E)+i\hbar \gamma \bigl]^2\} },
\end{equation}
where
\begin{equation}
    Q_{+}(z,E) = \sqrt{(E+\hbar z)^2-\Delta^2},
\end{equation}
and
\begin{equation}
    A_{+}(z,E)=E(E+\hbar z)+\Delta^2,
\end{equation}
here $\Delta$ is the temperature-dependent superconducting gap, which can be calculated from BCS theory:
\begin{equation}
    \Delta = c_1 k_B T_C \sqrt{1-\frac{T}{T_C}}\Bigl(c_2 + c_3 \frac{T}{T_C}\Bigl),
\end{equation}
and $c_1=1.764$, $c_2=0.9963$, and $c_3=0.7735$.

From data shown in Figure~\ref{fig_supp_BimontePressure} the effective Casimir pressure between NbTiN films at a separation of 190~nm, near $T_C$, can be expressed as
\begin{equation}\label{eq:realCasimirPressure}
    P_{\rm Cas}(d)\Bigl|_{d=190\rm nm} \approx \frac{-1.081\times 10^{-24}}{d^{3.507}}~[\rm Pa],
\end{equation}
with the separation $d$ measured in meters.
With this expression of the Casimir pressure, we can use FEM simulation to estimate how much the external Casimir pressure/force is acting on the membrane. In order to do that, we scale the Casimir pressure acting on the membrane by a factor $\beta$ close to unity, i.e. $P_{\rm sim}=(1+\beta)\cdot P_{\rm Cas}(d)$ with $\vert\beta\vert <0.001$, and found the linear conversion factors from the the frequency shift $\Delta \omega^2$ to variations in the Casimir force $\Delta F_{\rm Cas}$, the Casimir pressure $\Delta P_{\rm Cas}$, and membrane center static deflection $\Delta z_{\rm Cas}$. The conversion factors are $7.83\times 10^{-16}~\rm N/Hz^2$, $1.55\times 10^{-9}~\rm Pa/Hz^2$, and $6.28\times 10^{-19}~\rm m/Hz^2$, for force, pressure, and center deflection, respectively. From the simulation results, one can calculate that for the membrane suspends over 190~nm gap, the total Casimir force is around $-2.028\times 10^{-7}$~N, and the total center deflection is around $-153$~pm, indicating that the level of parallelism achieved in our device is state-of-the-art.

\begin{figure}[h]
\centering
\includegraphics[width=1\textwidth]{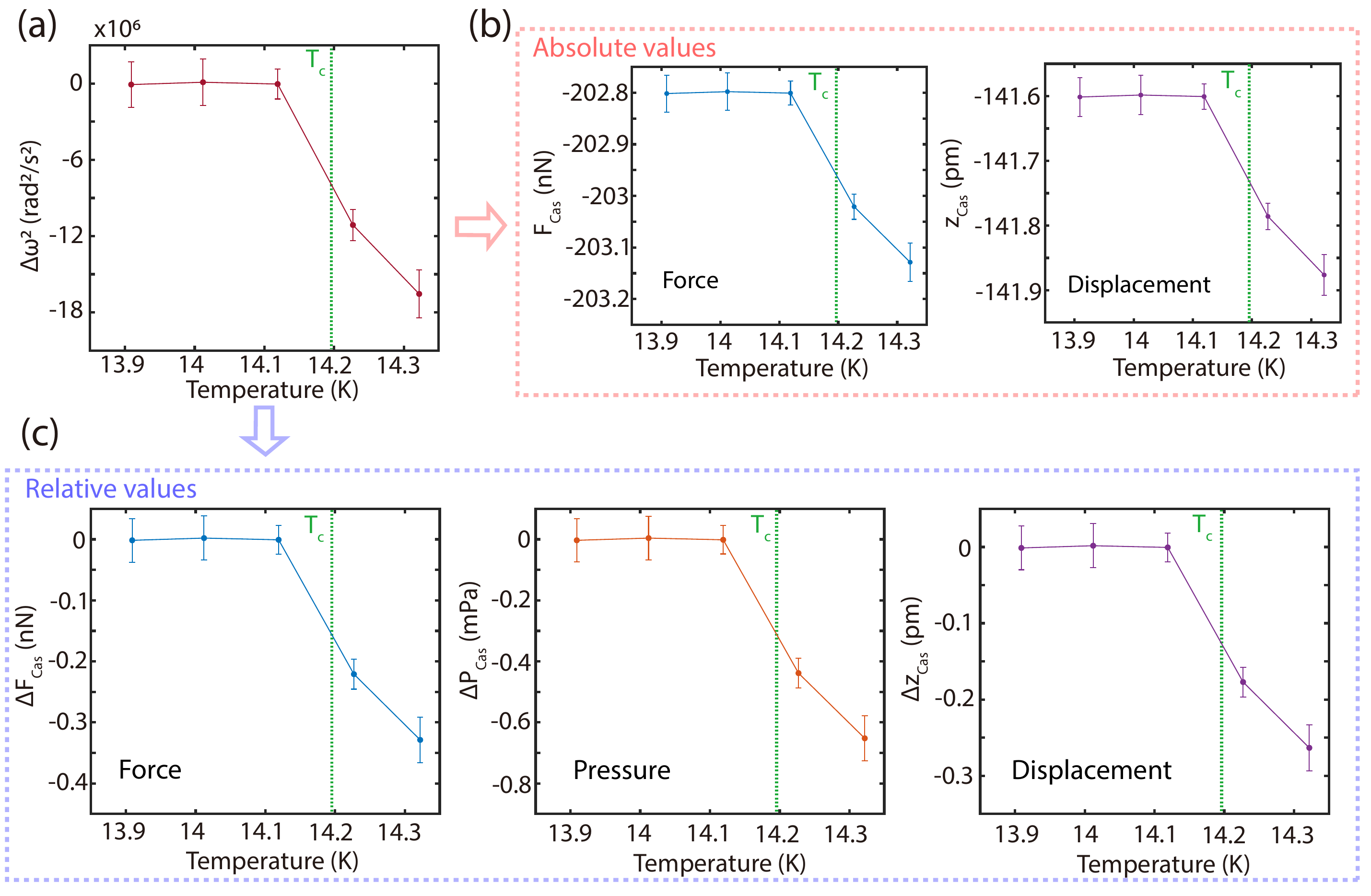}
\caption{Conversion of the shift of frequency square $\Delta \omega^2$ to parameters associated with the external Casimir force, using FEM simulation. \textbf{(a)} Experimentally measured shift of frequency square $\Delta \omega^2$ around $T_C$ of small-gap membrane, after calibrating out the elastic force dependence on temperature. \textbf{(b)} Absolute values of the Casimir force variation and center deflection, corresponding to the value of $\Delta \omega^2$. \textbf{(c)} Relative values of the Casimir force, Casimir pressure, and static deflection of the center of the membrane, corresponding to the variation of the measured frequency square value $\Delta \omega^2$.}
\label{fig_supp_BimontePressure_PressureChange}
\end{figure}

\newpage
~
\newpage

\subsection*{Supplementary Information H2: The \texorpdfstring{$n=0$}{n=0} classical contribution: Drude vs plasma models}

In this Section we study the $n=0$ contribution to the Lifshitz formula ~\ref{eq:bimontePressure}. This term represents a classical contribution to the Casimir pressure, which becomes the dominant contribution for separations $d \gg \lambda_T$, where $\lambda_T=\hbar c/ k_B T$ is the thermal length. 
The precise form of this contribution has been a matter of significant controversy within the Casimir community for more than two decades,  with experts divided between the Drude and plasma models. The Drude model accounts for electron scattering and dissipation, offering a more intuitive physical interpretation, while the plasma model, despite neglecting dissipation, aligns more closely with experimental observations. The key divergence lies in the treatment of the zero-frequency Matsubara term in the Lifshitz formula: the Drude approach predicts a vanishing transverse electric (TE) contribution, whereas the plasma model yields a finite term that significantly influences the force magnitude. This debate not only reflects discrepancies between theory and experiment but also challenges our fundamental understanding of the interplay between dissipation and quantum fluctuations.

The zero-frequency ($n=0$)  term in the discrete Matsubara frequency sum represents  a static contribution to  Lifshitz formulae Equations~\ref{eq:bimontePressure} and \ref{eq:bimontePressureGradient}. It is well-established that this single term plays a dominant role in determining the temperature dependence of the Casimir pressure \cite{brevik2006thermal,svetovoy2006casimir,bordag2009advances}.
The $n=0$ term can be separated into TM and TE polarization contributions:
\begin{equation}\label{eq:Pzero}
  P_{\rm cl} \equiv   P(d,T)\Bigl|_{n=0} = P_{\rm TM}^{(0)} + P_{\rm TE}^{(0)} ~,
\end{equation}
where
\begin{equation}
    P_{\alpha}^{(0)}= -\frac{k_B T}{2\pi} \int^{\infty}_{0} dk_{\bot} k_{\bot} ^2 \Bigl[ \frac{e^{2dk_{\bot}}}{r_{\alpha}^2 (0, k_{\bot} )-1} \Bigl]^{-1}~,
\end{equation}
where $\alpha=\rm TE,TM$, and we set $r_{\alpha}^2 (0, k_{\bot} ) =\lim_{\xi \rightarrow 0} r_{\alpha}^2 (i\xi, k_{\bot} )$. 
The TM component $P_{\rm TM}^{(0)}$ presents no difficulties; in the case of a conductor,  $\lim_{\xi \rightarrow 0} \epsilon ({\rm i} \xi)= \infty$, and  and it is easily demonstrated (see Eq. (\ref{eq:TMterm})) that $ r_{\rm TM}(0, k_{\perp}) = 1$. Evaluation of Equation~\ref{eq:Pzero} results into
\begin{equation}
    P_{\rm TM}^{(0)} = -\frac{k_B T}{2\pi d^3} \frac{\zeta (3)}{4}~,
\end{equation}
where $\zeta(3)=1.202$ is Riemann's zeta function.
It is observed that the TM contribution exhibits universality, as its magnitude remains invariant across all conductors, irrespective of their superconducting or normal state. The TM contribution, by its nature, cannot induce a discontinuous jump in the Casimir pressure (or its gradient) at the superconducting transition, a feature that appears to be presented in our data.

The TE contribution $P_{\rm TE}^{(0)}$, on the other hand, presents a considerably more intricate analysis. For normal (non-magnetic) conductors, the permittivity exhibits $1/\xi$ singularity as $\xi$ approaches zero, according to Equation~\ref{eq:Drude}. Consequently, one have $\lim_{\xi\rightarrow 0}s=\lim_{\xi\rightarrow 0}\sqrt{\epsilon(\text{i}\xi)\cdot \xi^2/c^2+k_{\bot}^2}\approx k_{\bot}$, which implies that the TE reflection coefficient (Equation~\ref{eq:TEterm}) approaches zero as frequency vanishes:
\begin{equation}
    r_{\text{TE}}(0,k_{\bot}) = 0, \quad\quad \text{(Drude model approach)}~.\label{Drudepres}
\end{equation}
The vanishing of $r_{\rm TE}$ means that the TE term does not contribute to the $n=0$ term:
\begin{equation}
    P_{\rm TE}^{(0)} = 0, \quad\quad \text{(Drude model approach)}~.
\end{equation}
Several room-temperature Casimir experiments \cite{decca2007novel,chang2012gradient,banishev2013demonstration,bimonte2016isoelectronic,bimonte2021measurement} with metallic surfaces have shown discrepancies with theoretical predictions based on the Lifshitz formula. Surprisingly, it has been demonstrated that the aforementioned discrepancies are resolvable through the adoption of the following expression for the static TE reflection coefficient:
\begin{equation}\label{eq:plasmaapproach}
    r_{\rm TE} (0,k_{\bot}) = \frac{k_{\bot}-\sqrt{\Omega^2/c^2+k_{\bot}^2} }{k_{\bot}+\sqrt{\Omega^2/c^2+k_{\bot}^2}}, \quad\quad \text{(plasma model approach)}~.
\end{equation}
Since the above formula for $r_{\rm TE}$ matches the $\xi\rightarrow 0$ limit of Equation~\ref{eq:TEterm}, when the plasma model is used for permittivity:
\begin{equation}\label{eq:PlasmaModel}
    \epsilon (i\xi) = 1+\frac{\Omega^2}{\xi^2},
\end{equation}
the approach shown in Equation~\ref{eq:plasmaapproach}, is known as the plasma-BCS approach. For a recent review of the problem of thermal Casimir effect and related experiments, see \cite{mostepanenko2021casimir}.

\begin{figure}[h]
\centering
\includegraphics[width=1\textwidth]{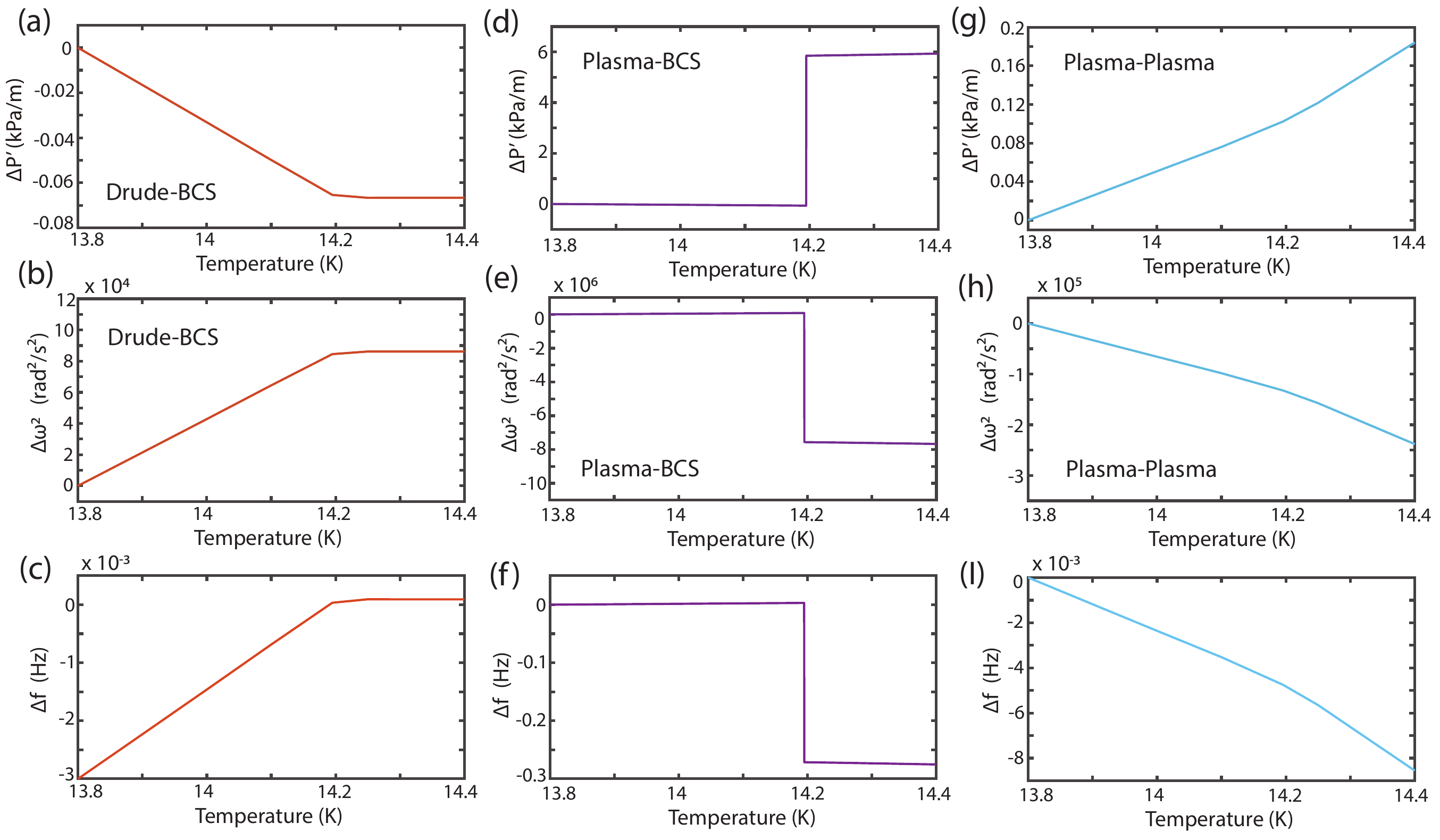}
\caption{ Theoretical predictions for the Casimir pressure gradient variation $\Delta P'$ and the associated membrane frequency square shift $\Delta \omega^2$ across the superconducting transition. Panels~\textbf{(a)}, \textbf{(b)} and \textbf{(c)} show curves computed using BCS model in the superconducting state ($T<T_C$) and the Drude model in the normal state ($T>T_C$). Panels~\textbf{(d)}, \textbf{(e)} and \textbf{(f)} show curves computed using BCS model in the superconducting state ($T<T_C$) and the plasma model in the normal state.($T>T_C$). Panels~\textbf{(g)}, \textbf{(h)} and \textbf{(I)} show curves computed using plasma model both in the superconducting state ($T<T_C$) and in the normal state.($T>T_C$).}
\label{fig_supp_BimontePressure_DrudePlasma}
\end{figure}

The analysis thus far has concerned the normal state of our membrane  ($T>T_c$). We now proceed to consider its superconducting state  ($T<T_c$). In the superconducting state, the permittivity includes both a Drude component and a BCS component, as detailed in Equation~\ref{eq:DrudeBCS}. The BCS contribution is proportional to the function $g(\xi;T)$, which, for small $\xi$, has the following expansion:
\begin{equation}\label{eq:gfunctionsmall}
    g(\xi;T)=\bar \Omega^2 (T) + B(T)\cdot \xi\cdot \log(\frac{\Delta}{\hbar \xi})+ o(\xi),
\end{equation}
where $B(T)=(\sigma_0 \Delta(T)/2T)\cdot\text{sech}^2(\Delta(T)/2T)$, $\Delta(T)$ is the BCS gap, $\sigma_0$ is the dc conductivity in the normal state, 
\begin{equation}
    \bar \Omega^2 (T) = \tilde \omega^2_s (T)\cdot \Omega^2. 
\end{equation}
Here $\tilde \omega_s (T)$ represents the (normalized) effective superfluid plasma frequency \cite{bimonte2010optical}:
\begin{equation}
    \tilde  \omega_s (T) =\frac{\pi}{2\eta}\text{tanh}\Bigl(\frac{1}{4 \eta\cdot t_{\rm red}}\Bigl) -\frac{1}{\eta^2} \int^{\infty}_0 dx \frac{\text{tanh}\bigl(E(x)/2t_{\rm red} \bigl)}{E(x) (4x^2+1)},
\end{equation}
where $\eta=\hbar\gamma/2\Delta(T)$, $t_{\rm red}=k_BT/\hbar \gamma$, and $E(x) = \sqrt{x^2+1/(4\eta^2)}$. It is noted that $0<\tilde\omega_s (T)<1$, implying $\bar \Omega(T)<\Omega$. Equations~\ref{eq:DrudeBCS} and \ref{eq:gfunctionsmall} show that $\epsilon_{\rm BCS}$ exhibits a $1/\xi^2$ singularity as $\xi\rightarrow0$,  similar to the plasma model  Equation~\ref{eq:PlasmaModel}, which results in the following expression of the TE reflection coefficient of a BCS superconductor:
\begin{equation}\label{eq:BCSplasma}
        r_{\rm TE} (0,k_{\bot}) = \frac{k_{\bot}-\sqrt{\bar \Omega(T)^2/c^2+k_{\bot}^2} }{k_{\bot}+\sqrt{\bar\Omega(T)^2/c^2+k_{\bot}^2}}, \quad\quad \text{(BCS model for $T<T_c$)}~.
\end{equation}

\begin{table*}[ht]
\centering
\caption{\label{tab:CasimirCase}
Alternative approaches for calculating the $n=0$ contribution to the Casimir force. } 
{\renewcommand{\arraystretch}{2.5}%
\begin{tabular}{|c|c|c|c|}
  \hline\hline
 \textbf{Approach}   & \textbf{Drude-BCS} & \textbf{Plasma-BCS} & \textbf{Plasma-Plasma}   \\
   \hline \hline
 Temperature & \multicolumn{3}{|c|}{$T>T_C$}  \\ \hline
  Model  & Drude & Plasma & Plasma \\ \hline
  $r_{\rm TM}(0,k_{\bot})$  & $r_{\rm TM}=1$ & $r_{\rm TM}=1$ & $r_{\rm TM}=1$ \\ \hline
  $r_{\rm TE}(0,k_{\bot})$  & $r_{\rm TE}=0$  & $r_{\rm TE}=\frac{k_{\bot}-\sqrt{\Omega^2/c^2 + k^2_{\bot}}}{k_{\bot}+\sqrt{\Omega^2/c^2 + k^2_{\bot}}}$ & $r_{\rm TE}=\frac{k_{\bot}-\sqrt{\Omega^2/c^2 + k^2_{\bot}}}{k_{\bot}+\sqrt{\Omega^2/c^2 + k^2_{\bot}}}$ \\ [0.8em] \hline \hline
  Temperature & \multicolumn{3}{|c|}{$T<T_C$}  \\ \hline
  Model  & BCS & BCS & Plasma \\ \hline
  $r_{\rm TM}(0,k_{\bot})$  & $r_{\rm TM}=1$ & $r_{\rm TM}=1$ & $r_{\rm TM}=1$ \\ \hline
  $r_{\rm TE}(0,k_{\bot})$  & $ r_{\rm TE}=\frac{k_{\bot}-\sqrt{\bar\Omega^2/c^2 + k^2_{\bot}}}{k_{\bot}+\sqrt{\bar\Omega^2/c^2 + k^2_{\bot}}}$  & $ r_{\rm TE}=\frac{k_{\bot}-\sqrt{\bar\Omega^2/c^2 + k^2_{\bot}}}{k_{\bot}+\sqrt{\bar\Omega^2/c^2 + k^2_{\bot}}} $ & $r_{\rm TE}=\frac{k_{\bot}-\sqrt{\Omega^2/c^2 + k^2_{\bot}}}{k_{\bot}+\sqrt{\Omega^2/c^2 + k^2_{\bot}}}$ \\ [0.8em] \hline \hline
\end{tabular}}
\end{table*}

\newpage

\subsection*{Supplementary Information H3: Three different approaches for the Casimir effect between superconductors}

For our computations, we considered three different approaches to calculate the Casimir pressure. In all approaches, the positive Matsubara modes ($n>0$) were computed using identical expressions for the permittivity of NbTiN: the Drude model Equation~(\ref{eq:Drude}) for $T>T_c$, and the BCS formula Equation~(\ref{eq:DrudeBCS}) for $T<T_C$.  Additionally, for $n=0$, we set  $r_{\rm TM} (0,k_{\bot})=1$ in all cases. The three approaches differ in the expressions used for the static TE reflection coefficient $r_{\rm TE} (0,k_{\bot})$, as summarized in Table~(\ref{tab:CasimirCase}):\\

\noindent
{\bf Drude-BCS approach}: in this approach, $r_{\rm TE} (0,k_{\bot})=0$ for $T>T_C$  (Equation~(\ref{Drudepres})), and the BCS expression Equation~(\ref{eq:BCSplasma}) is used for $T<T_c$. As shown in panel (a) of Fig.~(\ref{fig_supp_BimontePressure_DrudePlasma}) this approach predicts a smooth decrease in the pressure gradient as the temperature increases through $T_C$, failing to explain the observed jump in 
$\Delta P'$. \\

\noindent
{\bf Plasma-BCS approach}: in this approach, for $T>T_c$ we use the plasma-model formula for $r_{\rm TE} (0,k_{\bot})$ (Equation~(\ref{eq:plasmaapproach})), and for $T<T_c$ the BCS expression Equation~(\ref{eq:BCSplasma}) is used. 
The plasma frequency $\Omega=5.33$ eV/$\hbar$ used corresponds to the optical response of NbTiN in the infrared region. As depicted in panel (c) of Fig.~(\ref{fig_supp_BimontePressure_DrudePlasma}) this approach does predict a jump in $\Delta P'$,  consistent with our experimental observations.
\\

\noindent
{\bf Plasma-Plasma approach}: in this approach, the plasma-model formula for $r_{\rm TE} (0,k_{\bot})$ (Equation~(\ref{eq:plasmaapproach})) is used both for $T>T_c$ and $T<T_c$. As shown in panel (e) of Fig.~(\ref{fig_supp_BimontePressure_DrudePlasma}) this approach predicts a smooth increase in the pressure gradient as the temperature increases through $T_C$, failing to explain the observed jump in 
$\Delta P'$. \\

\noindent
Figure~\ref{fig_supp_BimontePressure_n0change}\textbf{(a)}, \textbf{(b)} and \textbf{(c)} present the experimental data alongside theoretical predictions from the Drude-BCS and Plasma-BCS approaches for the variations in pressure gradient ($\Delta P'$), membrane angular frequency square shift $\Delta \omega^2$ and membrane linear frequency shift $\Delta f$, respectively. The Drude-BCS approach is clearly inconsistent with the data, whereas the Plasma-BCS approach predicts jumps in $\Delta P'$ and $\Delta \omega^2$ that are approximately half the experimental values, suggesting its potential to explain the observed phenomena. It is noted that the Plasma-BCS approach can fully reproduce the experimental data with a plasma frequency of $\Omega=12.7$ eV/$\hbar$. As our current understanding of the plasma approach to the Casimir effect provides no strong justification for limiting $\Omega$ to the material's infrared optical response, the potential for employing a different value cannot be dismissed.

Our observation of a sudden resonance frequency shift during the membrane’s superconducting transition, if attributable to the Casimir force, would be the first experimental evidence of the superconducting transition’s influence on the Casimir effect. However, interpreting the data in this manner reveals two unexpected features. First, the sign of the frequency shift indicates a decrease in the Casimir pressure gradient during the transition, which contradicts the intuition that a superconducting surface, acting as a near-ideal mirror, should increase the Casimir force. Second, and perhaps more significantly, the abrupt frequency shift at the critical temperature seems to be inconsistent with the second-order nature of the superconducting transition. Since the Casimir force is the negative derivative of the Helmholtz free energy ${\cal F}$ of the system with respect to separation \cite{bimonte2017nonequilibrium}, $F_{\rm Cas}=-\partial {\cal F}/\partial d$, $F_{\rm cas}$ is expected to be continuous across the transition. Furthermore, the smooth confluence of the BCS permittivity with the Drude permittivity at the superconducting transition temperature provides another reason to expect a continuous change in the Casimir force. The Drude-BCS theory of the Casimir effect in superconductors, as outlined in \cite{bimonte2019casimir}, cannot replicate the observed jump in the Casimir pressure, predicting instead a smooth increase in Casimir pressure. Conversely, the Plasma-BCS approach  semi-quantitatively reproduces the observed behavior. The established agreement of the plasma approach with room-temperature Casimir experiments \cite{decca2007novel,chang2012gradient,banishev2013demonstration,bimonte2016isoelectronic,bimonte2021measurement} suggests its expected applicability to the normal state of the membrane.

\begin{figure}[h]
\centering
\includegraphics[width=1\textwidth]{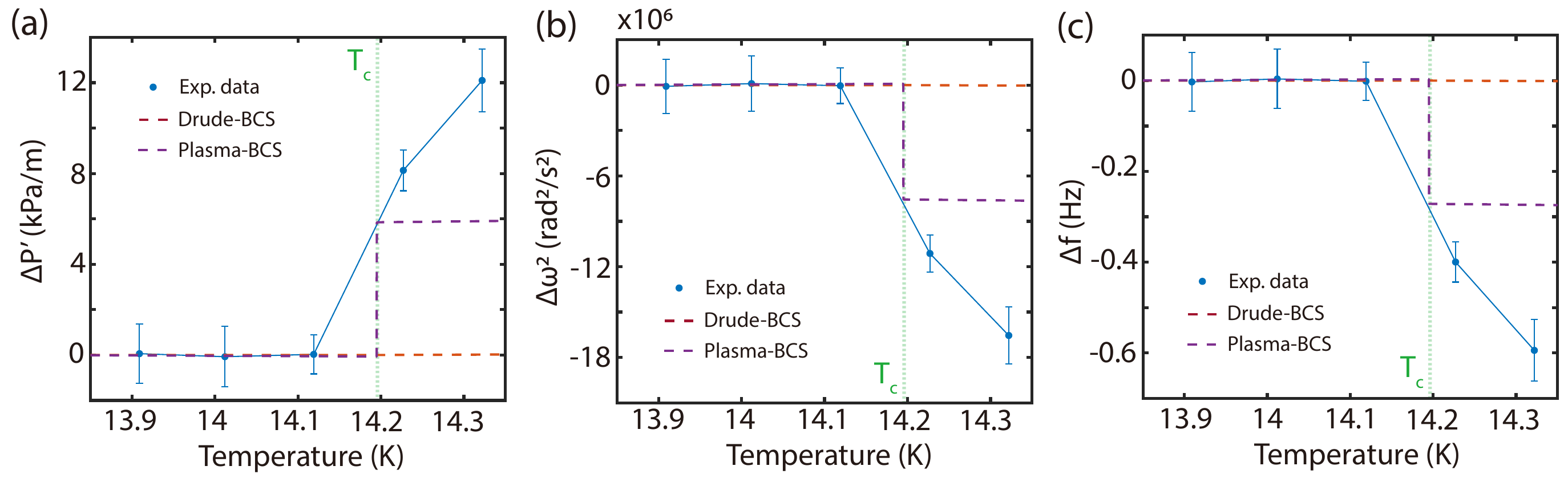}
\caption{Comparison of the experimental data and the theoretical analysis, on the abrupt changes in the panel \textbf{(a)} pressure gradient, in panel \textbf{(b)} membrane angular frequency square shift, and in panel \textbf{(c)} membrane linear frequency shift. The plasma-BCS approach seems capable of explaining the abrupt changes occurring at $T_C$.}
\label{fig_supp_BimontePressure_n0change}
\end{figure}


\newpage

\section*{Supplementary Information I: Quality factor of nanomembranes as a function of temperature}

In this section, we showed the mechanical quality factors of small- and big-gap membranes as a function of temperature, in Figure~\ref{fig_supp_QfactorBigT} (over a large temperature range) and Figure~\ref{fig_supp_QfactorSmallT} (across $T_C$).

\begin{figure*}[h]
\centering
\includegraphics[width=0.7\textwidth]{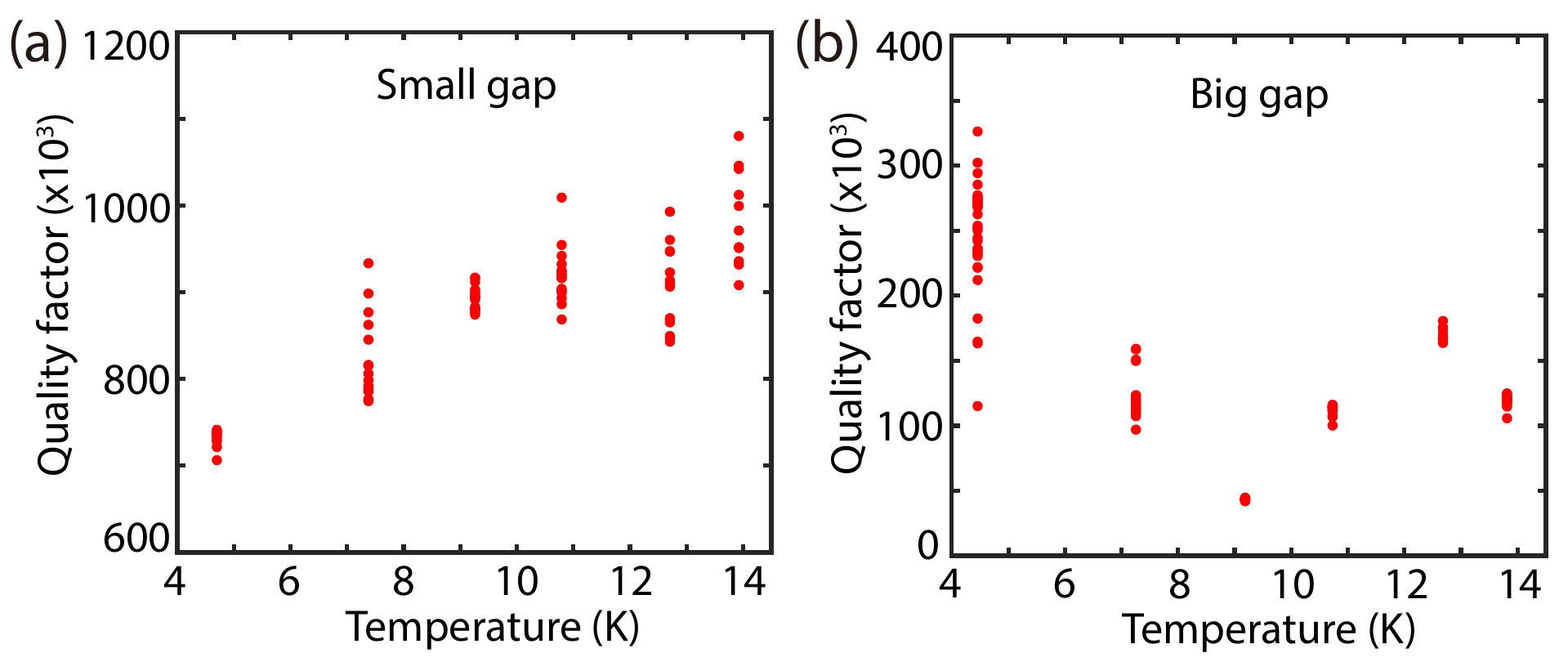}
\caption{Quality factor of $709\times 709~\mu\text{m}^2$ nanomembranes over small- (Panel~\textbf{(a)}) and big-gap (Panel~\textbf{(b)}) Casimir cavities, as a function of temperature. The temperature range is from 4.45~K to 14~K. For both gap sizes, every temperature dataset has at least 10 data points.}
\label{fig_supp_QfactorBigT}
\end{figure*}

\begin{figure*}[h]
\centering
\includegraphics[width=1\textwidth]{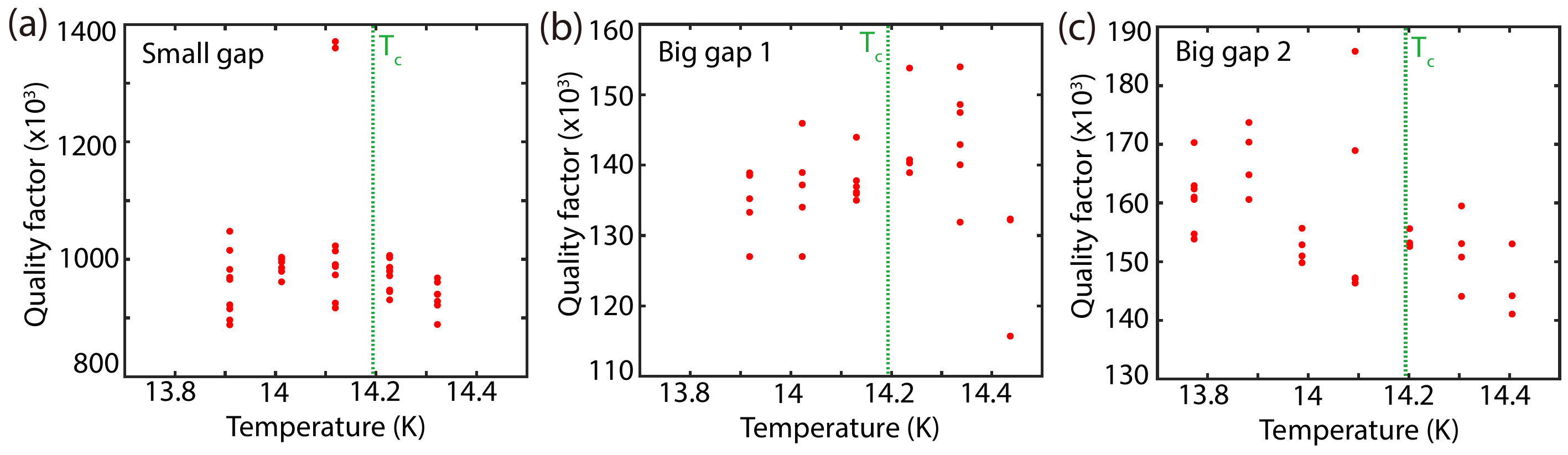}
\caption{Quality factor of $709\times 709~\mu\text{m}^2$ nanomembranes over small- (Panel~\textbf{(a)}) and big-gap (Panels~\textbf{(b)} and \textbf{(c)}) Casimir cavities, as a function of temperature. The temperature range is near the superconducting phase transition temperature $T_c=14.2$~K.}
\label{fig_supp_QfactorSmallT}
\end{figure*}

\newpage

\section*{Supplementary Information J: Logscale comparison on various Casimir force measurements}

\begin{figure*}[h]
\centering
\includegraphics[width=0.65\textwidth]{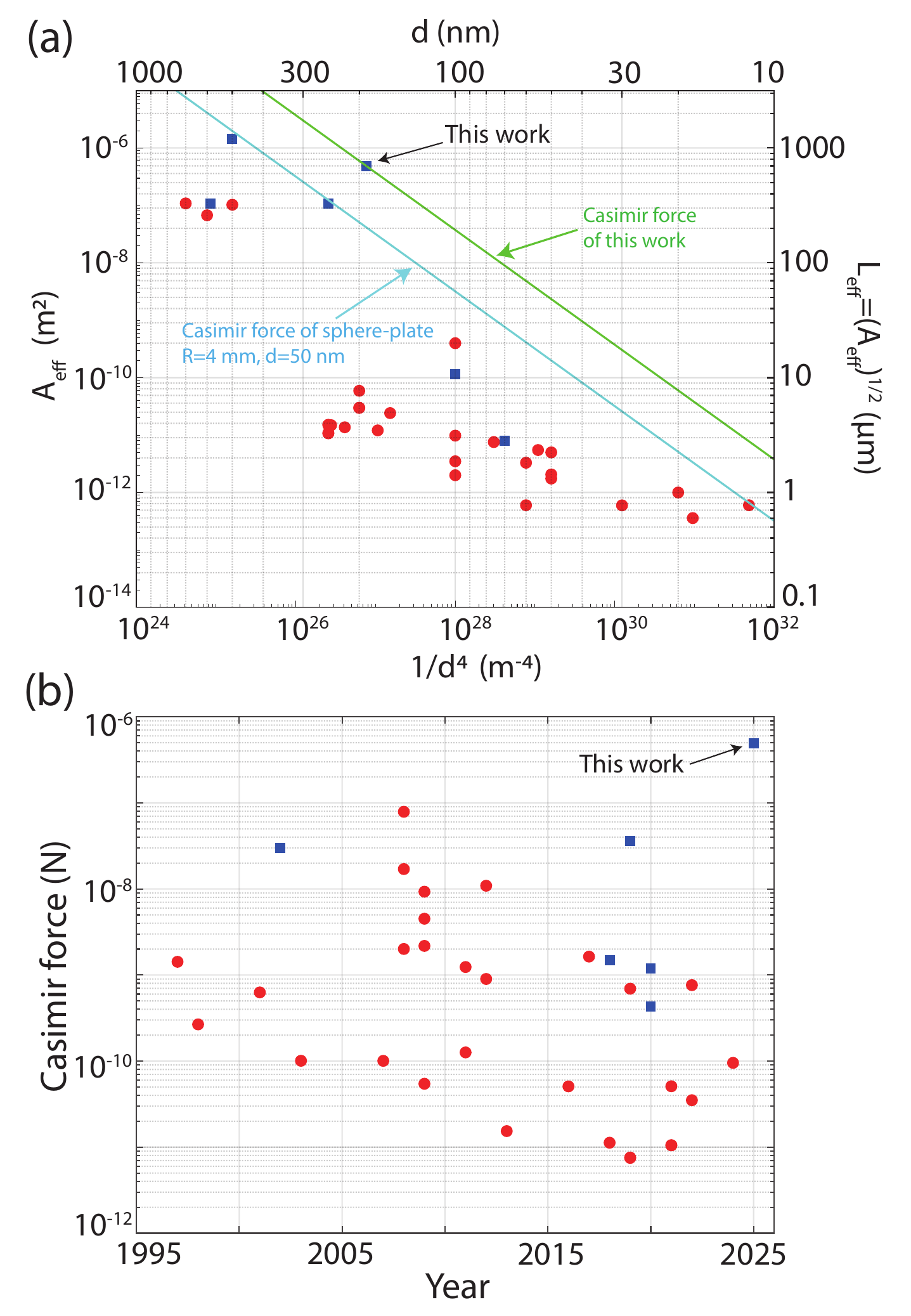}
\caption{Studies on Casimir effects. \textbf{(a)} The results are sorted out according to their minimum separation and effective areas. Data points further to the top right correspond a stronger Casimir force. \textbf{(b)} The results are sorted out according to their publication year, the vertical axis represents their corresponding Casimir force shown in log scale. The blue squares represent plate-plate experiments, and the red circles represent sphere-plate experiments.}
\label{fig_supp_CasimirYearLog}
\end{figure*}

\begin{table*}[h]\label{tab:CasmirPlateplate}
\centering
\caption{
Casimir experiments with plate-plate configuration. ( * radius of circular drum)} 
{\renewcommand{\arraystretch}{1.3}%
\begin{tabular}{c|c|c|c|c|c}
  \hline\hline
 Ref & Size (um) & Area (m$^2$) &  Min Sep (nm) &  Force (N) & Year  \\
   \hline 
\cite{bressi2002measurement} & $1200\times 1200$ & $1.44\times 10^{-6}$ & 500 & $2.99718\times 10^{-8}$ & 2002\\ \hline 
\cite{norte2018platform} & $384\times0.3$ & $1.152\times 10^{-10}$ & 100 & $1.49859\times 10^{-9}$ & 2018\\ \hline 
\cite{fong2019phonon} & $330\times330$ & $1.089\times 10^{-7}$ & 250 & $3.62659\times 10^{-8}$ & 2019\\ \hline 
\cite{perez2020system} & $20\times0.40$ & $8.0\times 10^{-12}$ & 70 & $4.3344\times 10^{-10}$ & 2020\\ \hline 
\cite{pate2020casimir} & R=185 $^*$ & $1.07518\times 10^{-7}$ & 585 & $1.19423\times 10^{-9}$ & 2020\\ \hline 
This work & $709\times 709$ & $4.9\times 10^{-7}$ & 190 & $4.89117\times 10^{-7}$ & 2025 \\
  \hline\hline
  Average (except this work) &  &  &  & $1.3873\times 10^{-8}$ &  \\ \hline
  Median (except this work) &  &  &  & $1.4986\times 10^{-9}$ &  \\
  \hline\hline
\end{tabular}}
\end{table*}

\begin{table*}\label{tab:CasmirSphereplate}
\centering
\caption{Casimir experiments with sphere-plate configuration. The effective area of the sphere-plate configuration is defined as $A_{\rm eff}=R/d_{\rm min}$, where $R$ is the radius of the sphere, and $d_{\rm min}$ is the minimum separation. The Casimir force calculated in this table is assuming the configurations are made of perfectly conducting plates, for a better comparison of their geometrical sizes, regardless of the materials.}
\vspace{1mm}
{\renewcommand{\arraystretch}{1.3}%
\begin{tabular}{c|c|c|c|c|c}
  \hline\hline
 Ref & Radius (um) & Effective Area (m$^2$) &  Min Sep (nm) & Force (N) & Year  \\
   \hline 
\cite{lamoreaux1997demonstration} & 113000 & $6.78\times 10^{-8}$ & 600 & $1.42528\times 10^{-9}$ & 1997\\ \hline
\cite{mohideen1998precision} & 98 & $9.8\times 10^{-12}$ & 100 & $2.66995\times 10^{-10}$ & 1998\\ \hline
\cite{chan2001quantum} & 100 & $7.57\times 10^{-12}$ & 75.7 & $6.28042\times 10^{-10}$ & 2001\\ \hline
\cite{decca2003measurement} & 296 & $5.92\times 10^{-11}$ & 200 & $1.00804\times 10^{-10}$ & 2003\\ \hline
\cite{decca2007tests} & 151.3 & $2.4208\times 10^{-11}$ & 160 & $1.00636\times 10^{-10}$ & 2007\\ \hline
\cite{van2008measurement} & 50 & $6.0\times 10^{-13}$ & 12 & $7.8832\times 10^{-8}$ & 2008\\ \hline
\cite{munday2008measurements} & 19.9 & $5.97\times 10^{-13}$ & 30 & $2.00801\times 10^{-9}$ & 2008\\ \hline
\cite{van2008influence} & 50 & $1.0\times 10^{-12}$ & 20 & $1.70277\times 10^{-8}$ & 2008\\ \hline
\cite{jourdan2009quantitative} & 20 & $2.0\times 10^{-12}$ & 100 & $5.44887\times 10^{-11}$ & 2009\\ \hline
\cite{de2009halving} & 100 & $5.0\times 10^{-12}$ & 50 & $2.17955\times 10^{-9}$ & 2009\\ \hline
\cite{masuda2009limits} & 207000 & $1.035\times 10^{-7}$ & 500 & $4.51167\times 10^{-9}$ & 2009\\ \hline
\cite{munday2009measured} & 19.9 & $3.582\times 10^{-13}$ & 18 & $9.29634\times 10^{-9}$ & 2009\\ \hline
\cite{torricelli2011casimir} & 10 & $6.0\times 10^{-13}$ & 60 & $1.26131\times 10^{-10}$ & 2011\\ \hline
\cite{sushkov2011observation} & 156000 & $1.092\times 10^{-7}$ & 700 & $1.2391\times 10^{-9}$ & 2011\\ \hline
\cite{chang2012gradient} & 41.3 & $2.065\times 10^{-12}$ & 50 & $9.00153\times 10^{-10}$ & 2012\\ \hline
\cite{garcia2012casimir} & 4000 & $4.0\times 10^{-10}$ & 100 & $1.08977\times 10^{-8}$ & 2012\\ \hline
\cite{banishev2013demonstration} & 61.7 & $1.36974\times 10^{-11}$ & 222 & $1.53639\times 10^{-11}$ & 2013\\ \hline
\cite{bimonte2016isoelectronic} & 149.3 & $2.986\times 10^{-11}$ & 200 & $5.08448\times 10^{-11}$ & 2016\\ \hline
\cite{eerkens2017investigations} & 100 & $5.5\times 10^{-12}$ & 55 & $1.63753\times 10^{-9}$ & 2017\\ \hline
\cite{xu2018reducing} & 60.8 & $1.4896\times 10^{-11}$ & 245 & $1.12637\times 10^{-11}$ & 2018\\ \hline
\cite{liu2019examining} & 43.446 & $1.08615\times 10^{-11}$ & 250 & $7.57541\times 10^{-12}$ & 2019\\ \hline
\cite{stange2019building} & 55 & $3.3\times 10^{-12}$ & 60 & $6.93722\times 10^{-10}$ & 2019\\ \hline
\cite{liu2019precision} & 43 & $1.075\times 10^{-11}$ & 250 & $7.49765\times 10^{-12}$ & 2019\\ \hline
\cite{liu2021demonstration} & 60.35 & $1.50875\times 10^{-11}$ & 250 & $1.05229\times 10^{-11}$ & 2021\\ \hline
\cite{liu2021experimental} & 60.35 & $1.50875\times 10^{-11}$ & 250 & $1.05229\times 10^{-11}$ & 2021\\ \hline
\cite{bimonte2021measurement} & 149.7 & $2.994\times 10^{-11}$ & 200 & $5.0981\times 10^{-11}$ & 2021\\ \hline
\cite{xu2022non} & 69.1 & $1.20925\times 10^{-11}$ & 175 & $3.51269\times 10^{-11}$ & 2022\\ \hline
\cite{xu2022observation} & 35 & $1.75\times 10^{-12}$ & 50 & $7.62842\times 10^{-10}$ & 2022\\ \hline
\cite{xu2024observation} & 35 & $3.5\times 10^{-12}$ & 100 & $9.53552\times 10^{-11}$ & 2024\\ \hline
  \hline\hline
Average &   &  &   & $4.5856\times 10^{-9}$ &  \\ \hline
Median &   &  &   & $2.6700\times 10^{-10}$ &  \\ 
  \hline\hline
\end{tabular}}
\end{table*}




\begin{thebibliography}{118}%
\makeatletter
\providecommand \@ifxundefined [1]{%
 \@ifx{#1\undefined}
}%
\providecommand \@ifnum [1]{%
 \ifnum #1\expandafter \@firstoftwo
 \else \expandafter \@secondoftwo
 \fi
}%
\providecommand \@ifx [1]{%
 \ifx #1\expandafter \@firstoftwo
 \else \expandafter \@secondoftwo
 \fi
}%
\providecommand \natexlab [1]{#1}%
\providecommand \enquote  [1]{``#1''}%
\providecommand \bibnamefont  [1]{#1}%
\providecommand \bibfnamefont [1]{#1}%
\providecommand \citenamefont [1]{#1}%
\providecommand \href@noop [0]{\@secondoftwo}%
\providecommand \href [0]{\begingroup \@sanitize@url \@href}%
\providecommand \@href[1]{\@@startlink{#1}\@@href}%
\providecommand \@@href[1]{\endgroup#1\@@endlink}%
\providecommand \@sanitize@url [0]{\catcode `\\12\catcode `\$12\catcode
  `\&12\catcode `\#12\catcode `\^12\catcode `\_12\catcode `\%12\relax}%
\providecommand \@@startlink[1]{}%
\providecommand \@@endlink[0]{}%
\providecommand \url  [0]{\begingroup\@sanitize@url \@url }%
\providecommand \@url [1]{\endgroup\@href {#1}{\urlprefix }}%
\providecommand \urlprefix  [0]{URL }%
\providecommand \Eprint [0]{\href }%
\providecommand \doibase [0]{https://doi.org/}%
\providecommand \selectlanguage [0]{\@gobble}%
\providecommand \bibinfo  [0]{\@secondoftwo}%
\providecommand \bibfield  [0]{\@secondoftwo}%
\providecommand \translation [1]{[#1]}%
\providecommand \BibitemOpen [0]{}%
\providecommand \bibitemStop [0]{}%
\providecommand \bibitemNoStop [0]{.\EOS\space}%
\providecommand \EOS [0]{\spacefactor3000\relax}%
\providecommand \BibitemShut  [1]{\csname bibitem#1\endcsname}%
\let\auto@bib@innerbib\@empty

\bibitem{casimir1948attraction}
H.~B. Casimir,
\newblock In \emph{Proc. Kon. Ned. Akad. Wet.}, volume~51. \textbf{1948} 793.

\bibitem{lamoreaux1997demonstration}
S.~K. Lamoreaux,
\newblock \emph{Physical Review Letters} \textbf{1997}, \emph{78}, 1 5.

\bibitem{bardeen1957theory}
J.~Bardeen, L.~N. Cooper, J.~R. Schrieffer,
\newblock \emph{Physical review} \textbf{1957}, \emph{108}, 5 1175.

\bibitem{gong2022electrically}
T.~Gong, B.~Spreng, M.~Camacho, I.~Liberal, N.~Engheta, J.~N. Munday,
\newblock \emph{Physical Review A} \textbf{2022}, \emph{106}, 6 062824.

\bibitem{torricelli2010switching}
G.~Torricelli, P.~Van~Zwol, O.~Shpak, C.~Binns, G.~Palasantzas, B.~Kooi, V.~Svetovoy, M.~Wuttig,
\newblock \emph{Physical Review A—Atomic, Molecular, and Optical Physics} \textbf{2010}, \emph{82}, 1 010101.

\bibitem{bressi2002measurement}
G.~Bressi, G.~Carugno, R.~Onofrio, G.~Ruoso,
\newblock \emph{Physical review letters} \textbf{2002}, \emph{88}, 4 041804.

\bibitem{fong2019phonon}
K.~Y. Fong, H.-K. Li, R.~Zhao, S.~Yang, Y.~Wang, X.~Zhang,
\newblock \emph{Nature} \textbf{2019}, \emph{576}, 7786 243.

\bibitem{vsivskins2020magnetic}
M.~{\v{S}}i{\v{s}}kins, M.~Lee, S.~Ma{\~n}as-Valero, E.~Coronado, Y.~M. Blanter, H.~S. van~der Zant, P.~G. Steeneken,
\newblock \emph{Nature communications} \textbf{2020}, \emph{11}, 1 2698.

\bibitem{suchoi2017damping}
O.~Suchoi, E.~Buks,
\newblock \emph{Europhysics Letters} \textbf{2017}, \emph{117}, 5 57008.

\bibitem{binek2017elastic}
C.~Binek,
\newblock \emph{Scientific Reports} \textbf{2017}, \emph{7}, 1 1.

\bibitem{garcia2012casimir}
D.~Garcia-Sanchez, K.~Y. Fong, H.~Bhaskaran, S.~Lamoreaux, H.~X. Tang,
\newblock \emph{Physical Review Letters} \textbf{2012}, \emph{109}, 2 027202.

\bibitem{liu2020elimination}
M.~Liu, R.~Schafer, J.~Xu, U.~Mohideen,
\newblock \emph{Modern Physics Letters A} \textbf{2020}, \emph{35}, 03 2040001.

\bibitem{garrett2020measuring}
J.~L. Garrett, J.~Kim, J.~N. Munday,
\newblock \emph{Physical Review Research} \textbf{2020}, \emph{2}, 2 023355.

\bibitem{bimonte2019casimir}
G.~Bimonte,
\newblock \emph{Physical Review A} \textbf{2019}, \emph{99}, 5 052507.

\bibitem{norte2018platform}
R.~A. Norte, M.~Forsch, A.~Wallucks, I.~Marinkovi{\'c}, S.~Gr{\"o}blacher,
\newblock \emph{Physical review letters} \textbf{2018}, \emph{121}, 3 030405.

\bibitem{eerkens2017investigations}
H.~Eerkens,
\newblock Ph.D. thesis, Ph. D. thesis, Leiden University, \textbf{2017}.

\bibitem{rodriguez2015classical}
A.~W. Rodriguez, P.-C. Hui, D.~P. Woolf, S.~G. Johnson, M.~Lon{\v{c}}ar, F.~Capasso,
\newblock \emph{Annalen der Physik} \textbf{2015}, \emph{527}, 1-2 45.

\bibitem{andrews2015quantum}
R.~Andrews, A.~Reed, K.~Cicak, J.~Teufel, K.~Lehnert,
\newblock \emph{Nature communications} \textbf{2015}, \emph{6}, 1 10021.

\bibitem{tang2017measurement}
L.~Tang, M.~Wang, C.~Ng, M.~Nikolic, C.~T. Chan, A.~W. Rodriguez, H.~B. Chan,
\newblock \emph{Nature Photonics} \textbf{2017}, \emph{11}, 2 97.

\bibitem{wang2021strong}
M.~Wang, L.~Tang, C.~Ng, R.~Messina, B.~Guizal, J.~Crosse, M.~Antezza, C.~T. Chan, H.~B. Chan,
\newblock \emph{Nature communications} \textbf{2021}, \emph{12}, 1 600.

\bibitem{perez2020system}
D.~P{\'e}rez-Morelo, A.~Stange, R.~W. Lally, L.~K. Barrett, M.~Imboden, A.~Som, D.~K. Campbell, V.~A. Aksyuk, D.~J. Bishop,
\newblock \emph{Microsystems \& nanoengineering} \textbf{2020}, \emph{6}, 1 115.

\bibitem{pate2020casimir}
J.~M. Pate, M.~Goryachev, R.~Y. Chiao, J.~E. Sharping, M.~E. Tobar,
\newblock \emph{Nature Physics} \textbf{2020}, \emph{16}, 11 1117.

\bibitem{de2025measurement}
M.~H. de~Jong, E.~Korkmazgil, L.~Banniard, M.~A. Sillanp{\"a}{\"a}, L.~M. de~L{\'e}pinay,
\newblock \emph{arXiv preprint arXiv:2501.13759} \textbf{2025}.

\bibitem{bimonte2008low}
G.~Bimonte, D.~Born, E.~Calloni, G.~Esposito, U.~Huebner, E.~Il'ichev, L.~Rosa, F.~Tafuri, R.~Vaglio,
\newblock \emph{Journal of Physics A: Mathematical and Theoretical} \textbf{2008}, \emph{41}, 16 164023.

\bibitem{van2008measurement}
P.~Van~Zwol, G.~Palasantzas, M.~Van De~Schootbrugge, J.~T.~M. De~Hosson,
\newblock \emph{Applied Physics Letters} \textbf{2008}, \emph{92}, 5.

\bibitem{schmid2011damping}
S.~Schmid, K.~Jensen, K.~Nielsen, A.~Boisen,
\newblock \emph{Physical Review B—Condensed Matter and Materials Physics} \textbf{2011}, \emph{84}, 16 165307.

\bibitem{sansa2016frequency}
M.~Sansa, E.~Sage, E.~C. Bullard, M.~G{\'e}ly, T.~Alava, E.~Colinet, A.~K. Naik, L.~G. Villanueva, L.~Duraffourg, M.~L. Roukes, et~al.,
\newblock \emph{Nature nanotechnology} \textbf{2016}, \emph{11}, 6 552.

\bibitem{azgin2011resonant}
K.~Azgin, C.~Ro, A.~Torrents, T.~Akin, L.~Valdevit,
\newblock In \emph{2011 IEEE 24th International Conference on Micro Electro Mechanical Systems}. IEEE, \textbf{2011} 545--548.

\bibitem{salimi2024squeeze}
M.~Salimi, R.~V. Nielsen, H.~B. Pedersen, A.~Dantan,
\newblock \emph{Sensors and Actuators A: Physical} \textbf{2024}, \emph{374} 115450.

\bibitem{bimonte2025proposal}
G.~Bimonte,
\newblock \emph{Physical Review B} \textbf{2025}, \emph{111}, 17 174512.

\bibitem{klimchitskaya2023casimir}
G.~L. Klimchitskaya, V.~M. Mostepanenko,
\newblock \emph{Physics} \textbf{2023}, \emph{5}, 4 952.

\bibitem{klimchitskaya2022theory}
G.~Klimchitskaya, V.~Mostepanenko,
\newblock \emph{Physical Review A} \textbf{2022}, \emph{105}, 1 012805.

\bibitem{villarreal2019casimir}
C.~Villarreal, S.~F. Caballero-Benitez,
\newblock \emph{Physical Review A} \textbf{2019}, \emph{100}, 4 042504.

\bibitem{kempf2008casimir}
A.~Kempf,
\newblock \emph{Journal of Physics A: Mathematical and Theoretical} \textbf{2008}, \emph{41}, 16 164038.

\bibitem{orlando2018correlation}
M.~Orlando, A.~Rouver, J.~Rocha, A.~Cavichini,
\newblock \emph{Physics Letters A} \textbf{2018}, \emph{382}, 22 1486.

\bibitem{strongin1968enhanced}
M.~Strongin, O.~Kammerer, J.~Crow, R.~Parks, D.~Douglass~Jr, M.~Jensen,
\newblock \emph{Physical Review Letters} \textbf{1968}, \emph{21}, 18 1320.

\bibitem{strongin1968superconductive}
M.~Strongin, O.~Kammerer,
\newblock \emph{Journal of Applied Physics} \textbf{1968}, \emph{39}, 6 2509.

\bibitem{avino2018archimedes}
S.~Avino, A.~Basti, E.~Calloni, S.~Caprara, M.~de~Laurentis, R.~de~Rosa, L.~Errico, G.~Esposito, F.~Frasconi, G.~Gagliardi, et~al.,
\newblock In \emph{Gravitational-waves Science\&Technology Symposium}. \textbf{2018} 020.

\bibitem{calloni2002vacuum}
E.~Calloni, L.~Di~Fiore, G.~Esposito, L.~Milano, L.~Rosa,
\newblock \emph{Physics Letters A} \textbf{2002}, \emph{297}, 5-6 328.

\bibitem{fulling2007does}
S.~A. Fulling, K.~A. Milton, P.~Parashar, A.~Romeo, K.~Shajesh, J.~Wagner,
\newblock \emph{Physical Review D—Particles, Fields, Gravitation, and Cosmology} \textbf{2007}, \emph{76}, 2 025004.

\bibitem{bimonte2006energy}
G.~Bimonte, E.~Calloni, G.~Esposito, L.~Rosa,
\newblock \emph{Physical Review D—Particles, Fields, Gravitation, and Cosmology} \textbf{2006}, \emph{74}, 8 085011.

\bibitem{quach2015gravitational}
J.~Q. Quach,
\newblock \emph{Physical Review Letters} \textbf{2015}, \emph{114}, 8 081104, \\Note: While the published version of this work suggested that gravitational Casimir forces could exceed electromagnetic Casimir forces, follow-up analysis and private communication with the authors indicate that key aspects of the model remain unresolved. Current estimates, based on revised assumptions about reflectivity and cutoff frequencies, suggest the effect may be several orders of magnitude smaller and perhaps immeasurable with modern experimental techniques.

\bibitem{elbertse2026detection}
R.~Elbertse, M.~Xu, A.~Ke{\c{s}}kekler, S.~Otte, R.~Norte,
\newblock \emph{arXiv preprint arXiv:2601.01074} \textbf{2026}.

\bibitem{bimonte2017nonequilibrium}
G.~Bimonte, T.~Emig, M.~Kardar, M.~Kr{\"u}ger,
\newblock \emph{Annual Review of Condensed Matter Physics} \textbf{2017}, \emph{8}, 1 119.

\bibitem{andrews2014bidirectional}
R.~W. Andrews, R.~W. Peterson, T.~P. Purdy, K.~Cicak, R.~W. Simmonds, C.~A. Regal, K.~W. Lehnert,
\newblock \emph{Nature physics} \textbf{2014}, \emph{10}, 4 321.

\bibitem{Youssefi2025}
A.~Youssefi, M.~Chegnizadeh, M.~Scigliuzzo, T.~J. Kippenberg,
\newblock \emph{arXiv preprint} \textbf{2025}, \emph{arXiv:2501.03211}.

\bibitem{seis2022ground}
Y.~Seis, T.~Capelle, E.~Langman, S.~Saarinen, E.~Planz, A.~Schliesser,
\newblock \emph{Nature communications} \textbf{2022}, \emph{13}, 1 1507.

\bibitem{halg2021membrane}
D.~H{\"a}lg, T.~Gisler, Y.~Tsaturyan, L.~Catalini, U.~Grob, M.-D. Krass, M.~H{\'e}ritier, H.~Mattiat, A.-K. Thamm, R.~Schirhagl, et~al.,
\newblock \emph{Physical Review Applied} \textbf{2021}, \emph{15}, 2 L021001.




\bibitem{revie2008corrosion}
R.~W. Revie,
\newblock \emph{Corrosion and corrosion control: an introduction to corrosion science and engineering},
\newblock John Wiley \& Sons, \textbf{2008}.

\bibitem{zhang2018characterization}
L.~Zhang, L.~You, L.~Ying, W.~Peng, Z.~Wang,
\newblock \emph{Physica C: Superconductivity and its applications} \textbf{2018}, \emph{545} 1.

\bibitem{maier1995new}
D.~Maier-Schneider, J.~Maibach, E.~Obermeier,
\newblock \emph{Journal of microelectromechanical systems} \textbf{1995}, \emph{4}, 4 238.



\bibitem{khestanova2018unusual}
E.~Khestanova, J.~Birkbeck, M.~Zhu, Y.~Cao, G.~Yu, D.~Ghazaryan, J.~Yin, H.~Berger, L.~Forro, T.~Taniguchi, et~al.,
\newblock \emph{Nano letters} \textbf{2018}, \emph{18}, 4 2623.

\bibitem{schmid2016fundamentals}
S.~Schmid, L.~G. Villanueva, M.~L. Roukes,
\newblock \emph{Fundamentals of nanomechanical resonators}, volume~49,
\newblock Springer, \textbf{2016}.

\bibitem{derjaguin1934untersuchungen}
B.~Derjaguin,
\newblock \emph{Kolloid-Zeitschrift} \textbf{1934}, \emph{69} 155.



\bibitem{behunin2012modeling}
R.~Behunin, F.~Intravaia, D.~Dalvit, P.~M. Neto, S.~Reynaud,
\newblock \emph{Physical Review A—Atomic, Molecular, and Optical Physics} \textbf{2012}, \emph{85}, 1 012504.

\bibitem{zhang2020radiative}
C.~Zhang, M.~Giroux, T.~A. Nour, R.~St-Gelais,
\newblock \emph{Physical Review Applied} \textbf{2020}, \emph{14}, 2 024072.

\bibitem{smith1976thermal}
T.~Smith, T.~Finlayson,
\newblock \emph{Journal of Physics F: Metal Physics} \textbf{1976}, \emph{6}, 5 709.

\bibitem{simpson1978thermal}
M.~Simpson, T.~Smith,
\newblock \emph{J. Low Temp. Phys.;(United States)} \textbf{1978}, \emph{32}, 1.

\bibitem{nelson2013atomic}
N.~Nelson-Fitzpatrick, C.~Guthy, S.~Poshtiban, E.~Finley, K.~D. Harris, B.~J. Worfolk, S.~Evoy,
\newblock \emph{Journal of Vacuum Science \& Technology A} \textbf{2013}, \emph{31}, 2.

\bibitem{zhang2020mechanical}
Y.~Zhang, D.~Sun, J.~Cheng, J.~K.~H. Tsoi, J.~Chen,
\newblock \emph{Regenerative biomaterials} \textbf{2020}, \emph{7}, 1 119.



\bibitem{mattis1958theory}
D.~C. Mattis, J.~Bardeen,
\newblock \emph{Physical Review} \textbf{1958}, \emph{111}, 2 412.

\bibitem{tinkham2004introduction}
M.~Tinkham,
\newblock \emph{Introduction to superconductivity},
\newblock Courier Corporation, \textbf{2004}.

\bibitem{hong2013terahertz}
T.~Hong, K.~Choi, K.~Ik~Sim, T.~Ha, B.~Cheol~Park, H.~Yamamori, J.~Hoon~Kim,
\newblock \emph{Journal of Applied Physics} \textbf{2013}, \emph{114}, 24.

\bibitem{bimonte2010optical}
G.~Bimonte, H.~Haakh, C.~Henkel, F.~Intravaia,
\newblock \emph{Journal of Physics A: Mathematical and Theoretical} \textbf{2010}, \emph{43}, 14 145304.


\bibitem{brevik2006thermal}
I.~Brevik, S.~A. Ellingsen, K.~A. Milton,
\newblock \emph{New Journal of Physics} \textbf{2006}, \emph{8}, 10 236.

\bibitem{svetovoy2006casimir}
V.~Svetovoy, R.~Esquivel,
\newblock \emph{Journal of physics A: mathematical and general} \textbf{2006}, \emph{39}, 21 6777.

\bibitem{bordag2009advances}
M.~Bordag, G.~L. Klimchitskaya, U.~Mohideen, V.~M. Mostepanenko,
\newblock \emph{Advances in the Casimir effect}, volume 145,
\newblock OUP Oxford, \textbf{2009}.

\bibitem{decca2007novel}
R.~Decca, D.~L{\'o}pez, E.~Fischbach, G.~Klimchitskaya, D.~Krause, V.~Mostepanenko,
\newblock \emph{The European Physical Journal C} \textbf{2007}, \emph{51} 963.

\bibitem{chang2012gradient}
C.-C. Chang, A.~Banishev, R.~Castillo-Garza, G.~Klimchitskaya, V.~Mostepanenko, U.~Mohideen,
\newblock \emph{Physical Review B—Condensed Matter and Materials Physics} \textbf{2012}, \emph{85}, 16 165443.

\bibitem{banishev2013demonstration}
A.~Banishev, G.~Klimchitskaya, V.~Mostepanenko, U.~Mohideen,
\newblock \emph{Physical Review Letters} \textbf{2013}, \emph{110}, 13 137401.

\bibitem{bimonte2016isoelectronic}
G.~Bimonte, D.~L{\'o}pez, R.~S. Decca,
\newblock \emph{Physical Review B} \textbf{2016}, \emph{93}, 18 184434.

\bibitem{bimonte2021measurement}
G.~Bimonte, B.~Spreng, P.~A. Maia~Neto, G.-L. Ingold, G.~L. Klimchitskaya, V.~M. Mostepanenko, R.~S. Decca,
\newblock \emph{Universe} \textbf{2021}, \emph{7}, 4 93.

\bibitem{mostepanenko2021casimir}
V.~M. Mostepanenko,
\newblock \emph{Universe} \textbf{2021}, \emph{7}, 4 84.




\bibitem{mohideen1998precision}
U.~Mohideen, A.~Roy,
\newblock \emph{Physical Review Letters} \textbf{1998}, \emph{81}, 21 4549.

\bibitem{chan2001quantum}
H.~B. Chan, V.~A. Aksyuk, R.~N. Kleiman, D.~J. Bishop, F.~Capasso,
\newblock \emph{Science} \textbf{2001}, \emph{291}, 5510 1941.

\bibitem{decca2003measurement}
R.~Decca, D.~L{\'o}pez, E.~Fischbach, D.~Krause,
\newblock \emph{Physical review letters} \textbf{2003}, \emph{91}, 5 050402.

\bibitem{decca2007tests}
R.~Decca, D.~L{\'o}pez, E.~Fischbach, G.~Klimchitskaya, D.~Krause, V.~Mostepanenko,
\newblock \emph{Physical Review D—Particles, Fields, Gravitation, and Cosmology} \textbf{2007}, \emph{75}, 7 077101.


\bibitem{munday2008measurements}
J.~Munday, F.~Capasso, V.~A. Parsegian, S.~M. Bezrukov,
\newblock \emph{Physical Review A—Atomic, Molecular, and Optical Physics} \textbf{2008}, \emph{78}, 3 032109.

\bibitem{van2008influence}
P.~Van~Zwol, G.~Palasantzas, J.~T.~M. De~Hosson,
\newblock \emph{Physical Review B—Condensed Matter and Materials Physics} \textbf{2008}, \emph{77}, 7 075412.

\bibitem{jourdan2009quantitative}
G.~Jourdan, A.~Lambrecht, F.~Comin, J.~Chevrier,
\newblock \emph{Europhysics Letters} \textbf{2009}, \emph{85}, 3 31001.

\bibitem{de2009halving}
S.~De~Man, K.~Heeck, R.~Wijngaarden, D.~Iannuzzi,
\newblock \emph{Physical review letters} \textbf{2009}, \emph{103}, 4 040402.

\bibitem{masuda2009limits}
M.~Masuda, M.~Sasaki,
\newblock \emph{Physical review letters} \textbf{2009}, \emph{102}, 17 171101.

\bibitem{munday2009measured}
J.~N. Munday, F.~Capasso, V.~A. Parsegian,
\newblock \emph{Nature} \textbf{2009}, \emph{457}, 7226 170.

\bibitem{torricelli2011casimir}
G.~Torricelli, I.~Pirozhenko, S.~Thornton, A.~Lambrecht, C.~Binns,
\newblock \emph{Europhysics Letters} \textbf{2011}, \emph{93}, 5 51001.

\bibitem{sushkov2011observation}
A.~Sushkov, W.~Kim, D.~Dalvit, S.~Lamoreaux,
\newblock \emph{Nature Physics} \textbf{2011}, \emph{7}, 3 230.

\bibitem{xu2018reducing}
J.~Xu, G.~Klimchitskaya, V.~Mostepanenko, U.~Mohideen,
\newblock \emph{Physical Review A} \textbf{2018}, \emph{97}, 3 032501.

\bibitem{liu2019examining}
M.~Liu, J.~Xu, G.~Klimchitskaya, V.~Mostepanenko, U.~Mohideen,
\newblock \emph{Physical Review B} \textbf{2019}, \emph{100}, 8 081406.

\bibitem{stange2019building}
A.~Stange, M.~Imboden, J.~Javor, L.~K. Barrett, D.~J. Bishop,
\newblock \emph{Microsystems \& nanoengineering} \textbf{2019}, \emph{5}, 1 14.

\bibitem{liu2019precision}
M.~Liu, J.~Xu, G.~Klimchitskaya, V.~Mostepanenko, U.~Mohideen,
\newblock \emph{Physical Review A} \textbf{2019}, \emph{100}, 5 052511.

\bibitem{liu2021demonstration}
M.~Liu, Y.~Zhang, G.~Klimchitskaya, V.~Mostepanenko, U.~Mohideen,
\newblock \emph{Physical Review Letters} \textbf{2021}, \emph{126}, 20 206802.

\bibitem{liu2021experimental}
M.~Liu, Y.~Zhang, G.~Klimchitskaya, V.~Mostepanenko, U.~Mohideen,
\newblock \emph{Physical Review B} \textbf{2021}, \emph{104}, 8 085436.

\bibitem{xu2022non}
Z.~Xu, X.~Gao, J.~Bang, Z.~Jacob, T.~Li,
\newblock \emph{Nature nanotechnology} \textbf{2022}, \emph{17}, 2 148.

\bibitem{xu2022observation}
Z.~Xu, P.~Ju, X.~Gao, K.~Shen, Z.~Jacob, T.~Li,
\newblock \emph{Nature Communications} \textbf{2022}, \emph{13}, 1 6148.

\bibitem{xu2024observation}
Z.~Xu, P.~Ju, K.~Shen, Y.~Jin, Z.~Jacob, T.~Li,
\newblock \emph{arXiv preprint arXiv:2403.06051} \textbf{2024}.


\end{thebibliography}


\end{widetext}

\end{document}